\documentclass[12pt]{article}
\usepackage{amsfonts,amsthm,amsmath,amssymb,upgreek,bm}
\usepackage{dsfont}
\usepackage[paper=letterpaper,margin=.85in]{geometry}
\usepackage{graphicx}
\usepackage{units}
\usepackage{float}
\usepackage[export]{adjustbox}
\usepackage{bbold}
\usepackage[T1]{fontenc}
\usepackage[shortlabels]{enumitem}

\usepackage{color}
\usepackage[dvipsnames]{xcolor}
\definecolor{darkblue}{rgb}{0.1,0.1,.7}
\definecolor{purple}{rgb}{0.6,0,0.6}
\definecolor{orange}{rgb}{0.9,0.6,0}
\usepackage[colorlinks, linkcolor=darkblue, citecolor=darkblue, urlcolor=darkblue, linktocpage]{hyperref} 
\usepackage[square, comma, compress,numbers]{natbib}
\usepackage{physics}
\usepackage{tensor}
\usepackage{psfrag}
\usepackage{footmisc}
\usepackage{url}
\usepackage{mathtools}
\usepackage{tikz}
    \usepackage{amssymb,amsfonts,amsmath}
    \usepackage{tkz-euclide}
        \usetikzlibrary{arrows,calc,patterns}
\usepackage{pgfplots}
\usepackage[]{amsmath}
\usepackage[]{graphicx}
\usepackage[]{latexsym}
\usepackage[utf8]{inputenc}
\usepackage{geometry}
\usepackage{amscd}
\usepackage[all,cmtip]{xy}
\usepackage{mathrsfs}
\usepackage{hyperref}
\usepackage{cleveref}
\crefname{equation}{Eq.}{Eqs.}
\Crefname{equation}{Eq.}{Eqs.}

\usepackage[most]{tcolorbox}
\usepackage{tikz}
\usepackage{subcaption}

\numberwithin{equation}{section}
\newcommand{\be}{\begin{equation}}
\newcommand{\ee}{\end{equation}}

\usetikzlibrary{arrows,positioning}

\def\O{{\cal O}}

\def\SL2{\widetilde{SL}(2,\mathbb R)}

\def\la{\langle}
\def\ra{\rangle}

\numberwithin{equation}{section}

\newcommand {\bes} {\begin {equation*}}
\newcommand {\ees} {\end {equation*}}
\newcommand {\beq} {\begin {equation}}
\newcommand {\eeq} {\end {equation}}
\newcommand {\bea} {\begin {eqnarray}}
\newcommand {\ea} {\end {eqnarray}}
\newcommand {\eea} {\end {eqnarray}}

\newcommand{\SO}{\mathrm{SO}}
\newcommand{\SU}{\mathrm{SU}}

\newcommand{\bbT}{\mathbb{T}}

\newcommand{\CF}{\mathcal{F}}
\newcommand{\CL}{\mathcal{L}}
\renewcommand{\CD}{\mathcal{D}}
\newcommand{\CK}{\mathcal{K}}

\newcommand{\cY}{\mathcal{Y}}

\renewcommand{\dd}{\mathrm{d}}
\newcommand{\symp}[2]{\left\langle #1,#2\right\rangle}

\numberwithin{equation}{section}

\renewcommand{\i}{{\rm i}}

\newcommand{\CN}{\mathcal{N}}

\def\<{\langle}
\def\>{\rangle}

\tikzset{
    >=stealth',
    punkt/.style={
           rectangle,
           rounded corners,
           draw=black, very thick,
           text width=15em,
           minimum height=2em,
           text centered},
    pil/.style={
           ->,
           thick,
           shorten <=2pt,
           shorten >=2pt,}
}

 \def\ie{\begin{equation}\begin{aligned}}
\def\fe{\end{aligned}\end{equation}}

\begin{document}
\thispagestyle{empty}
\vspace*{2.5cm}
\begin{center}

{\LARGE\bfseries   
Proving the OSV conjecture \\ with the gravitational path integral
}

\begin{center}
\vspace{1cm}
 {\bf Nikolay Bobev${}^1$, Jan Boruch${}^{2,3}$, Luca V. Iliesiu${}^2$, Dawid Maskalaniec${}^2$}\\
\end{center}
\bigskip 
${}^1$ Institute for Theoretical Physics and Leuven Gravity Institute, KU Leuven,\\ 
Celestijnenlaan 200D, B-3001 Leuven, Belgium\\[0.15cm]
${}^2$ Leinweber Institute for Theoretical Physics and Department of Physics, \\
University of California, Berkeley, CA 94720, USA\\[0.15cm]
${}^3$ Leinweber Institute for Theoretical Physics at Stanford,\\
382 Via Pueblo, Stanford, CA 94305, USA

\vspace{1.5cm}
{\bf Abstract}
\end{center}

\begin{quotation}
\noindent The OSV conjecture relates the %
helicity supertrace of black holes in $4d$  $\mathcal N=2$ supergravity obtained through a Calabi--Yau compactification of Type IIA string theory to the modulus squared of the topological string partition function evaluated at the attractor point in moduli space for a given set of magnetic black hole charges and electric chemical potentials. We find Euclidean black hole saddles that compute this index to all orders in the 4d supergravity derivative expansion and calculate their action using equivariant localization. 
The gravitational on-shell action localizes entirely to the north and south poles of each black hole, with the two contributions determined by the holomorphic and anti-holomorphic topological string free energies, respectively, evaluated at the attractor values of the moduli. For single-center black hole saddles, this provides a proof of the OSV conjecture to all orders in the topological string coupling. The same localization formula applies also to all saddles that contribute to the helicity supertrace, such as multi-center black holes, which leads to a non-perturbative refinement of the conjecture.

\end{quotation}

\setcounter{page}{0}
\setcounter{tocdepth}{2}
\setcounter{footnote}{0}
\newpage

\newpage
\tableofcontents
\section{Introduction}
\label{sec:intro}

The gravitational path integral (GPI) has become a precision tool for understanding quantum properties of black holes. In non-supersymmetric settings, it has been used to probe fine-grained statistics of the spectrum of black holes \cite{Cotler:2016fpe,Saad:2019lba,Penington:2019kki,Almheiri:2019qdq} and provide approximate counts of the number of black hole microstates \cite{Iliesiu:2020qvm,Balasubramanian:2022gmo}. In the presence of supersymmetry it has been used to reproduce the exact degeneracies of $1/8$-BPS black holes in $\mathcal N=8$ supergravity \cite{Dabholkar:2010uh,Dabholkar:2011ec,Iliesiu:2022kny},\footnote{There are however still several assumptions and subtleties about the GPI that need to be addressed \cite{Sen:2023dps}. } matching previous microstate counts from string theory \cite{Strominger:1996sh,Maldacena:1999bp}. However, in both settings, the GPI is insensitive to the interplay between the possible UV corrections to the gravitational theory and the sum over spacetimes that is expected to be an essential part of the GPI. In the non-supersymmetric setup, such UV effects, which are captured by higher-derivative corrections to the gravitational action, are not under sufficient calculational control to be systematically included in the analysis, whereas in $\mathcal N=8$ supergravity the higher-derivative corrections coming from string theory truncate at tree level \cite{Dabholkar:2011ec}. A useful middle ground to systematically study the role of UV effects and their interplay with the sum over geometries in the GPI is provided by $1/2$-BPS black holes in $\mathcal N=2$ supergravity obtained through a Calabi--Yau (CY) compactification of Type IIA string theory. While in this setting an exact count of black hole microstates is as yet unknown, Ooguri, Strominger and Vafa (OSV) conjectured \cite{Ooguri:2004zv} an exact formula for the black hole partition function in terms of the A-model topological string partition function that should include all stringy effects,
    \begin{align}
        Z_\text{BH}^{\text{OSV}}( \Gamma^I, \mu^I; \Omega_\infty) &\equiv \text{Tr}_{\substack{\text{Fixed }\Gamma^I }} (-1)^F (J_3)^2 e^{-\beta \{Q, Q^\dagger \} + \beta \mu^I \Gamma_I}\nonumber \\ 
        &\overset{?}{=} |Z_\text{top}(t^A, g_{\rm top})|^2 \,.
        \label{eq:black-hole-partition-function-intro}
    \end{align}
The first line above defines an indexed black hole partition function, i.e. a helicity supertrace~\cite{Bachas:1997mc,Kiritsis:1997gu}, in an ensemble where the magnetic charges $\Gamma^{I}$ and electric chemical potentials $\mu^{I}$ of the black hole are fixed, where bosonic and fermionic black hole microstates contribute with opposite signs, and where the vector multiplet scalar moduli are fixed at infinity.\footnote{This is done by fixing a normalized period vector $\Omega|_{r\to \infty} = \Omega_\infty$. Additionally, note that the second line of~\eqref{eq:black-hole-partition-function-intro} is independent of $\Omega_\infty$ as one might expect of an index where wall crossing is neglected \cite{Denef:2000nb,Denef:2007vg}.}   The second line is given by the squared modulus of the A-model topological string partition function $Z_\text{top}(t^A, g_{\rm top})$,  evaluated at the attractor values \cite{Ferrara:1995ih,Strominger:1996kf,Ferrara:1996dd} of the scalar moduli $t^{A}$, which in turn are fixed in terms of the black hole charges and chemical potentials. Since $Z_\text{top}(t^A, g_{\rm top})$ can be expressed in a genus expansion in $ g_{\rm top}$ in terms of topological invariants on the CY \cite{Bershadsky:1993cx,Gopakumar:1998ii,Gopakumar:1998jq}, the conjecture would determine the indexed black hole partition function to arbitrary orders in the topological string coupling $g_{\rm top}$ with each order specified exactly in terms of the inverse string tension $\alpha'$. Since only BPS states contribute to the trace in the first line of \eqref{eq:black-hole-partition-function-intro} and because the action of $\{Q, Q^\dagger\}$ on such states vanishes, we expect the trace to be independent of $\beta$. For this reason, neither   $Z_\text{BH}^{\text{OSV}}( \Gamma^I, \mu^I; \Omega_\infty)$ nor $Z_\text{top}(t^A, g_{\rm top})$ has a $\beta$ label. 

In this paper we will use the GPI to provide a proof of this conjecture to all orders in the topological string coupling and to leading order in $G_N$.  We will do this by explicitly finding the 4d supergravity saddles that compute the indexed partition function \eqref{eq:black-hole-partition-function-intro} to all orders in the supergravity derivative expansion, generalizing solutions previously found in a two-derivative theory~\cite{Iliesiu:2021are, Boruch:2023gfn,Boruch:2025biv}. A key aspect that facilitates our analysis is the fact that the gravitational saddles are regular supersymmetric Euclidean backgrounds with an $S^1_{\beta} \times S^2$ asymptotic boundary and finite inverse temperature $\beta$. This should be contrasted with the Lorentzian extremal geometries that have infinite $\beta$ and whose entropy was historically considered when computing the black hole partition function that enters the OSV relation \cite{LopesCardoso:1998tkj,LopesCardoso:1999cv,LopesCardoso:1999fsj,Sen:2005wa,Mohaupt:2000mj}. This entropy is supposed to approximate the logarithm of the sum of the bosonic and fermionic black hole degeneracies. In contrast, our finite-temperature GPI explicitly computes the difference between these degeneracies, as required by the indexed partition function in the conjecture. The explicit evaluation of the GPI relies on two structural features of 4d conformal supergravity and its supersymmetric solutions. First, as we review below, we use the results in \cite{Adhikari:2025eok} where it was shown that the two-derivative Euclidean Poincar\'e supergravity solutions of~\cite{Boruch:2023gfn,Boruch:2025biv} are supersymmetric off-shell backgrounds of conformal supergravity to all orders in the derivative expansion. Second, we use the formalism of equivariant localization to compute the on-shell action of these saddles \cite{BenettiGenolini:2019jdz,BenettiGenolini:2024lbj,BenettiGenolini:2026qdm,Sparks:2026taw} in terms of the prepotential of the 4d $\mathcal{N}=2$ supergravity theory evaluated at the finite $\beta$ attractor values of the vector multiplet scalars. To arrive at this result we use the important property of the finite $\beta$ Euclidean saddles in~\cite{Boruch:2023gfn,Boruch:2025biv} that the Killing vector constructed out of the Killing spinor on this background has only isolated fixed points, i.e. nuts. For a single supersymmetric black hole the Euclidean finite $\beta$ saddle has two isolated fixed points, the north and south poles of the horizon, which reduces the action of the saddle to a sum of two copies of the supergravity prepotential. It is well-known that the 4d supergravity higher-derivative prepotential is encoded in the free energy of the topological string theory on the CY manifold~\cite{Bershadsky:1993cx,Antoniadis:1993ze}. This thus allows us to derive the relation between the black hole and topological string path integrals~\eqref{eq:black-hole-partition-function-intro} to all orders in the topological string coupling for the contribution of a single black hole saddle to the GPI. It is worth noting that a conceptually similar derivation of the OSV relation in terms of ``north/south pole localization'' factorization was discussed in~\cite{Beasley:2006us,Guica:2007wd} based on previous results in~\cite{Beasley:2005iu}. Our analysis has the novel feature that we work with finite $\beta$ Euclidean supersymmetric solutions which are regular and thus allow for the application of the equivariant localization methods of~\cite{BenettiGenolini:2024lbj,BenettiGenolini:2026qdm,Sparks:2026taw} which in turn eschews the subtleties associated with the GPI in the infinite $\beta$ limit. 

An important aspect of the analysis of the GPI pertains to the choice of ensemble. Two natural boundary conditions of the 4d gravitational path integral are to either fix the electric and magnetic charges of the black hole, i.e. be in the canonical ensemble, or work in a mixed ensemble with fixed magnetic charges and electric chemical potentials. The OSV relation in~\eqref{eq:black-hole-partition-function-intro} is related to the mixed ensemble but with an important subtlety. The helicity supertrace in~\eqref{eq:black-hole-partition-function-intro} is weighted by~$e^{-\beta \{Q,Q^{\dagger}\}}$, while the Euclidean GPI with Dirichlet boundary conditions is weighted by the Hamiltonian $H=\{Q,Q^{\dagger}\}+|\mathcal{Z}|$ where $\mathcal{Z}$ is the central charge of the superalgebra with supercharges $Q$ and $Q^\dagger$. The GPI in the mixed ensemble therefore has a non-trivial dependence on $\beta$ which stems from the fact that $\mathcal{Z}$ is related to the ADM mass of a given gravitational saddle. As we discuss in detail below, this difference between the mixed and the OSV ensemble in~\eqref{eq:black-hole-partition-function-intro} can be remedied by adding appropriate boundary terms in the evaluation of the gravitational action, or equivalently, by changing ensembles via a Laplace transformation and passing through the canonical ensemble.

\begin{figure}[t!]
\centering
\begin{tikzpicture}[
    x=1cm, y=1cm,
    line cap=round,
    line join=round,
    font=\normalsize,
    connection/.style={
        black, line width=0.7pt,
        dash pattern=on 3pt off 4pt
    },
    surface/.style={
        black, line width=0.7pt
    },
    equator/.style={
        black, line width=0.45pt
    },
    hidden/.style={
        black!60, line width=0.4pt,
        dash pattern=on 2pt off 2pt
    },
    rotaxis/.style={
        black!80, line width=0.5pt
    },
    spinarrow/.style={
        black, line width=0.6pt, -stealth,
        preaction={draw=white, line width=2.2pt, -}
    },
    label/.style={inner sep=2pt}
]
\definecolor{northcolor}{RGB}{235,55,40}
\definecolor{southcolor}{RGB}{20,105,235}
 
\def\figspin{1}
\def\figoblateness{1.25}
 
\def\kerrhorizon#1#2#3#4{%
    \pgfmathsetmacro{\klen}{sqrt((#3-#1)^2+(#4-#2)^2)}%
    \pgfmathsetmacro{\kang}{atan2(#4-#2,#3-#1)}%
    \pgfmathsetmacro{\kpol}{0.5*\klen}%
    \pgfmathsetmacro{\keq}{\figoblateness*\kpol}%
    \pgfmathsetmacro{\ktilt}{0.17*\keq}%
    \pgfmathsetmacro{\kturn}{\kang-90}%
    \begin{scope}[shift={(#1,#2)}, rotate=\kturn]
        \draw[rotaxis] (0,0) -- (0,-0.30);
 
        \shade[inner color=white, outer color=black!22]
            (0,\kpol) ellipse[x radius=\keq cm, y radius=\kpol cm];
        \draw[surface]
            (0,\kpol) ellipse[x radius=\keq cm, y radius=\kpol cm];
 
        \draw[hidden] (0,0) -- (0,\klen);
 
        \draw[hidden]
            (\keq,\kpol) arc[start angle=0, end angle=180,
                x radius=\keq cm, y radius=\ktilt cm];
        \draw[equator]
            (-\keq,\kpol) arc[start angle=180, end angle=360,
                x radius=\keq cm, y radius=\ktilt cm];
 
        \pgfmathsetmacro{\astart}{90+15*\figspin}%
        \pgfmathsetmacro{\aend}{90+240*\figspin}%
        \pgfmathsetmacro{\tx}{-\figspin*0.26*sin(\aend)}%
        \pgfmathsetmacro{\ty}{\figspin*0.09*cos(\aend)}%
        \pgfmathsetmacro{\tn}{sqrt(\tx*\tx+\ty*\ty)}%
        \draw[rotaxis] (0,\klen) -- (0,\klen+0.66);
        \draw[spinarrow]
            ({0.26*cos(\astart)},{\klen+0.36+0.09*sin(\astart)})
            arc[start angle=\astart, end angle=\aend,
                x radius=0.26cm, y radius=0.09cm]
            -- ++({0.07*\tx/\tn},{0.07*\ty/\tn});
    \end{scope}%
}
 
\def\figpoledata{a/0.55/1.05/0.20/2.40,
              b/2.60/2.05/2.67/2.95,
              c/3.35/0.15/4.75/0.75}
 
\def\drawpoles#1#2{%
    \foreach \name/\xs/\ys/\xn/\yn in \figpoledata{
        \pgfmathsetmacro{\pang}{atan2(\yn-\ys,\xn-\xs)}%
        \pgfmathsetmacro{\panchorN}{\pang+180}%
        \fill[northcolor] (\xn,\yn) circle[radius=2.1pt];
        \fill[southcolor] (\xs,\ys) circle[radius=2.1pt];
        \node[label, northcolor, anchor=\panchorN,
              shift={(\pang:#1)}] at (\xn,\yn)
            {$Z_{\mathrm{top}}(t_{\name})$};
        \node[label, southcolor, anchor=\pang,
              shift={(\panchorN:#2)}] at (\xs,\ys)
            {$\overline{Z}_{\mathrm{top}}(\bar t_{\name})$};
    }%
}
 
\begin{scope}[xshift=-1.0cm]
    \foreach \name/\xs/\ys/\xn/\yn in \figpoledata{
        \draw[connection] (\xs,\ys) -- (\xn,\yn);
    }
    \drawpoles{0.12}{0.12}
\end{scope}
 
\node[font=\Large] at (5.45,1.45) {$=$};
 
\begin{scope}[xshift=6.85cm]
    \foreach \name/\xs/\ys/\xn/\yn in \figpoledata{
        \kerrhorizon{\xs}{\ys}{\xn}{\yn}
    }
    \drawpoles{0.72}{0.36}
\end{scope}

\end{tikzpicture}
\caption{A cartoon of a time-slice for the finite-temperature rotating saddles relevant to the index. Focusing on the example of three black holes, the left figure represents the time-slice in base-space coordinates, showing the positions of the north poles (\textit{red}) and south poles (\textit{blue}) where the on-shell action gets localized to the topological string free energy as indicated through~\eqref{eq:OSV-refinement-intro}. The right figure is a representation of the time-slice that emphasizes that the dotted lines connecting the poles in base-space are in fact the Euclidean horizons with finite proper area. All black holes rotate simultaneously around different axes of rotation such that fermionic fields are all periodic when going around the thermal circle.}
\label{fig:north_south_top}
\end{figure}
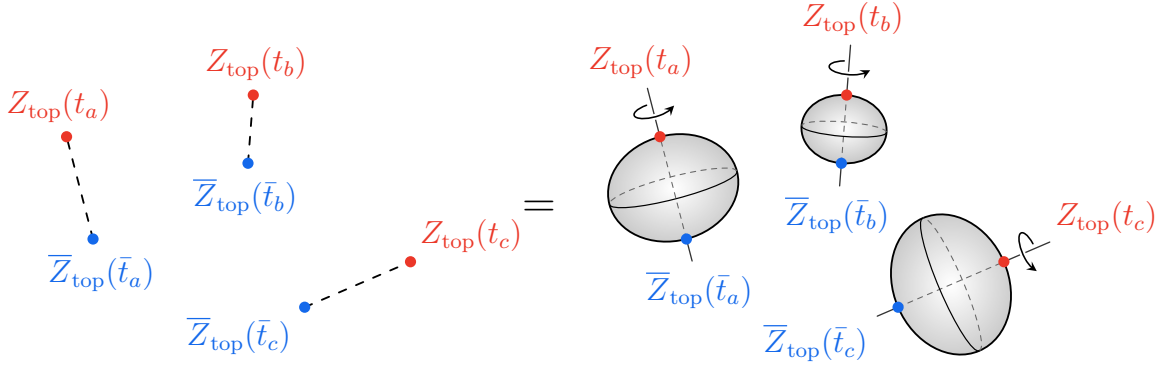

We are not limited to studying the contribution of a single-center black hole saddle. The GPI also allows us to analyze all possible saddles that can contribute to the indexed partition function even if such saddles give a non-perturbatively small contribution to the GPI, suppressed by $e^{-\#S}$ compared to the leading saddle, where $S$ is the entropy associated with the leading saddle and $\#$ is a positive constant.\footnote{We often fix the 4d Newton constant $G_N=1$ in this work. Therefore, the relevant large parameter that controls the semiclassical expansion of the GPI is the black hole entropy $S$. In a slight abuse of notation, we sometimes refer to $1/G_N$ as the large semiclassical parameter.}  Such contributions can come from saddles whose 4d Abelian gauge field configurations can only be specified using additional patches that are responsible for the proper quantization of electric charges in \eqref{eq:black-hole-partition-function-intro}, or from saddles that represent the contribution of multi-center black hole configurations. We can compute the on-shell action of such configurations using equivariant localization in terms of the higher-derivative supergravity prepotential, which in turn leads to a result in terms of the topological string free energy to all orders in the topological string coupling. This allows us to formulate a non-perturbative refinement of the OSV conjecture,
\begin{equation}
\text{OSV refinement:}~~ Z_\text{BH}^{\text{OSV}}(\{\Gamma^I_a\}, \mu^I, \{n^I_a\}; \Omega_\infty) \underset{\substack{\text{saddle-point}\\{\text{approx.}}}}{=} \prod_{a=1}^n Z_\text{top}(t_{N,a}, g_{\text{top}, N, a}) \bar{Z}_\text{top}(t_{S,a}, g_{\text{top}, S, a})\,, 
\label{eq:OSV-refinement-intro}
\end{equation}
where the topological string partition function is evaluated at, 
\begin{align}
    t_{N, a}^A&=\frac{\Gamma^A_a
                       +\frac{\i \beta \mu^A}{\pi} - 2n^A_a}{\Gamma^0_a
                       +\frac{\i \beta \mu^0}{\pi} - 2n^0_a} ,
    &g_{{\rm top},N,a}&=\frac{4\pi\i}{\Gamma^0_a
                       +\frac{\i \beta \mu^0}{\pi} - 2n^0_a}\,,\nonumber
    \\
    t_{S,a}^A&=\frac{\Gamma^A_a
                       -\frac{\i \beta \mu^A}{\pi} + 2n^A_a}{\Gamma^0_a
                       -\frac{\i \beta \mu^0}{\pi} + 2n^0_a} ,
    &g_{{\rm top},S,a}&=\frac{-4\pi\i}{\Gamma^0_a
                       -\frac{\i \beta \mu^0}{\pi} + 2n^0_a},
    \label{eq:OSV-topological-moduli-general-intro}
\end{align}
Above, $Z_\text{BH}^{\text{OSV}}(\{\Gamma^I_a\}, \mu^I, \{n^I_a\}; \Omega_\infty) $ represents the possible contribution of each saddle to indexed black hole partition function $Z_\text{BH}^{\text{OSV}}( \Gamma^I, \mu^I; \Omega_\infty) $ in \eqref{eq:black-hole-partition-function-intro} where $\{\Gamma^I_a\}$ represents the distribution of charges among $n$ black holes in the ensemble with $\sum_{a=1}^n \Gamma_a^I = \Gamma^I$ and $n_a^I$ is a vector of integers that labels the distinct gauge field configurations for the $n$ black hole solutions.\footnote{Whether or not a saddle contributes depends on the value of the scalar moduli at infinity, fixed by $\Omega_\infty$, which determines the chamber in which the helicity supertrace is evaluated. Determining which saddles actually contribute to $ Z_\text{BH}^{\text{OSV}}( \Gamma^I, \mu^I; \Omega_\infty) $ is an interesting question that goes beyond the scope of this paper.} Each topological string partition function is evaluated at the value of the scalar moduli at the north pole and south pole of each black hole, with these moduli values in turn fixed by the Euclidean attractor mechanism \cite{Boruch:2023gfn,Boruch:2025biv}. This refinement can thus be interpreted as a generalization of OSV that applies saddle-by-saddle at least to all orders in the topological string coupling. This refinement applies and is especially relevant when the single-center configuration does not exist or is subdominant in  $Z_\text{BH}^{\text{OSV}}( \Gamma^I, \mu^I; \Omega_\infty)$.

Several generalizations, clarifications, and non-perturbative refinements of the OSV conjecture have appeared previously, see~\cite{Pioline:2006ni,Guica:2007wd,Denef:2007vg} for reviews and further references. Here we do not attempt a historical account of the large body of work pertaining to the OSV conjecture and briefly comment on some of these generalizations. In non-compact Calabi--Yau geometries, exact descriptions in terms of $q$-deformed two-dimensional Yang--Mills theory organize exponentially suppressed corrections into multi-center or ``baby-universe'' sectors involving products of topological string wavefunctions \cite{Aganagic:2004js,Dijkgraaf:2005bp}. These constructions exploit a decoupling limit in which four-dimensional gravity is non-dynamical, and the would-be black holes are more properly regarded as BPS particles. For the special compact Calabi--Yau manifolds $K3\times T^2$ and $T^6$, which give theories with enhanced $\mathcal N=4$ and $\mathcal N=8$ supersymmetry, respectively, Shih and Yin directly evaluated the OSV black hole partition function from the exact microscopic degeneracies, under the restriction of vanishing D6-brane charge \cite{Shih:2005he}. They found a sum over integer shifts of the electric potentials whose summand takes the form of \eqref{eq:OSV-refinement-intro} for $n=1$, multiplied by a nontrivial measure factor, together with corrections non-perturbative in $g_{\rm top}$ that can be summed exactly in the $\mathcal N=8$ case. This provides a close microscopic precedent for the shifted-saddle sector of our proposal. However, their calculation did not identify the shifted terms with distinct Euclidean GPI saddles and did not include the multi-center contributions in \eqref{eq:OSV-refinement-intro}, which do not contribute to the indexed partition function in $\mathcal N=8$ supergravity. For generic compact Calabi--Yau manifolds, Denef and Moore proposed a different refinement based on the microscopic factorization of BPS indices into split D6--$\overline{\mathrm{D6}}$ configurations described by attractor-flow trees \cite{Denef:2007vg}.   Our proposal is instead a direct statement about the GPI in a compact Calabi--Yau compactification: it assigns a product of north- and south-pole topological string partition functions to each admissible Euclidean single- or multi-black-hole saddle. The refinement is also different from the spin-refined OSV conjecture of \cite{Aganagic:2012si}, which is only formulated for non-compact Calabi--Yau manifolds.

The remainder of this paper is organized as follows. In Sec.~\ref{sec:new-attractor-section}, we review the elements of Euclidean conformal supergravity needed for our analysis, find the metric and field configurations for the gravitational index saddles in the higher-derivative theory, derive the new attractor mechanism, and determine the regularity conditions for the gauge fields. In Sec.~\ref{sec:equivariant-localization-section}, we review the formalism of equivariant localization in conformal supergravity and use it to compute the on-shell action of these saddles in the canonical and mixed ensembles. In Sec.~\ref{sec:gravitational-index-to-all-orders}, we combine these results to prove the original OSV conjecture to all orders in the topological string coupling for the unshifted single-center saddle. In Sec.~\ref{sec:gravitational-index-to-all-refinement}, we formulate our non-perturbative refinement of OSV and discuss both the shifted saddles required by electric-charge quantization and the contributions of multi-center black hole saddles. We conclude in Sec.~\ref{sec:discussion} with a discussion of our results and future directions. Appendices~\ref{app:conventions}--\ref{sec:enigmas} contain, respectively, our conventions, the derivation of Poincaré supergravity from conformal supergravity, a detailed analysis of the electromagnetic boundary terms for dyonic black holes in the canonical ensemble, an equivariant localization rederivation of the attractor mechanism and of the gauge-field regularity conditions that clarifies which of them hold off shell, and an explicit example of an enigmatic two-center saddle in the OSV ensemble.

\section{The new attractor mechanism in a higher-derivative theory}

\label{sec:new-attractor-section}

\subsection{Conformal supergravity}
\label{sec:conformal-sugra-review}

In this subsection, we review the superconformal multiplets needed below, construct the higher-derivative action, and specify our gauge-fixing conventions. To be concrete, we will focus on the action with Dirichlet boundary conditions, which in the context of our black hole computation will be used to compute the indexed partition functions in a mixed ensemble with fixed electric chemical potentials and magnetic charges. We also pay particular attention to the Dirichlet variational problem for the vector-multiplet scalars, which requires an additional boundary term contributing to the on-shell action. This term was absent in previous analyses \cite{deWit:2017cle,Adhikari:2025eok}, and although we focus on asymptotically flat solutions in ungauged supergravity, it may be relevant for asymptotically AdS solutions of gauged supergravity as well. 

Euclidean four-dimensional $\mathcal N=2$ conformal supergravity is obtained by gauging the superconformal algebra \cite{deWit:2017cle,Hegde:2023jmp}, whose bosonic subgroup is the product of the conformal group $\SO(5,1)$ and the R-symmetry group $\SU(2)_{\rm R}\times \SO(1,1)_{\rm R}.$ The resulting theory contains the gauge fields associated with local superconformal symmetry, together with a number of auxiliary fields. Ordinary Poincar\'e supergravity is recovered by coupling appropriate compensating multiplets, gauge-fixing the additional superconformal symmetries, and, in the on-shell formulation, integrating out the auxiliary fields \cite{deWit:1980lyi,Bergshoeff:1980is,deWit:1984rvr} (see \cite{Freedman:2012zz,Lauria:2020rhc} for a review). In what follows, we use the off-shell formalism because both the higher-derivative BPS solutions of \cite{Adhikari:2025eok} and the equivariant localization formulae of \cite{BenettiGenolini:2026qdm}, which we use below, are formulated directly in conformal supergravity. The on-shell formulation is less convenient in the presence of higher-derivative interactions, since eliminating the auxiliary fields generically induces additional interactions and modifies the supersymmetry transformations \cite{Chemissany:2012pf}.

Before reviewing the field content of the theory, let us note that our conventions differ slightly from those of \cite{deWit:2017cle,Adhikari:2025eok,BenettiGenolini:2026qdm}. These works use notation adapted to the Euclidean theory obtained by timelike dimensional reduction of five-dimensional Lorentzian supergravity, whose reality conditions differ from those obtained by Wick rotation of the physical four-dimensional Lorentzian theory. We instead use conventions adapted to the latter theory, which can be obtained from a spacelike reduction of the Wick-rotated five-dimensional theory. Accordingly, to translate the results of \cite{deWit:2017cle,Adhikari:2025eok,BenettiGenolini:2026qdm} to our variables, one first uplifts to five dimensions, performs the five-dimensional Wick rotation, and then reduces back to four dimensions. The resulting dictionary is summarized in App.~\ref{app:conventions}. We stress that the choice of reality conditions specifies the integration contour in the gravitational path integral. As discussed, for example, in \cite{Kolanowski:2026gii,Kolanowski:2026ylt}, the path integral may receive contributions from complex saddles that do not lie on its original integration contour. Thus, we formulate the Euclidean superconformal theory on the complexified field space. In particular, the vector multiplet scalar fields $X^I$ and $\bar X^I$ are treated as independent variables when solving the Euclidean supersymmetry equations. Importantly, the change of conventions described above does not modify the complexified superconformal transformation rules or the corresponding BPS equations. We can therefore use the results from \cite{Adhikari:2025eok,BenettiGenolini:2026qdm}.

Throughout this and the following sections, we restrict to bosonic field configurations and set all fermionic fields to zero. 

\paragraph{Supersymmetric multiplets: }
The fields of the theory are organized into superconformal multiplets. The fundamental multiplet is the Weyl multiplet, whose bosonic field content is
\begin{equation}
    \mathbb W=
    \left(
        e_\mu{}^a,\,
        \omega_\mu{}^{ab},\,
        b_\mu,\,
        f_\mu{}^a,\,
        A_\mu,\,
        \mathcal V_\mu{}^i{}_j,\,
        T_{ab},\,
        D
    \right).
    \label{eq:Weyl_multiplet_field_content}
\end{equation}
Here $e_\mu{}^a$ is the vielbein, $\omega_\mu{}^{ab}$ is the spin connection, $b_\mu$ and $f_\mu{}^a$ are the gauge fields associated with local dilatations and special conformal transformations, respectively, $\mathcal V_\mu{}^i{}_j$ and $A_\mu$ are the $\SU(2)_{\rm R}$ and $\SO(1,1)_{\rm R}$ connections, while $T_{ab}$ and $D$ are an auxiliary two-form, expressed in the vielbein basis, and an auxiliary scalar, respectively. The 2-form $T_{ab}$ can be decomposed into its self-dual (SD) part, $T_{ab}^+$, and anti-self-dual (ASD) part $T_{ab}^-.$ The connections $\omega_\mu{}^{ab}$ and $f_\mu{}^a$ are not independent. They are fixed by the conventional curvature constraints.

We restrict the matter sector to $n_v$ Abelian vector multiplets since these play an essential role in the OSV conjecture.\footnote{Nevertheless, we briefly discuss hypermultiplets and $\mathbb{T}$-log multiplets, which are important for the general construction of conformal supergravity actions but do not contribute to the supersymmetric observables of interest here.} The superconformal formulation also requires one additional Abelian vector multiplet, which serves as a compensator. We therefore introduce $n_v+1$ vector multiplets. Their bosonic fields are:
\begin{equation}
    \mathbb V^I
    =
    \left(
        X^I,\,
        \bar X^I,\,
        W_\mu^I,\,
        Y^{I,ij}
    \right),\qquad I=0,1,\dots,n_v.
    \label{eq:vector_multiplet_field_content}
\end{equation}
Here $X^I$ and $\bar X^I$ are scalar fields, $W_\mu^I$ is an Abelian gauge
field with field strength
\begin{equation}
    F_{\mu\nu}^I=2\partial_{[\mu}W_{\nu]}^I
    = \partial_{\mu}W_{\nu}^I
    - \partial_{\nu}W_{\mu}^I
    \label{eq:field_strengths}
    ,
\end{equation}
and $Y^{I,ij}=Y^{I,ji}$ is an auxiliary $\SU(2)_{\rm R}$ triplet. Upon imposing reality conditions arising from the Lorentzian four-dimensional $\CN=2$ supergravity, $X^I$ and $\bar X^I$ become complex conjugates.

To gauge-fix the $\SU(2)_{\rm R}$ symmetry, we also introduce a single compensating hypermultiplet \cite{deWit:1998zg,deWit:1999fp}. The bosonic fields of a hypermultiplet,
\be 
\mathbb H = \left( 
A_i^\alpha 
\right) ,
\ee
are the four scalars $A_i{}^\alpha$, where $i=1,2$ is an $\SU(2)_{\rm R}$ index and $\alpha=1,2$ is an $\mathrm{Sp}(1)$ index. The scalars obey the pseudo-reality condition
\begin{equation}
    A_i{}^\alpha\varepsilon^{ij}\varepsilon_{\alpha\beta} = A_\beta{}^j \equiv \left(A_j{}^\beta\right)^* ,
\end{equation}
where $\varepsilon_{\alpha\beta}$ is the invariant antisymmetric tensor of $\mathrm{Sp}(1)$. We use conventions where $\varepsilon_{12} = 1 = -\varepsilon_{21}$ and $\varepsilon^{12} = -1 = \varepsilon^{21}$, so that $\varepsilon^{ij}$ is the inverse of $\varepsilon_{ij}$. It is also convenient to define the hyper-K\"ahler potential,
\begin{equation}
    \chi
    =
    \frac12\varepsilon^{ij}\varepsilon_{\alpha\beta}
    A_i{}^\alpha A_j{}^\beta .
\end{equation}

\paragraph{Constructing the action: }To construct superconformally invariant actions, we also introduce chiral and anti-chiral multiplets. An anti-chiral (chiral) multiplet $\Phi$ ($\bar\Phi$) is defined by requiring that its lowest scalar component transforms into a spinor of negative (positive) chirality. In the bosonic truncation we write
\begin{equation}
    \Phi= ( A,B^{ij},F_{ab}^{+},C),\qquad \bar{\Phi} = (\bar{A},\bar{B}^{ij},F^-_{ab},\bar{C}),
\end{equation}
where $A, \bar{A},C,\bar{C}$ are scalars,
$B^{ij}=B^{(ij)}$ and $\bar{B}^{ij}=\bar{B}^{(ij)}$ are $\SU(2)_{\rm R}$ triplets, and the two-forms $F_{ab}^{-}$ and $F_{ab}^+$ are anti-self-dual and self-dual, $F_{ab}^\pm= \pm \frac{1}{2} \varepsilon_{abcd}F^{\pm cd}$. A field $\varphi$ of Weyl weight $w$ transforms under local dilatations parametrized by $\Lambda_D$ as $\delta_D\varphi=w\Lambda_D\varphi$. We refer to the Weyl weight of $A$, $w[A]$, as the Weyl weight of the multiplet $\Phi.$ 
By coupling anti-chiral and chiral multiplets of weight $w[\Phi]=2$ to the background Weyl multiplet of conformal supergravity one can construct superconformal Lagrangians, whose bosonic parts are
\begin{equation}
    \CL[\Phi;T^-] = C + \frac{1}{16}A\,(T_{ab}^-)^2,\qquad \bar{\CL}[\bar\Phi;T^+] = \bar{C} + \frac{1}{16}\bar{A}\,(T^+_{ab})^2, 
    \label{eq:chiral_antichiral_actions_weight_two}
\end{equation}
where $T^\pm$ are the auxiliary two-forms of the Weyl multiplet \eqref{eq:Weyl_multiplet_field_content}.

The actions of superconformal gravity will now follow from constructing chiral multiplets in terms of the supergravity fields \eqref{eq:Weyl_multiplet_field_content}, \eqref{eq:vector_multiplet_field_content}.  
First, an Abelian vector multiplet $\mathbb V^I$ gives rise to an anti-chiral and chiral multiplet of Weyl weight one, whose lowest components are
\begin{align}
    A|_{\mathbb V^I} = X^I,\qquad \bar{A}|_{\mathbb V^I} = \bar{X}^I.
\end{align}
Second, the Weyl multiplet determines an anti-chiral and chiral Weyl-squared multiplet of Weyl weight two, which we denote by $\mathbb W^2$ and $\overline{\mathbb W}^2$.
Their lowest bosonic components are:
\begin{align}
    A|_{\mathbb W^2} = T^+_{ab}T^{+ab},\qquad \bar{A}|_{\overline{\mathbb{W}}^2} = T^-_{ab}T^{-ab}.
\end{align}
Now, the most general action of the type \eqref{eq:chiral_antichiral_actions_weight_two}, constructed from the vector multiplets and the Weyl multiplet, is obtained  by introducing functions $\mathcal F (X^I,A|_{\mathbb W^2})$ and $\bar{\mathcal F}(\bar X^I,\bar{A}\big|_{\overline{\mathbb W}^2})$ with transformation properties 
\begin{equation}
    \mathcal F(\lambda X^I,\lambda^2 A|_{\mathbb W^2}) =
    \lambda^2
    \mathcal F(X^I,A|_{\mathbb W^2}),\qquad \bar{\mathcal F}(\lambda\bar X^I,\lambda^2 \bar{A}\big|_{\overline{\mathbb W}^2})= \lambda^2 \bar{\mathcal F}(\bar X^I,\bar{A}\big|_{\overline{\mathbb W}^2}).
    \label{eq:prepotentials_transformation_properties}
\end{equation}
These functions, known as prepotentials \cite{Bergshoeff:1980is,LopesCardoso:1998tkj}, can now be used to construct weight-two chiral multiplets
\begin{equation}
\Phi|_{\mathcal F} = (A|_{\CF}, B^{ij}|_{\CF},F^+_{ab}|_{\CF}, C|_{\CF}) , 
\qquad 
\bar\Phi|_{\bar{\mathcal F}} = (\bar A|_{\bar \CF}, \bar B^{ij}|_{\bar \CF},F^-_{ab}|_{\bar \CF}, \bar C|_{\bar \CF}) ,
\end{equation}
with lowest components given by 
\begin{equation}
    A|_{\mathcal F}
    =
    \mathcal F (X^I,A|_{\mathbb W^2}), \qquad \bar{A}|_{\bar \CF} =\bar{\mathcal F}(\bar X^I,\bar{A}\big|_{\overline{\mathbb W}^2}) ,
\end{equation}
and the higher components determined by product rules for chiral multiplets (see Sec.~4.1 of \cite{deWit:2017cle}).
Since $X^I$ has Weyl weight one and $A|_{\mathbb W^2}$ has Weyl weight two, \eqref{eq:prepotentials_transformation_properties} guarantees consistency with superconformal invariance.

With these definitions, we can write the bosonic part of the action for vector and Weyl multiplets as:
\begin{equation}
    I_{\mathbb{VW}} = -\frac{\i}{16\pi}\int_M\left(\CL[\Phi|_{\CF};T^-] - \bar{\CL}[\bar \Phi|_{\bar \CF};T^+]\right){\rm vol}_M,\label{eq:IVW}
\end{equation}
where ${\rm vol}_M$ is the volume form on the Euclidean spacetime $M$. To obtain the standard Einstein--Hilbert term in the action, we need to add the compensating-hypermultiplet action $I_{\mathbb H}$ to $I_{\mathbb{VW}}$ (see Eq.~(2.8) in \cite{Bobev:2021oku} and also \cite{Adhikari:2025eok}), as otherwise the auxiliary field $D$ from the Weyl multiplet would act as a Lagrange multiplier setting the Einstein--Hilbert term in the action to zero. The hypermultiplet action is given by
\begin{equation}
    I_{\mathbb{H}} = -\frac{1}{8\pi}\int \left\{\chi\left(-\frac16R+\frac12D\right) - \frac12\varepsilon^{ij}\varepsilon_{\alpha\beta}(\CD_\mu A_i{}^\alpha)(\CD^\mu A_j{}^\beta)\right\}{\rm vol}_M\label{eq:Ihyper}
\end{equation}
where $\CD_{\mu}$ is the covariant derivative without the $f_\mu{}^a$ connection corresponding to special conformal transformations, and $R$ is the Ricci scalar associated with the metric $g_{\mu\nu} = \eta_{ab}e^a_\mu e^b_\nu$.

\paragraph{Boundary terms: }As can be seen from Eqs.~\eqref{eq:2derivL} and  \eqref{eq:2derivLbar}\footnote{See also Eq.~(6.3) of \cite{deWit:2017cle}.}, the action $I_{\mathbb{VW}}$ contains kinetic terms for scalars of the form
\begin{equation}
    I_{\mathbb{VW}} \supset -\frac{\i}{8\pi} \int_M \left( \mathcal F_I \Box_c \bar X^I - \bar{\mathcal F}_I \Box_c X^I \right) {\rm vol}_M,
\end{equation}
where $\Box_c$ is the superconformal Laplacian acting on scalars $X^I$, $\bar{X}^I$ as $\CD_\mu \CD^\mu X^I - \bigl(\tfrac16R + D\bigr)X^I.$ Consequently, the variation of $I_{\mathbb{VW}}$ contains the boundary contribution
\begin{equation}
    \delta I_{\mathbb{VW}} \supset -\frac{\i}{8\pi} \int_{\partial M} \dd^3x\,\sqrt{h}\,n^\mu \left( \mathcal F_I\mathcal D_\mu\delta\bar X^I - \bar{\mathcal F}_I\mathcal D_\mu\delta X^I \right),
\end{equation}
up to terms proportional to $\delta X^I$ and $\delta\bar X^I$. Here $n^\mu$ is the unit normal to the boundary $\partial M$, and $h$ is the determinant of the induced metric on the boundary. Therefore, the action $I_{\mathbb{VW}}$ does not by itself give a well-defined variational principle upon imposing Dirichlet boundary conditions
\begin{equation}
    \delta X^I\big|_{\partial M}
    = 0=
    \delta\bar X^I\big|_{\partial M}.
\end{equation}
To fix this, we must add the following boundary term
\begin{equation}
    I^X_{\partial M} = \frac{\i}{8\pi} \int_{\partial M}\dd^3x\,\sqrt{h}\,n^\mu \left(\mathcal F_I\mathcal D_\mu\bar X^I - \bar{\mathcal F}_I\mathcal D_\mu X^I \right).\label{eq:new-scalar-bdry-term}
\end{equation}

\paragraph{Total action:}
Combining all the contributions above, the bosonic action with Dirichlet boundary conditions for hyper, Weyl and vector multiplet fields is:
\begin{equation}
    I_{\rm Dir} = I_{\mathbb{VW}} +I_{\mathbb H} +I_{\partial M}^X +I_{\rm GHY}^{\rm HD}.\label{eq:Itot_dirichlet}
\end{equation}
Here $I_{\rm GHY}^{\rm HD}$ denotes the gravitational boundary terms required for a well-defined metric variational problem. At the two-derivative level this reduces to the standard Gibbons--Hawking--York boundary term \cite{York:1972sj,Gibbons:1976ue}. In the presence of higher-derivative interactions it is, in general, supplemented by additional boundary terms depending on the higher-curvature couplings \cite{Myers:1987yn,Dyer:2008hb,Teimouri:2016ulk}. We will not need their explicit form below.

The action \eqref{eq:Itot_dirichlet} that we analyze in this section contains only F-terms, i.e.\ chiral densities built from the prepotentials $\CF$ and $\bar\CF$. In principle, one can also add D-terms (full-superspace integrals), such as those built from kinetic multiplets \cite{deWit:2010za}. These modify the equations of motion, but not the off-shell supersymmetry conditions used below. In this section, we assume that D-terms do not affect the form of the supersymmetric saddles in any way relevant for their on-shell action: they change neither the fixed points of the Killing vector, nor the values of $X^I$ and $A|_{\mathbb W^2}$ at these fixed points, nor the asymptotic falloffs. In Subsec.~\ref{sec:on-shell-action-in-flatspace-SUGRA}, we confirm this through a direct analysis of D-terms, which shows that they leave these fixed-point data unchanged and that their contribution to the on-shell action vanishes.

\paragraph{Electromagnetic duality:} In addition to the
electric field strengths \eqref{eq:field_strengths}, 
we define the dual field strengths $G_I$ through their self-dual and anti-self-dual components $G_I^{\pm}$, 
\begin{align}
    G_{I\,\mu\nu}^{+} & = -\frac{16\pi \i}{\sqrt{g}}\frac{\delta I_{\mathbb{VW}}}{\delta F^{I+\mu\nu}} = \mathcal F_{IJ} \left(
        F^{J+}_{\mu\nu}
        -\tfrac14\bar X^J T^{+}_{\mu\nu}
    \right)
    +\mathcal F_{IA}
    F^{+}_{\mu\nu}\big|_{\mathbb W^2}
    +\tfrac14\bar{\mathcal F}_I T^{+}_{\mu\nu},\label{eq:Gminus}
    \\[0.2cm]
    G_{I\,\mu\nu}^{-}
    &= +\frac{16\pi \i}{\sqrt{g}}\frac{\delta I_{\mathbb{VW}}}{\delta F^{I-\mu\nu}} =
    \bar{\mathcal F}_{IJ}
    \left(
        F^{J-}_{\mu\nu}
        -\tfrac14 X^J T^{-}_{\mu\nu}
    \right)
    +\bar{\mathcal F}_{IA}
    F^{-}_{\mu\nu}\big|_{\overline{\mathbb W}^{\,2}}
    +\tfrac14\mathcal F_I T^{-}_{\mu\nu},\label{eq:Gplus}
\end{align}
where $\mathcal{F}_I = \partial_{X^I}\mathcal{F}$, $\mathcal{F}_{IJ} = \partial_{X^I} \partial_{X^J} \mathcal{F}$, and analogously for $\bar{\mathcal{F}}_{I \dots}$.
We can organize the field strengths $F^I$, $G_I$ into symplectic vectors $\mathbb{F}^{\pm} = (F^{I\pm},G_{I}^{\pm})$, which transform in the fundamental representation under the electromagnetic duality group ${\rm Sp}(2n_v+2)$. The symplectic vector of magnetic and electric charges, $\Gamma = (\Gamma^I,\Gamma_I)$, is defined as the integral of $\mathbb{F} = \mathbb{F}^{+} + \mathbb{F}^-$ over a sphere at asymptotic spatial infinity,
\begin{equation}
    \Gamma = \frac{1}{4\pi}\int_{S^2_{\infty}} \mathbb{F}.
\end{equation}
Similarly, we introduce the holomorphic and anti-holomorphic period vectors:
\begin{equation}
    \Omega = (X^I,\CF_I),
    \qquad \bar{\Omega} = (\bar{X}^I,\bar{\CF}_I).\label{eq:period_vectors}
\end{equation}
For two symplectic vectors, $A = (A^I,A_I)$ and $B = (B^I,B_I)$, we define the symplectic product by
\begin{equation}
    \la A,B\ra =A^I B_I - A_I B^I.
\end{equation}
We can then define the generalized K\"ahler potential by
\begin{equation}
    e^{-\CK} = \i\la\Omega,\bar{\Omega}\ra,
\end{equation}
and the central charge function by
\begin{equation}
    \mathcal{Z}(\Gamma,\Omega) = e^{\CK/2}\symp{\Gamma}{\Omega}.
\end{equation}

\paragraph{Gauge-fixing:} Lastly, to remove the redundant gauge freedom associated with the additional superconformal symmetries, one has to impose a number of gauge-fixing conditions. In particular, special conformal transformations are fixed by setting $b_\mu=0$. The local $\SU(2)_{\rm R}$ symmetry is fixed using the compensating hypermultiplet scalars via the condition
\begin{equation}
    A_i{}^\alpha=\sqrt{-\chi}\,\delta_i{}^\alpha.
\end{equation}
Local dilatations are fixed by setting the hyper-K\"ahler potential to $\chi=-2$. The compensating hypermultiplet also fixes $S$-supersymmetry through the corresponding hyperino gauge condition.

The remaining $\SO(1,1)_{\rm R}$ symmetry acts with opposite weights on the chiral and anti-chiral vector-multiplet scalars. In the two-derivative theory, one often uses this freedom to impose $X^0=\bar X^0$. However, for the general higher-derivative theory, and in particular when discussing its half-BPS configurations, we will refrain from imposing this gauge and keep $X^0$ and $\bar X^0$ independent.

\subsection{The gravitational index saddle}
\label{sec:gravitational-index-saddle}

In this subsection we begin the discussion of stationary half-BPS configurations considered in~\cite{Adhikari:2025eok}. The defining feature of these configurations is that they admit a symplectic-Majorana Killing spinor $\epsilon^i$ satisfying the embedding condition
\begin{equation}
    \gamma_0\epsilon_+^i
    =
    \i f\,\epsilon_-^i ,
    \label{eq:Killing-spinor-embedding}
\end{equation}
where the function $f$ depends on the choice of the $\SO(1,1)_{\rm R}$ gauge and
\begin{equation}
    \epsilon_\pm^i = \frac12(1\pm\gamma_5)\epsilon^i
    ,
\end{equation}
denote the two chiral projections of the Euclidean supersymmetry parameter. The condition~\eqref{eq:Killing-spinor-embedding} relates the two chiral components of $\epsilon^i$ and therefore preserves one half of the supersymmetries. Our Euclidean spinor and gamma-matrix conventions follow Sec.~3 of \cite{deWit:2017cle}.

These configurations correspond to multi-center rotating black hole saddles of Israel--Wilson--Perj\'es type \cite{Israel:1972vx,Perjes:1971gv}, and provide a higher-derivative generalization of the finite-temperature index saddles constructed in \cite{Iliesiu:2021are,Boruch:2023gfn,Boruch:2025biv,Chen:2024gmc}. 
We begin by reviewing general aspects of these solutions and analyzing their asymptotic expansion at infinity. Then, we impose appropriate regularity conditions on these saddles, including smoothness of the geometry and finiteness of the near-horizon region which in turn ensures a finite value of $\beta$. The resulting geometries satisfy the new form of the attractor mechanism \cite{Boruch:2023gfn}.

\paragraph{Weyl and hypermultiplet fields:} The spacetime geometry derived in \cite{Adhikari:2025eok} is an $S^1$-fibration over an $\mathbb{R}^3$ base space, with Euclidean time $\tau$ parametrizing the circle. We denote the corresponding connection 1-form by $\omega$. The supersymmetry conditions further imply that the three-dimensional base is flat, and thus, the metric takes the form
\begin{equation}\label{12BPS_metric}
    \dd s^2 = e^{2U}(\dd\tau+\omega)^2 +e^{-2U}\dd\vec{x}^{\,2}, \qquad \omega=\omega_m\dd x^m,
\end{equation}
where $\vec{x}=(x^1,x^2,x^3)$. The function $f$ entering the embedding condition \eqref{eq:Killing-spinor-embedding} is independent of time, $f =f(\vec{x})$. The remaining fields in the Weyl multiplet are fixed algebraically or differentially in terms of the two geometric data $U$ and $\omega$, together with the function $f$. 

 A convenient choice of vielbein is
\begin{equation}
    e^0=e^U(\dd\tau+\omega),
    \qquad
    e^p=e^{-U}\dd x^p,
    \qquad p=1,2,3.\label{halfBPS_on_frame}
\end{equation}
We denote flat four-dimensional indices by $a,b=0,1,2,3$, curved four-dimensional indices by $\mu,\nu=\tau,1,2,3$, flat spatial indices by $p,q=1,2,3$, and curved spatial indices by $m,n=1,2,3$. Spatial indices $m,n$ are raised and lowered with the metric on the three-dimensional base, $\delta_{mn}$, without the overall factor $e^{-2U}$. Reference \cite{Adhikari:2025eok} presents the results using the orthonormal frame components, but in what follows, we shall translate them into coordinate components.\footnote{The basis dual to \eqref{halfBPS_on_frame} is
\begin{equation}
    E_0=e^{-U}\partial_\tau,
    \qquad
    E_p=e^U\left(\partial_p-\omega_p\partial_\tau\right).
\end{equation}
For a general covariant tensor,
\begin{equation}
    \mathcal T_{a_1\ldots a_k}
    =
    E_{a_1}{}^{\mu_1}\cdots E_{a_k}{}^{\mu_k}
    \mathcal T_{\mu_1\ldots\mu_k}.
\end{equation}
In particular, for any two-form $\alpha$,
\begin{equation}
    \alpha_{p0}=\delta_p{}^m \alpha_{m\tau},
    \qquad
    \alpha_{m\tau}=\delta_m{}^p \alpha_{p0}.
\end{equation}
Hence the factors of $e^U$ cancel in the mixed time-space components, and there is no additional term proportional to $\omega_m$. The same relation holds separately for the self-dual and anti-self-dual parts of a two-form.} 

The $\SO(1,1)_{\rm R}$ gauge field is given by\footnote{This follows from Eq.~(4.57) of \cite{Adhikari:2025eok}, namely,
\begin{equation}
    R_{pq}(A) = 2e_{p}^m e^n_{q}\partial_{[m}A_{n]} = e^{4U}\left[2(\partial_{[p}U)R(\omega)_{q]} + \partial_{[p}R(\omega)_{q]}\right].
\end{equation}}:
\begin{equation}
    A_\tau = \partial_\tau\Lambda,\quad A_m = \partial_m \Lambda + \frac{1}{2}\,e^{2U}R(\omega)_m.
\end{equation}
Then Eq.~(4.53) of \cite{Adhikari:2025eok} implies that:
\begin{equation}
    f^{-1}\partial_m f = -\partial_m \Lambda.
\end{equation}
The $\SU(2)_{\rm R}$ field strength vanishes,
\begin{equation}
    R(\mathcal{V})_{\mu\nu}{}^i{}_j = 2\partial_{[\mu}\mathcal{V}_{\nu]}{}^i{}_j + \mathcal{V}_{[\mu}{}^i{}_{k}\mathcal{V}_{\nu]}{}^k{}_j = 0,
\end{equation}
and we can choose a gauge where $\mathcal{V}_{\mu}{}^i{}_j = 0$. This is compatible with equations of motion for the hypermultiplet scalars, since:
\begin{equation}
    D = \frac{1}{3}R = \frac{2}{3}e^{2U}\left[\Delta U - (\partial_m U)(\partial^m U) - \frac{1}{4}e^{4U}R(\omega)_mR(\omega)^m\right],
\end{equation}
where $\Delta$ is the Laplacian on the base space ($\Delta = \partial_m \partial^m$ in Cartesian coordinates). The gauge field associated with local special conformal transformations is (this follows from Eq.~(4.29) in~\cite{Adhikari:2025eok} in the gauge where $\chi = {\rm const.}$):
\begin{equation}
    f_\mu{}^a = -\frac{3}{4}(D e_\mu^a - 2D e_\mu^a \delta^{a0}) + \tilde{R}(A)_{\mu 0}\delta^{a0}-\tilde{R}(A)_{\mu}{}^a,
\end{equation}
where $\tilde{R}(A)_{\mu\nu}$ is the Hodge dual of $R(A)_{\mu\nu}$. The gauge field corresponding to local Weyl (dilatation) transformations, $b_\mu$, is gauge-fixed to zero, $b_{\mu} = 0$. The two-form $T_{ab}$ is given by:
\begin{align}
    T_{m\tau}^{+} &= 4f^{-1} e^U\,T_{m}^+,\qquad T_{m}^+\dd x^m = -\dd U + \frac{1}{2}e^{2U}R(\omega)_m\dd x^m,\label{Tplus_sol}
    \\
    T^-_{m\tau} &= 4fe^U\,T^-_{m},\qquad\quad T_m^- \dd x^m = \dd U + \frac{1}{2}e^{2U}R(\omega)_m\dd x^m.\label{Tminus_sol}
\end{align}
The components $T_{mn}^{\pm}$ are fixed by (anti-)self-duality conditions. $R(\omega)_m \dd x^m = \star_3 \dd \omega$ is the Hodge dual of $\dd\omega$, computed with respect to the $\mathbb{R}^3$ base space.

The equation of motion for the $D$ field determines the lapse function $U$:
\begin{align}
    \nonumber e^{-\CK} + \frac{1}{2}\chi =&\, 128\,\i\,e^{3U}\partial_m\left[e^{-U}\partial^m(\CF_A - \bar\CF_A)\right]
    \\
    \nonumber &\,-32\,\i\,e^{6U}R(\omega)_m R(\omega)^m(\CF_A - \bar\CF_A)
    \\
    &\,-64\,\i\,e^{4U}R(\omega)_m\partial^m(\CF_A + \bar\CF_A ).\label{Deom}
\end{align}
The equation of motion for the $\SO(1,1)_{\rm R}$ gauge field $A_\mu$ determines $\omega=\omega_m\dd x^m$ through
\begin{align}
    \nonumber &\i(\,fX^I + f^{-1}\bar{X}^I)\overset{\leftrightarrow}{\partial_m}(f\CF_I + f^{-1} \bar{\CF}_I) - \frac{\chi}{2}e^{2U}R(\omega)_m
    \\
    &=-128\,\i\, e^{2U}\partial^n\left[2(\partial_{[n}U)\partial_{m]}(\CF_A + \bar\CF_A) + \partial_{[n}(e^{2U}R(\omega)_{m]}(\CF_A - \bar\CF_A))\right]. \label{eq:SO11_EOM}
\end{align}
Here and below, subscripts $I$ and $A$ on the prepotentials denote derivatives with respect to $X^I$ and $A|_{\mathbb W^2}$, respectively, and analogously in the anti-chiral sector.

\paragraph{Vector multiplet fields: }Let us now turn to the vector-multiplet sector. The auxiliary $\SU(2)_{\rm R}$ triplets are set to zero,
\begin{equation}
    Y^{I,ij} = 0.
\end{equation}
Let us define the rescaled period vectors as
\begin{equation}
    \mathcal Y=\i\,e^{-U}f\,\Omega = (\cY^I,\CF_I(\cY^J,\Upsilon)),
    \qquad
    \bar{\mathcal Y}=-\i\,e^{-U}f^{-1}\bar\Omega = (\bar{\cY}^I,\bar{\CF}_I(\bar{\cY}^J,\bar\Upsilon)),\label{eq:rescaled_periods}
\end{equation}
where
\begin{align}
    \mathcal Y^I&=\i\,e^{-U}f^{}X^I,\qquad\quad \;\;\Upsilon = -e^{-2U}f^2A\big|_{\mathbb W^2} = -64\,T_m^{+}T^{+ m},\label{eq:rescaled_hol_vars}
    \\[0.2cm]
    \bar {\mathcal Y}^I&=-\i\,e^{-U}f^{-1}\bar X^I,\qquad \bar\Upsilon = -e^{-2U}f^{-2} \bar A\big|_{\overline{\mathbb W}^{\,2}}=-64\,T_m^{-}T^{-m}.\label{eq:rescaled_antihol_vars}
\end{align}
The vector-multiplet field strengths are given by
\begin{align}
    \mathbb F^+_{m\tau}
    &=
    -\i e^{2U}\left[
        \partial_m\mathcal Y
        +(\partial_mU)(\mathcal Y+\bar{\mathcal Y})
        -\frac12e^{2U}R(\omega)_m
        (\mathcal Y+\bar{\mathcal Y})
    \right],\label{eq:symp-field-strength-m}
    \\
    \mathbb F^-_{m\tau}
    &=
    -\i e^{2U}\left[
        \partial_m\bar{\mathcal Y}
        +(\partial_mU)(\mathcal Y+\bar{\mathcal Y})
        +\frac12e^{2U}R(\omega)_m
        (\mathcal Y+\bar{\mathcal Y})
    \right].\label{eq:symp-field-strength-p}
\end{align}
The remaining components of $\mathbb{F}^{\pm}$ are fixed by the (anti-)self-duality conditions. The Maxwell equations and Bianchi identities can be written as:
\begin{equation}
    \CD^a(\mathbb{F}^{+}_{ab} - \mathbb{F}^-_{ab}) = 0.
\end{equation}
They imply that the vector multiplet scalars $X^I$, $\bar{X}^I$ satisfy the generalized stabilization equations,
\begin{equation}
    H = \i\,(\mathcal{Y} - \bar {\mathcal{Y}}),\label{eq:stab_eqs}
\end{equation}
where $H = (H^I, H_I)$ is a symplectic vector of harmonic functions on the $\mathbb{R}^3$ base space \cite{Behrndt:1997ny,Denef:2000nb,Bates:2003vx,LopesCardoso:2000qm}. A non-extremal supersymmetric solution corresponding to a configuration of $n$ black holes is specified by:
\begin{equation}
    H = h + \sum_{a=1}^n\frac{\gamma_{N,a}}{|\vec{x} - \vec{x}_{N,a}|} + \sum_{a=1}^n\frac{\gamma_{S,a}}{|\vec{x} - \vec{x}_{S,a}|},\label{eq:harmonic_fun}
\end{equation}
where $\vec{x}_{N,a}$ and $\vec{x}_{S,a}$ are the locations of the north and south poles of the $n$ horizons. Later, it will prove convenient to parametrize the residues $\gamma_{N,a}, \gamma_{S,a}$ by black hole charges $\Gamma_a$ and dipole moments $\delta_{a} = (\delta_a^I,\delta_{I,a})$, so that:
\begin{equation}
    \gamma_{N,a} = \frac{\Gamma_a}{2} + \i\delta_a,\qquad \gamma_{S,a} = \frac{\Gamma_a}{2} - \i\delta_a.\label{eq:dipole_moments}
\end{equation}
Finally, adding the self-dual and anti-self-dual components of the field strengths gives
\begin{align}
    \mathbb{F}_{m\tau} &= \mathbb{F}_{m\tau}^{+}+\mathbb{F}_{m\tau}^{-} = -\i\,\partial_m \left[e^{2U}\left(\mathcal{Y}+\bar{\mathcal Y}\right)
    \right] = \i\,\partial_m\Phi,
\end{align}
where $\Phi = (\Phi^I,\Phi_I)$ is the symplectic vector of electric and magnetic potentials,
\begin{align}
    \Phi = -e^{2U}\left(\mathcal{Y}+\bar{\mathcal Y}\right).\label{eq:el_pot}
\end{align}

\subsection{Asymptotic expansion}
\label{par:asymptotic_expansion}

For the purpose of imposing the boundary conditions and evaluating the on-shell action in the later sections, we now carefully analyze the asymptotic expansions of a general multi-center half-BPS solution. 
At $r=|\vec{x}|\to\infty$, the harmonic functions behave as
\begin{equation}
    H=h+\frac{\Gamma}{r}+\O(r^{-2}),
\end{equation}
where
\begin{equation}
    \Gamma = (\Gamma^I,\Gamma_I)
    =
    \sum_{a=1}^{n}\Gamma_a = \sum_{a=1}^n(\Gamma_a^I,\Gamma_{a,I}).
\end{equation}
Below we verify explicitly that $\Gamma^I$ and $\Gamma_I$ are the total magnetic and electric charges measured at asymptotic infinity.

Asymptotic flatness requires
\begin{equation}
    U(\vec{x})
    =
    \frac{U^{(1)}(\theta,\varphi)}{r}
    +\O(r^{-2}).
\end{equation}
The function $U^{(1)}(\theta,\varphi)$ is determined by the leading terms in the large-$r$ expansion of the equation of motion of the $D$ field, \eqref{Deom}. Moreover, the one-form $\omega$ must remain finite at infinity. At leading order in the large-$r$ expansion it can possess only angular components, and therefore
\begin{equation}
    (\dd\omega)_{\theta\varphi}=\O(1),
    \qquad
    (\dd\omega)_{r\theta}=\O(r^{-1})
    =
    (\dd\omega)_{r\varphi}.
\end{equation}
Recall that $R(\omega)=\star_3\dd\omega$. Hence
\begin{equation}
    R(\omega)\wedge\dd\omega
    =
    \|\dd\omega\|^2
    r^2\dd r\wedge\dd\theta\wedge\sin\theta\,\dd\varphi
    +\text{subleading}.
\end{equation}
Since the inverse metric in the angular directions scales as
$g^{\hat m\hat n}=\O(r^{-2})$, where $\hat m,\hat n=\theta,\varphi$, we have 
\begin{equation}
    \|\dd\omega\|^2
    =
    \frac12(\dd\omega)_{mn}(\dd\omega)^{mn}
    =
    \O(r^{-4}).
\end{equation}
Consequently, the leading terms in the equation above scale as $\O(r^{-2})$, and therefore
\begin{equation}
    R(\omega)=\O(r^{-2}),
    \qquad
    \|\dd U\|=\O(r^{-2}).
\end{equation}
It follows that the right-hand side of \eqref{Deom} vanishes at asymptotic infinity, implying $e^{-\CK} = -\frac{\chi}{2} = 1.$ Furthermore,
\begin{equation}
    A_m = \frac12e^{2U}R(\omega)_m+\partial_m\Lambda \xrightarrow{r\to\infty} \partial_m\Lambda,
\end{equation}
and hence, $f^{-1}\partial_m f=-\partial_m\Lambda$. We see that $f$ takes a finite value in a regular gauge.

Since $T_m^\pm\dd x^m=\O(r^{-2})$, we also have
\begin{equation}
    A\big|_{\mathbb W^2}=\O(r^{-4}),
    \qquad
    \bar A\big|_{\overline{\mathbb W}^{\,2}}=\O(r^{-4}),
    \qquad
    \Upsilon,\bar\Upsilon=\O(r^{-4}).
\end{equation}
The generalized stabilization equations \eqref{eq:stab_eqs} imply the asymptotic expansions
\begin{equation}
    \mathcal Y
    =
    \mathcal Y^{(0)}
    +\frac{1}{r}\mathcal Y^{(1)}
    +\O(r^{-2}),
    \qquad
    \bar{\mathcal Y}
    =
    \bar{\mathcal Y}^{(0)}
    +\frac{1}{r}\bar{\mathcal Y}^{(1)}
    +\O(r^{-2}),
\end{equation}
with $\mathcal{Y}^{(0)}$ and $\mathcal{Y}^{(1)}$ satisfying
\begin{equation}
        \i\left(
        \mathcal Y^{(0)}
        -
        \bar{\mathcal Y}^{(0)}
    \right)
    =
    h,
    \qquad
    \i\left(
        \mathcal Y^{(1)}
        -
        \bar{\mathcal Y}^{(1)}
    \right)
    =
    \Gamma .\label{y(0)y(1)_eqs}
\end{equation}
Denote
\begin{equation}
    \cY^I = y^{I(0)} + \frac{1}{r}y^{I(1)} + \O(r^{-2}),\qquad \bar \cY^I = \bar y^{I(0)} + \frac{1}{r}\bar y^{I(1)} + \O(r^{-2}).
\end{equation}
Since $\Upsilon,\bar\Upsilon=\O(r^{-4})$, the first two coefficients of $\mathcal{Y}$ and $\bar{\mathcal Y}$ are given by $y^{I(0)}$ and $y^{I(1)}$,
\begin{align}
    \mathcal Y^{(0)}
    &=
    \left(
        y^{I(0)},
        \CF_I(y^{(0)},0)
    \right),
    &
    \bar{\mathcal Y}^{(0)}
    &=
    \left(
        \bar y^{I(0)},
        \bar{\CF}_I(\bar y^{(0)},0)
    \right),
    \\
    \mathcal Y^{(1)}
    &=
    \left(
        y^{I(1)},
        \CF_{IJ}(y^{(0)},0)y^{J(1)}
    \right),
    &
    \bar{\mathcal Y}^{(1)}
    &=
    \left(
        \bar y^{I(1)},
        \bar{\CF}_{IJ}(\bar y^{(0)},0)\bar y^{J(1)}
    \right).
\end{align}
Therefore, Eqs.~\eqref{y(0)y(1)_eqs} uniquely determine $y^{I(0)}$ and $y^{I(1)}$ in terms of $h$ and $\Gamma$. Note that, since $e^U$ approaches $1$ at infinity, $\mathcal{Y}^{(0)}$ is directly related to the asymptotic value of the period vectors $\Omega$, which we will denote as $\Omega_{\infty} = \lim_{r\to \infty}\Omega$,
\begin{equation}
    \mathcal{Y}^{(0)} = \i f_{\infty}(\hat{x}) \Omega_{\infty}(\hat{x}),\qquad \bar{\mathcal{Y}}^{(0)} = -\i f_{\infty}(\hat{x})^{-1} \bar\Omega_{\infty}(\hat{x}),
\end{equation}
where $\hat{x} = \vec{x}/|\vec{x}|$ and $f_{\infty}(\hat{x}) = \lim_{r\to \infty}f(\vec{x})$.

Let us now analyze the equation of motion of the $\SO(1,1)_{\rm R}$ gauge field. In our conventions it can be written as
\begin{equation}
    \i\langle H,\partial_m H\rangle - \frac{\chi}{2}R(\omega)_m = -128\,\i\,\partial^nJ_{nm},
    \label{Aeom_v1}
\end{equation}
where
\begin{equation}
    J_{nm} = 2(\partial_{[n}U) \partial_{m]} \left(\CF_A+\bar{\CF}_A\right) + \partial_{[n}\big(e^{2U}R(\omega)_{m]} (\CF_A-\bar{\CF}_A)\big).\label{eq:Aeom_current}
\end{equation}
First, we consider the radial component, $m=r$, and denote angular indices by $\hat m,\hat n=\theta,\varphi$. Since $\CF_A$ and
$\bar{\CF}_A$ remain finite at large $r$, we have $J_{\hat n r}=\O(r^{-2}).$ Using the large-$r$ scaling of the inverse angular metric, $\partial^{\hat n} = g^{\hat n\hat p}\partial_{\hat p} = \O(r^{-2})\partial_{\hat p},$ we obtain $\partial^nJ_{nr} = \partial^{\hat n}J_{\hat n r} = \O(r^{-4}).$ Similarly, using $J_{\hat n\hat m}=\O(r^{-1})$ and $J_{r\hat m}=\O(r^{-2}),$ we find $\partial^nJ_{n\hat m} = \partial_rJ_{r\hat m} + g^{\hat n\hat p} \partial_{\hat p}J_{\hat n\hat m} +\O(r^{-3})  = \O(r^{-3}).$ We therefore conclude that, for every component $m$,
\begin{equation}
    \partial^nJ_{nm}=\O(r^{-3}).
\end{equation}
Consequently, at leading order Eq.~\eqref{Aeom_v1} reduces to $-\i\,\frac{\dd r}{r^2} \langle h,\Gamma\rangle - \frac{\chi}{2}R(\omega) = \O(r^{-3}).$ In the gauge $\chi=-2$, this gives
\begin{equation}
    R(\omega) = \i\langle h,\Gamma\rangle\,\frac{\dd r }{r^2}+\O(r^{-3}), \qquad \dd\omega = \star_3R(\omega) = \i\langle h,\Gamma\rangle\,{\rm vol}_{S^2} +\O(r^{-1}),
\end{equation}
where ${\rm vol}_{S^2}$ is the volume form on the unit round two-sphere. The NUT charge of the solution is given by the integral of $\dd \omega$ over the sphere at asymptotic infinity, $Q_{\rm NUT} = \frac{1}{4\pi}\int_{S^2_{\infty}} \dd\omega= \i \la h, \Gamma\ra.$ Thus, to ensure asymptotic flatness and remove the Dirac--Misner string at asymptotic infinity, we need to require
\begin{equation}
    \langle h,\Gamma\rangle=0.\label{eq:NUT=0}
\end{equation}
Consequently,
\begin{equation}
    \dd\omega=\O(r^{-1}),
    \qquad
    R(\omega)=\O(r^{-3}).
\end{equation}

At asymptotic infinity, the electromagnetic field strengths take the form
\begin{align}
    \mathbb{F}^{-} &= \mathbb{F}_{\tau r}^{-} \left(\dd\tau\wedge\dd r-r^2{\rm vol}_{S^2} \right),\qquad \mathbb{F}^{+} = \mathbb{F}_{\tau r}^{+} \left( \dd\tau\wedge\dd r+r^2{\rm vol}_{S^2} \right). \label{eq:full_as_F}
\end{align}
Since $R(\omega)=\O(r^{-3})$, their radial components are
\begin{align}
    \mathbb{F}_{\tau r}^{+}
    &=
    \i\left[
        \partial_r\mathcal{Y}
        +
        (\partial_rU)
        \left(\mathcal{Y}+\bar{ \mathcal{Y}}\right)
    \right]
    +\O(r^{-3}),
    \label{Fmrtau_asymp}
    \\
    \mathbb F_{\tau r}^{-}
    &=
    \i\left[
        \partial_r\bar{\mathcal{Y}}
        +
        (\partial_rU)
        \left(\mathcal{Y}+\bar{ \mathcal{Y}}\right)
    \right]
    +\O(r^{-3}).
    \label{Fprtau_asymp}
\end{align}
Using these expressions, we can directly evaluate the pullback of
$\mathbb F=\mathbb F^{+}+\mathbb F^{-}$ on the sphere at asymptotic infinity:
\begin{align}
    \iota_{S_\infty^2}^*\mathbb{F}
    &=
    r^2\partial_r
    \left[
        \i\left(\mathcal{Y}-\bar{\mathcal{Y}}\right)
    \right]
    {\rm vol}_{S^2}
    +\O(r^{-1})
    \nonumber\\
    &=
    r^2\partial_rH\,{\rm vol}_{S^2}
    +\O(r^{-1})
    = -\Gamma\,{\rm vol}_{S^2}
    +\O(r^{-1}).
    \label{eq:F_as_sphere}
\end{align}
Since ${\rm vol}_{\partial M} = -(\dd \tau + \omega)\wedge r^2\,{\rm vol}_{S^2} + \O(r^{-1}),$ orientation of $S^2_{\infty}$ is opposite to that of ${\rm vol}_{S^2}$. We therefore explicitly identify $\Gamma$ as the total electromagnetic charge,
\begin{equation}
    \Gamma = \frac{1}{4\pi}\int_{S^2_\infty}\mathbb{F}\,.\label{eq:EM-charge-integral}
\end{equation}

We denote the asymptotic values of the electric and magnetic potentials by $\phi = (\phi^I,\phi_I)$. From Eq.~\eqref{eq:el_pot}, we obtain
\begin{equation}
    \Phi \xrightarrow{r\to\infty} -\mathcal{Y}^{(0)}-\bar{\mathcal{Y}}^{(0)} \equiv\phi.
\end{equation}

Differentiating \eqref{Deom} with respect to $r$ and extracting the leading $\O(r^{-2})$ terms, we can express the coefficient $U^{(1)}$ in terms of $y^{I(0)},\bar y^{I(0)}$ and $y^{I(1)},\bar y^{I(1)}$:
\begin{align}
    2\partial_rU
    =
    \frac{\i}{r^2}\Big[
        &\bar{\CF}_I(\bar y^{(0)},0)y^{I(1)}
        -
        \CF_I(y^{(0)},0)\bar y^{I(1)}
        \nonumber\\
        &-
        \CF_{IJ}(y^{(0)},0)
        y^{I(1)}\bar y^{J(0)}
        +
        \bar{\CF}_{IJ}(\bar y^{(0)},0)
        \bar y^{I(1)}y^{J(0)}
    \Big]
    +\O(r^{-3}).
    \label{ADM_mass}
\end{align}
At $\Upsilon=\bar\Upsilon=0$, the prepotentials are homogeneous functions of degree two, and therefore
\begin{equation}
    \CF_I(y^{(0)},0)
    =
    \CF_{IJ}(y^{(0)},0)y^{J(0)},
    \qquad
    \bar{\CF}_I(\bar y^{(0)},0)
    =
    \bar{\CF}_{IJ}(\bar y^{(0)},0)\bar y^{J(0)}.
    \label{eq:Euler-asymp}
\end{equation}
Using \eqref{y(0)y(1)_eqs}, we find
\begin{equation}
    \phi_I \Gamma^I =  -\i\left[ \CF_I y^{I(1)} - \CF_I\bar y^{I(1)} + \bar{\CF}_Iy^{I(1)} -\bar{\CF}_I\bar y^{I(1)}\right].\label{eq:phiIP-expand}
\end{equation}
Similarly,
\begin{equation}
    \phi^I\Gamma_I = -\i\left[\CF_Iy^{I(1)} +\CF_{IJ}\bar y^{I(0)}y^{J(1)}-\bar{\CF}_{IJ}y^{I(0)}\bar y^{J(1)}-\bar{\CF}_I\bar y^{I(1)}\right].\label{eq:phiQ-expand}
\end{equation}
All derivatives of $\CF$ and $\bar{\CF}$ in the equations above are evaluated at $(y^{(0)},0)$ and $(\bar y^{(0)},0)$, respectively. Subtracting \eqref{eq:phiIP-expand} from  \eqref{eq:phiQ-expand} gives
\begin{equation}
    \symp{\phi}{\Gamma} = \i\left[\bar{\CF}_Iy^{I(1)} - \CF_I\bar y^{I(1)} - \CF_{IJ}y^{I(1)}\bar y^{J(0)} + \bar{\CF}_{IJ}\bar y^{I(1)}y^{J(0)} \right] = -2U^{(1)}.
    \label{eq:C-symplectic}
\end{equation}
On a constant-$\tau$ slice, the asymptotic spatial metric is
\begin{equation}
    g_{ij}^{(3)} = e^{-2U}\delta_{ij} +\O(r^{-4}) = \delta_{ij} - \frac{2U^{(1)}}{r}\delta_{ij} +\O(r^{-2}).
\end{equation}
Hence, the standard ADM surface integral gives
\begin{align}
    M_{\rm ADM} &= \frac{1}{16\pi} \lim_{r\to\infty} \int_{S^2_r} \left(\partial_j g_{ij}^{(3)} - \partial_i g_{jj}^{(3)} \right) n^i\dd S= -U^{(1)} = \frac{1}{2}\symp{\phi}{\Gamma}.
    \label{eq:MADM-pot-charge}
\end{align}
Since $\phi = -\i f_{\infty}\Omega_{\infty} + \i f_{\infty}^{-1}\bar{\Omega}_{\infty}$ and $h = -f_{\infty}\Omega_{\infty} - f_{\infty}^{-1}\bar{\Omega}_{\infty}$, the regularity condition \eqref{eq:NUT=0} implies that
\begin{equation}
    M_{\rm ADM} = \i f_{\infty} \symp{\Gamma}{\Omega_{\infty}} \equiv \i f_{\infty}\,\mathcal{Z}(\Gamma,\Omega_{\infty}).
\end{equation}
Equivalently, $M_{\rm ADM} = -\i f_{\infty}^{-1}\,\bar{\mathcal{Z}}(\Gamma,\bar\Omega_{\infty})$. For configurations that admit a continuation to the physical Lorentzian branch, $\bar{\mathcal{Z}}(\Gamma,\bar\Omega_{\infty}) = \mathcal{Z}(\Gamma,\Omega_{\infty})^*$. Thus, comparing the expressions for $M_{\rm ADM}$ in terms of $\mathcal{Z}$ and $\bar{\mathcal{Z}}$, we obtain:
\begin{equation}
    f_{\infty} \mathcal{Z}(\Gamma,\Omega_\infty) + f_{\infty}^{-1} \mathcal{Z}(\Gamma,\Omega_\infty)^* = 0.
\end{equation}
This equation has two solutions for $f_{\infty}$. Writing $\mathcal{Z}(\Gamma,\Omega_\infty) = |\mathcal{Z}(\Gamma,\Omega_\infty)|e^{\i\alpha_\infty}$, the two solutions are $f_{\infty} = \pm \i\,e^{-\i\alpha_\infty}.$ They give $\i f_{\infty} \mathcal{Z}(\Gamma,\Omega_\infty) = \mp |\mathcal{Z}(\Gamma,\Omega_\infty)|$. The two signs correspond to the different choices of time orientation. Choosing $f_\infty = -\i \,e^{-\i\alpha_{\infty}}$, we obtain
\begin{equation}
    M_{\rm ADM}(\Gamma, \Omega_\infty) = |\mathcal{Z}(\Gamma,\Omega_{\infty})|,
    \label{eq:ADM-mass}
\end{equation}
which matches the result in the two-derivative theory.

\subsection{The new attractor mechanism}
\label{sec:new-attractor}

In this subsection we analyze the behavior of the stationary half-BPS solutions near the poles of the harmonic functions \eqref{eq:harmonic_fun} and impose the appropriate smoothness conditions. 

First, we use the equation of motion of the $\SO(1,1)_{\rm R}$ connection to ensure the regularity of the one-form $\omega$. Imposing removability of Dirac-Misner strings near each fixed point yields a set of algebraic conditions relating the residues of the harmonic functions to the relative positions of the fixed points. These are the finite-temperature regularity conditions~\cite{Boruch:2025biv} which guarantee the absence of conical singularities at the poles. 
Next, we determine the local conditions required for the four-dimensional geometry and the supergravity fields to remain smooth at each pole of $H$. These conditions fix the allowed leading behavior of $U$, $\omega$, the Weyl-multiplet fields, and the function $f$. We will show that each pole corresponds to a fixed point of the Killing vector $\partial_\tau$ generating Euclidean time translations. 
Finally, by studying the generalized stabilization equations near the poles, we show that the attractor mechanism discovered in \cite{Boruch:2023gfn} in the two-derivative theory extends to arbitrary higher-derivative theories of the type considered here.  The attractor equations fix the black hole dipole moments $\delta_a = \frac{1}{2\i} ( \gamma_{N,a} - \gamma_{S,a})$ in terms of the charges $\Gamma_{a} =  \gamma_{N,a} + \gamma_{S,a}.$ In App.~\ref{app:off-shell-attractor}, using the equivariant formalism of \cite{BenettiGenolini:2026qdm} reviewed in Subsect.~\ref{sec:equivariant-closed-forms}, we provide an alternative derivation of the attractor mechanism, which also clarifies which of the attractor equations hold off shell.

We consider the stationary half-BPS geometry \eqref{12BPS_metric} with $n$ black holes, represented by a symplectic vector of harmonic functions $H$ with $2n$ poles:
\begin{equation}\label{eq:harmonic_fun_1}
    H(\vec{x}) = h +
    \sum_{a=1}^{n}
    \frac{\gamma_{N,a}}
    {|\vec{x}-\vec{x}_{N,a}|}
    +
    \sum_{a=1}^{n}
    \frac{\gamma_{S,a}}
    {|\vec{x}-\vec{x}_{S,a}|} ,
\end{equation}
where each black hole horizon is specified by a pair of poles, which we denote by $N_a$ and $S_a$ (north and south poles).
It is convenient to label the $2n$ poles collectively by
\begin{equation}
    \alpha=(a,\sigma),\qquad \sigma=N,S, 
    \qquad a= 1, \dots ,n ,
\end{equation}
and define
\begin{equation}
    \rho_\alpha=|\vec{x}-\vec{x}_\alpha|,
    \qquad
    R_{\alpha\hat{\alpha}}=|\vec{x}_\alpha-\vec{x}_{\hat{\alpha}}|.
\end{equation}
Near a generic pole $\vec{x}=\vec{x}_\alpha$, the harmonic functions therefore admit the expansion
\begin{equation}\label{eq:H-near-pole}
    H
    =
    \frac{\gamma_\alpha}{\rho_\alpha}
    +
    H_\alpha^{(0)}
    +
    \O(\rho_\alpha),
    \qquad
    H_\alpha^{(0)}
    =
    h
    +
    \sum_{\hat{\alpha}\neq\alpha}
    \frac{\gamma_{\hat{\alpha}}}{R_{\alpha\hat{\alpha}}}.
\end{equation}
For each black hole horizon, the north and south poles are connected by a Dirac--Misner string of the one-form $\omega$, specified through the equation of motion for the $\SO(1,1)_{\rm R}$ connection,
\begin{equation}
    - \frac{\chi}{2} \star_3 \dd \omega =
    \i\symp{\dd H}{H} + 128 \i \star_3 \dd j, 
    \qquad j \equiv \star_3 \left( \frac{1}{2} J_{mn}\dd x^m \wedge \dd x^n \right) ,
    \label{eq:A-eom}
\end{equation}
where $\dd = \dd x^m \partial_m$ and the current $J_{mn}$ is given by \eqref{eq:Aeom_current}. Note that even though the one-form $\omega$ might not exist globally, the current only depends on $R(\omega)$ and therefore can always be specified globally. 
The presence of Dirac--Misner strings is now probed by acting on both sides of the equation with $\dd \star_3$ and integrating over a closed region of the $\mathbb{R}^3$ base stretching from infinity and enclosing a single pole. Applying Stokes' theorem twice leads to 
\begin{equation}
    \frac{\chi}{2} \int_{\partial S \to 0} \omega_{\varphi} \dd \varphi = 4\pi \i \symp{\gamma_\alpha}{H^{(0)}_\alpha} ,
\end{equation}
where we used that $\dd \dd j = 0$, due to the global existence of the current. The nonvanishing value of the integral around the infinitesimal loop indicates the presence of the Dirac--Misner strings. 
To ensure now the lack of conical singularities at each pole, we have to impose that the Dirac--Misner strings of $\omega$ are only coordinate singularities \cite{Hartle:1972ya}. 
The Dirac--Misner strings can be removed by a coordinate transformation of the type $\tau \to \tau - \frac{\chi}{2} (\omega_\varphi|_{\partial S \to 0}) \varphi$, as long as the periods of the time and angular coordinate $\varphi$ around the string are consistent with each other, 
\begin{equation}
 4\pi\i\symp{\gamma_\alpha}{H^{(0)}_\alpha}
 +\frac{s_\alpha \beta \chi}{2}=0 ,\label{eq:veryverycomplicatedargument}
\end{equation}
where $s_{N,a} = +1$ and $s_{S,a} = -1$ for north and south poles, respectively. With this, locally in spherical coordinates $\theta_{\alpha},\varphi_{\alpha}$, in the gauge where the Dirac--Misner string stretches between north and south poles, we can write $\omega_{\varphi} = s_{\alpha} \frac{\beta}{4\pi}  (\cos\theta_{\alpha} - 1)$. In the gauge where $\chi = -2$ Eq.~\eqref{eq:veryverycomplicatedargument} becomes
\begin{equation}
    \symp{\gamma_\alpha}{h} +\sum_{\hat{\alpha}\neq\alpha}\frac{\symp{\gamma_\alpha}{\gamma_{\hat{\alpha}}}} {R_{\alpha\hat{\alpha}}} =\frac{s_\alpha\beta}{4\pi\i}. \label{eq:position}
\end{equation}
These are exactly the finite-temperature regularity conditions derived in  the two-derivative supergravity theory in \cite{Boruch:2023gfn,Boruch:2025biv}, now derived for any higher-derivative conformal supergravity theory. 
We find that, due to the global existence of the current \eqref{eq:Aeom_current}, the structural form of the regularity conditions remains the same in the higher-derivative theory.
Summing \eqref{eq:position} over all $2n$ poles, all pairwise terms cancel, and since $\sum_\alpha s_\alpha=0$ we find
\begin{equation}
    \symp{\Gamma}{h}=0. \label{eq:noNUT}
\end{equation}
This is again the total no-NUT condition \eqref{eq:NUT=0}, encountered above, guaranteeing that no Dirac--Misner strings reach infinity.  One linear combination of the $2n$ equations \eqref{eq:position} is therefore redundant.  The remaining equations constrain the $3(2n)-3$ relative coordinates of the poles, but in general do not fix them all \cite{Boruch:2025biv}. The unfixed relative-position parameters constitute the moduli space of the multi-center half-BPS solution.

Next, to ensure the smoothness of the geometry we also need to guarantee that the near-horizon throat caps off smoothly at a finite distance in the bulk, that is, we expect that near the poles the geometry locally looks like $\mathbb{R}^4$. For this, we need the local behavior of $U$ to be
\begin{align}
    e^{-2U}=\frac{L_{\alpha}}{\rho_\alpha}+\O (1), 
\qquad
    \omega=s_\alpha L_{\alpha}\cos\theta_\alpha\,\dd\varphi_\alpha+\O(\rho_{\alpha}^{\epsilon}), 
    \qquad \epsilon>0,
    \label{eq:omegalocal}
\end{align}
where we note that the pole of the harmonic function entering the metric as a first-order pole guarantees the finite length of the throat in the near-horizon region. 
We will now use \eqref{eq:omegalocal} to determine the local behaviors of all the other fields and subsequently verify their self-consistency with the remaining equations of motion \eqref{Deom}, \eqref{eq:SO11_EOM}. 

Starting with the local metric, using  \eqref{eq:omegalocal} and changing coordinates to $R_\alpha=2\sqrt{L_\alpha \rho_\alpha}$ and $\psi=\tau/L_\alpha$, \eqref{12BPS_metric} near the pole takes the form
\begin{equation}
    \dd s^2=\dd R_\alpha^2+\frac{R_\alpha^2}{4} \left[\dd\theta_\alpha^2+\sin^2\theta_\alpha\dd\varphi_\alpha^2  +(\dd\psi+s_\alpha\cos\theta_\alpha\dd\varphi_\alpha)^2\right] + \cdots .
 \label{eq:R4local}
\end{equation}
The expression in brackets is four times the metric on the unit three-sphere, written as a Hopf fibration. Smoothness at $R_\alpha=0$ therefore requires $\psi$ to have period $4\pi$, and hence
\begin{equation}
    \tau \sim \tau + \beta,\qquad \beta = 4\pi L_\alpha \equiv 4\pi L ,
\end{equation}
consistently with the regularity condition derived above \eqref{eq:position}. 
Since $\dd \omega = -s_\alpha L_\alpha\sin\theta_\alpha\,\dd\theta_\alpha\wedge \dd\varphi_\alpha$, we find
\begin{equation}
    R(\omega) = -s_\alpha L_\alpha\frac{\dd\rho_\alpha}{\rho_\alpha^2} + \cdots,
\end{equation}
where the ellipses denote terms that are regular at $\rho_\alpha = 0$. Together with $e^{2U}=\rho_\alpha/L_\alpha+\O(\rho_\alpha^2)$ and $\dd U=\dd\rho_\alpha/(2\rho_\alpha)+\O(1)$, \eqref{Tplus_sol} and \eqref{Tminus_sol} give
\begin{equation}
    T^+_m \dd x^m = -\frac{1 + s_\alpha}{2}\frac{\dd\rho_\alpha}{\rho_\alpha} + \O(1),\qquad T^-_m\dd x^m = \frac{1 - s_\alpha}{2}\frac{\dd\rho_\alpha}{\rho_\alpha} + \O(1).
\end{equation}
Recall that $s_{N,a} = +1$ at the north poles $\vec{x} = \vec{x}_{N,a}$ and $s_{S,a} = -1$ at the south poles $\vec{x} = \vec{x}_{S,a}$.\footnote{If the number of north poles were different from the number of south poles, the resulting regularity conditions would not be compatible with the zero NUT-charge condition \eqref{eq:NUT=0} and the spacetime would have Taub--NUT asymptotics. In the context of five-dimensional uplifts, one can also consider including unsplit centers in the geometry subject to the integrability condition \cite{Boruch:2025sie}. As we work in four dimensions, these will not be relevant for us in this paper.} Then, near the north pole
\begin{equation}
    {\Upsilon} \overset{\rho_{N,a}\to 0}{\sim} -\frac{64}{\rho_{N,a}^2} + \O(\rho_{N,a}^{-1}),
\end{equation}
and near the south pole
\begin{equation}
    \bar\Upsilon \overset{\rho_{S,a}\to 0}{\sim}-\frac{64}{\rho_{S,a}^2} + \O(\rho_{S,a}^{-1}).
\end{equation}
Requiring that ${A}|_{{\mathbb W}^2} = -f^{-2}e^{2U} \Upsilon$ and $\bar A|_{\overline{\mathbb{W}}^2} = -f^{2}e^{2U}\bar\Upsilon$ are finite at the north and south poles, respectively, implies that $f$ must scale as:
\begin{equation}
    f \overset{\rho_{N,a}\to 0}{\sim} \frac{1}{\sqrt{\rho_{N,a}}},\qquad f \overset{\rho_{S,a}\to 0}{\sim} \sqrt{\rho_{S,a}}.
\end{equation}
Recall that $T^+_{m\tau} = 4f^{-1}e^{U}T_m^+$ and $T^-_{m\tau} = 4f e^{U} T^-_m$, and since $T^{\pm}_{\mu\nu}$ must be regular at the north and south poles, we need
\begin{equation}
    T^-_{m} \overset{\rho_{N,a}\to 0}{\sim} \O(\rho^{1/2}_{N,a}),\qquad T^+_{m} \overset{\rho_{S,a}\to 0}{\sim} \O(\rho^{1/2}_{S,a}).
\end{equation}
Consequently,
\begin{equation}
    \bar\Upsilon \overset{\rho_{N,a}\to 0}{\sim} \O(\rho_{N,a}),\qquad \Upsilon \overset{\rho_{S,a}\to 0}{\sim} \O(\rho_{S,a}).
\end{equation}

From the embedding condition \eqref{eq:Killing-spinor-embedding}, we see that for a finite positive-chirality component $\epsilon_+^i$ of the Killing spinor $\epsilon^i$, the negative-chirality component $\epsilon_-^i$ must vanish near the north poles as, %
\begin{equation}
    \epsilon_-^i \overset{\rho_{N,a}\to 0}{\sim} \sqrt{\rho_{N,a}}.
\end{equation}
Analogously, assuming that $\epsilon_-^i$ is finite at the south poles, we obtain:
\begin{equation}
    \epsilon_+^i \overset{\rho_{S,a}\to 0}{\sim} \sqrt{\rho_{S,a}}.
\end{equation}
The Killing spinor $\epsilon^i$ has positive (negative) chirality at the north (south) poles. The Killing vector associated with $\epsilon^i$ is
\begin{equation}
    \xi^a = 2\i\bar{\epsilon}_{i+}\gamma^a \epsilon^i_-.
\end{equation}
Clearly, if either chiral component of the Killing spinor vanishes, $\xi^a$ also vanishes. We learn that the north poles and south poles are fixed points of the isometry generated by $\xi$. We will refer to the north poles $N_a$ as positive-chirality fixed points and to $S_a$ as negative-chirality fixed points.
This establishes that the north and south poles are nuts, in agreement with the two-derivative analysis of \cite{Boruch:2023gfn,Boruch:2025biv}. The geometry does not have any fixed surfaces for $\xi$ and therefore has no bolts.

Finally, we can now study the stabilization equations
\begin{equation}
    \i\left(\mathcal Y-\bar{\mathcal Y}\right)=H ,
    \label{eq:stab2}
\end{equation}
near a pole $\vec x=\vec x_\alpha$ of $H$. 
Similarly to what happens in the two-derivative case \cite{Boruch:2023gfn}, this will determine the coefficients $\gamma_\alpha$ in terms of the charges $\Gamma_a$ of individual black holes.
Consider first a positive-chirality fixed point $\alpha=(a,N)$. Recall that
\begin{equation}
    \mathcal Y=\i e^{-U}f\Omega,
    \qquad
    \bar{\mathcal Y}=-\i e^{-U}f^{-1}\,\bar\Omega.
\end{equation}
Since $e^{-U}\sim\sqrt{L/\rho_{N,a}}$ and $f\sim\rho_{N,a}^{-1/2}$, while the physical period vectors $\Omega$ and $\bar\Omega$ remain regular, the upper components of the rescaled period vectors behave as
\begin{equation}
    {\mathcal Y}^I
    =
    \frac{{\mathcal Y}_{*N_a}^I}{\rho_{N,a}}
    +\O(1),\qquad \bar{\mathcal Y}^I=\O(1),
    \label{north_Y_expansion}
\end{equation}
As shown above, the corresponding Weyl-squared fields satisfy
\begin{equation}
    \Upsilon
    =
    -\frac{64}{\rho_{N,a}^2}
    +\O(\rho_{N,a}^{-1}),
    \qquad
    \bar\Upsilon=\O(\rho_{N,a}).
    \label{north_upsilon_expansion}
\end{equation}
Substituting \eqref{north_Y_expansion} and \eqref{north_upsilon_expansion} into \eqref{eq:stab2}, and using the homogeneity of the prepotential, the leading terms in the $\rho_{N,a}\to0$ expansion give
\begin{equation}
    {\mathcal Y}_{*N_a}^I=-\i\,\gamma_{N,a}^I,
    \qquad
    {\mathcal F}_I
    \left({\mathcal Y}_{*N_a},-64\right)
    =
    -\i\,\gamma_{N,a,I}.
    \label{eq:attractor-N}
\end{equation}
Equivalently, using weighted homogeneity of $\CF$ once more,
\begin{equation}
    \gamma_{N,a,I}
    =
    {\mathcal F}_I(\gamma_{N,a}^J,64).
    \label{eq:attractor-N-charge}
\end{equation}
At a negative-chirality fixed point $\alpha=(a,S)$, the roles of the
chiral and anti-chiral variables are interchanged. We have
\begin{equation}
    {\mathcal{Y}}^I = \O(1),\qquad \bar{\mathcal{Y}}^I =  \frac{\bar{\mathcal Y}_{*\,S_a}^I}{\rho_{S,a}} + \O(1),\label{south_Y_expansion}
\end{equation}
together with
\begin{equation}
    {\Upsilon} = \O(\rho_{S,a}),\qquad\bar\Upsilon = -\frac{64}{\rho_{S,a}^2} + \O(\rho_{S,a}^{-1}).\label{south_upsilon_expansion}
\end{equation}
The leading terms in the stabilization equations then imply,
\begin{align}
    \bar{\mathcal Y}_{*\,S_a}^I = \i\,\gamma_{S,a}^I,\qquad \bar\CF_I(\bar{\mathcal Y}_{*\,S_a}^J,-64) = \i\,\gamma_{S,a\,I}.\label{eq:attractor-S}
\end{align}
Hence,
\begin{equation}
    \gamma_{S,a,I} = \bar \CF_I(\gamma_{S,a}^J,64).\label{eq:attractor-S-charge}
\end{equation}
These equations determine the north- and south-pole residues $\gamma_{N,a}^I$ and $\gamma_{S,a}^I$ in terms of black hole charges $\Gamma_a$ via the following system of equations
\begin{align}
    \Gamma^I_a &= \gamma_{N,a}^I + \gamma^I_{S,a},\label{eq:sum_upper_gamma}
    \\
    \Gamma_{a,I} &= {\CF}_I(\gamma_{N,a}^J,64) + \bar{\CF}_I(\gamma_{S,a}^J,64).\label{eq:sum_lower_gamma}
\end{align}
Then $\gamma_{N,a,I}$ and $\gamma_{S,a,I}$ are determined by \eqref{eq:attractor-N-charge} and \eqref{eq:attractor-S-charge}. We thus find that the finite-temperature attractor mechanism of \cite{Boruch:2023gfn} persists in a generic higher-derivative theory. The higher-derivative couplings modify the algebraic attractor equations through the dependence of the prepotential on the Weyl-squared field,\footnote{The analogous analysis for the attractor mechanism at extremality was discussed in \cite{LopesCardoso:1998tkj}.} but the essential attractor property is unchanged: the physical scalars at each horizon are determined solely by the corresponding local charges, and are independent of the asymptotic moduli, temperature, and the charges of the other black holes.

If the electric and magnetic charges are fixed, Eqs.~\eqref{eq:attractor-N-charge}, \eqref{eq:attractor-S-charge}, \eqref{eq:sum_upper_gamma} and \eqref{eq:sum_lower_gamma} fully specify the data required to solve the remaining equations of motion \eqref{Deom}, \eqref{eq:SO11_EOM} for $U$ and $\omega$, and therefore fully determine the metric of each saddle.
This is not however the relevant ensemble for OSV where instead of fixing the electric charge we want to fix the electric chemical potential: to do so, we proceed by explicitly writing the regular gauge field configurations for such saddles to determine how the residues $\gamma_{N,a}$ and $\gamma_{S,a}$ of the harmonic function $H$ are fixed as a function of the chemical potentials.

\subsection{Regularity conditions for the gauge fields and shifted saddles}\label{sec:shifted_saddles1}

We now investigate the regularity conditions for the vector multiplet gauge fields $W^I$. The action \eqref{eq:Itot_dirichlet}, with no electromagnetic boundary terms, naturally defines a variational problem with fixed electric potentials $\mu^I$ and magnetic charges $\Gamma^I$ at asymptotic infinity $\partial M = S^1_\beta \times S^{2}_{\infty}$. We stress that magnetic charges of individual black holes, $\Gamma_a^I$, are not specified by asymptotic boundary conditions, and therefore we need to sum over all allowed $\Gamma_a^I$. As discussed in App.~\ref{app:KerrNewman_puzzle}, for nonzero magnetic charge $\Gamma^I,$ the asymptotic gauge fields must be specified in patches. On the northern and southern hemispheres of $S^2_{\infty}$, the gauge field $W^I$ takes the form
\begin{equation}
    W_{{\rm N}/{\rm S}}^I = 2\i\mu^I\dd\tau + \Gamma^I(\cos\theta \mp 1)\,\dd\varphi + \O(r^{-1}),\label{eq:Wasymptotic}
\end{equation}
with the minus sign in the northern patch and the plus sign in the southern patch. Here $\theta,\varphi$ are the spherical coordinates on $S^2_{\infty}.$ The $\tau$-component of the gauge field in the bulk satisfying these boundary conditions is:
\begin{equation}
    W^I_\tau = 2\i\mu^I - \i(\phi^I - \Phi^I),
\end{equation}
with $\Phi^I$ given by \eqref{eq:el_pot}, and $\phi^I = \lim_{r\to \infty}\Phi^I.$

In corotating coordinates associated with the $a$-th black hole, the temporal circle is contractible on the horizon of the $a$-th black hole. In the static coordinates, this contractible cycle is a linear combination of the $\tau$-circle and the azimuthal $\varphi$-circle around the rotation axis of the $a$-th black hole. At the north and south poles of the $a$-th horizon, however, the $\varphi$-circle contracts, so the holonomy of a regular gauge field around this circle must be trivial there. It follows that, at either pole, the holonomy along the contractible corotating thermal circle is equal to the holonomy along the static $\tau$-circle. Since the former must be trivial by smoothness, the holonomy around the static $\tau$-circle is therefore trivial at every north and south pole $\alpha$. Let $q_I$ be a unit of electric charge flux of the $I$-th gauge field, and let $\Lambda$ be the lattice generated by the unit charge vectors $(q_0,0,\dots,0),\dots,(0,\dots,0,q_{n_v})$. Then, the holonomy of the gauge field around the temporal circle at $\alpha$ is:
\begin{equation}
    \mathcal W = \exp[-\frac{\i}{2} q_I\oint_{S^1_\beta} W^I] = \exp[\beta q_I (\mu^I - \tfrac{1}{2}(\phi^I - \Phi_\alpha^I))],
    \label{eq:holonomy}
\end{equation}
where $\Phi^I_{\alpha}$ is the value of the function $\Phi^I$ at the fixed point $\alpha$, $\Phi_\alpha = \Phi(\vec{x}_{\alpha})$. To have a regular gauge field configuration we need to require that $\mathcal W = 1$. Equivalently,
\begin{equation}
    \tfrac{1}{2} \beta\Phi^I_{\alpha} = \beta(\tfrac{1}{2}\phi^I - \mu^I) + 2\pi \i n_{\alpha}^I.\label{eq:gauge-field-reg-cond}
\end{equation}
Here $n_{\alpha}^I$ belongs to the lattice $\Lambda^*$ dual to the charge lattice $\Lambda$, i.e., $n^I_{\alpha} q_I \in \mathbb{Z}$ for all $q \in \Lambda$. Configurations with  $n_{\alpha}^I=0$ do not require additional patches to specify the gauge field configuration whereas in configurations with  $n_{\alpha}^I\neq 0$ we can define a new small patch surrounding the north and south poles where $W^I_\tau =0$. This patch can indeed be glued to the bulk gauge field, whose holonomy is given by \eqref{eq:holonomy}, through a large gauge transformation when  $n_{\alpha}^I \in \Lambda^*$.  

Using Eqs.~\eqref{eq:attractor-N}, \eqref{eq:attractor-S} and \eqref{eq:el_pot}, one can easily show that
\begin{equation}
    \Phi^I_{N,a} = \frac{4\pi \i}{\beta}\gamma_{N,a}^I,\qquad \Phi_{S,a}^I = -\frac{4\pi \i}{\beta}\gamma_{S,a}^I,
\end{equation}
and consequently,
\begin{align}
    \gamma_{N,a}^I &= n_{N,a}^I + \frac{\i \beta}{2\pi}(\mu^I - \tfrac{1}{2}\phi^I) ,
    \label{eq:shifted_gammaN}
    \\
    \gamma^I_{S,a} &= - n_{S,a}^I
    - \frac{\i \beta}{2\pi}(\mu^I - \tfrac{1}{2}\phi^I)
    .
    \label{eq:shifted_gammaS}
\end{align}

Taking the difference of the two equations, we find that the difference between $n_{N,a}^I$ and $n_{S,a}^I$ is given by the magnetic charge of the $a$-th black hole,
\begin{equation}
    \Gamma^I_a = \gamma_{S,a}^I + \gamma_{N,a}^I = n_{N,a}^I - n_{S,a}^I.
\end{equation}
This directly implies the Dirac quantization condition,
\begin{equation}
    q_I \Gamma^I_a = q_I (n^I_{N,a} - n^I_{S,a}) \in \mathbb{Z}.
\end{equation}
It also implies that once we fix the magnetic charge of each black hole, the quantized vectors $n^I_{N,a}$ and $ n^I_{S,a}$ cannot be chosen independently. Taking the sum of the two equations we can find the dipole charge $\delta^I_a$ of each black hole, 
\begin{equation}\label{eq:deltaaIrel}
    \delta_a^I = \frac{\beta}{2\pi}(\mu^I - \tfrac{1}{2}\phi^I) - \frac{\i}{2}(n_{N,a}^I + n_{S,a}^I).
\end{equation}
Together with \eqref{eq:attractor-N-charge}, \eqref{eq:attractor-S-charge}, this fully fixes the residues of the harmonic function $H$ and consequently the metric, scalar and gauge field configurations. To summarize, we constructed a family of regular saddles, labeled by the electric potentials $\mu^I$, the magnetic charges 
\begin{equation}
    \Gamma_a^I \in \Lambda^*,\qquad \Gamma^I = \sum_{a=1}^n \Gamma_a^I,
\end{equation}
shifts of electric potentials $n_{N,a}^I \equiv -n_{a}^I \in \Lambda^*$ (or equivalently, dipole charges $\delta_a^I$ such that $\delta^I_a -\delta_b^I \in \frac{\i}{2}\Lambda^*$), and the positions of the poles $\vec{x}_{\alpha}$. The positions $\vec{x}_{\alpha}$ are subject to the smoothness conditions~\eqref{eq:position}, but they may parametrize a continuous moduli space. The potentials $\phi^I$ are determined by the asymptotic boundary conditions for the vector multiplet scalars, and the residues $\gamma^I_{N,a}$, $\gamma^I_{S,a}$ are given by \eqref{eq:shifted_gammaN} and \eqref{eq:shifted_gammaS}. The parameters $\gamma_{N,a,I}$ and $\gamma_{S,a,I}$ are determined by the attractor equations \eqref{eq:attractor-N-charge} and \eqref{eq:attractor-S-charge}. We now proceed with the evaluation of the on-shell action on these regular Euclidean gravitational saddles.

\section{Equivariant localization of the on-shell action }
\label{sec:equivariant-localization-section}

\subsection{Equivariantly closed forms in conformal supergravity}
\label{sec:equivariant-closed-forms}

In this subsection we summarize the equivariant localization results of \cite{BenettiGenolini:2026qdm,Sparks:2026taw}. We also discuss a new equivariantly closed form arising from the scalar boundary term \eqref{eq:new-scalar-bdry-term}. This was not explicitly discussed in \cite{BenettiGenolini:2026qdm,Sparks:2026taw}.\footnote{Our calculations may also have applications and implications beyond the analysis of flat-space black holes in $\mathcal N=2$ SUGRA, in particular, for asymptotically AdS$_4$ backgrounds.}

On a generic supersymmetric bosonic background, we can define the following bilinears of the Killing spinor $\epsilon^i$,
\begin{equation}
    S_{\pm} = \bar{\epsilon}_{i\pm}\epsilon^i_{\pm},\quad \xi_a = 2\i\bar{\epsilon}_{i+}\gamma_a \epsilon_-^i,\quad \Xi_{\pm i}{}^j = \frac{1}{2}(\bar{\epsilon}_{i\pm}\gamma_{ab}\epsilon^j_{\pm})\,e^a \wedge e^b.\label{eq:kill_spin_bilinears}
\end{equation}
As shown in \cite{BenettiGenolini:2026qdm,Sparks:2026taw}, given a weight-two anti-chiral multiplet
\begin{equation}
    \Phi = (A,B^{ij},F^+,C),
\end{equation}
supersymmetry allows one to complete its Lagrangian four-form into an equivariantly closed polyform,
\begin{equation}
    \Psi = \Psi_4 + \Psi_2 + \Psi_0,\qquad (\dd-\iota_\xi)\Psi = 0.
\end{equation}
Here the subscripts $n=0,2,4$ denote the degrees of the forms constituting $\Psi$. In our normalization conventions for the Lagrangian, the forms $\Psi_n$ are given by:
\begin{align}
    \Psi_4 &= \frac{1}{2}\left(C + \frac{1}{16}A\,(T_{ab}^{-})^2\right)\rm{vol}_{M},\label{Psi4}
    \\
    (\Psi_2)_{ab} &= 2 S_{+}(F_{ab}^{+} + \frac{1}{2}A\,T^{-}_{ab}) - B^{ij} \varepsilon_{jk}\Xi_{+ i}{}^k{}_{ab},\label{Psi2}
    \\
    \Psi_0 &= -4A\,S_{+}^2.\label{Psi0}
\end{align}
Similarly, for a chiral multiplet
\begin{equation}
    \bar \Phi = (\bar A,\bar{B}^{ij},F^-,\bar C),
\end{equation}
we have $(\dd - \iota_\xi)\bar{\Psi} = 0$ with
\begin{align}
    \bar \Psi_4 &= \frac{1}{2}\left(\bar C + \frac{1}{16}\bar A\,(T_{ab}^{+})^2\right)\rm{vol}_{M},\label{barPsi4}
    \\
    (\bar \Psi_2)_{ab} &= 2 S_{-}(F_{ab}^{-} + \frac{1}{2}\bar A\,T^{+}_{ab}) - \bar B^{ij} \varepsilon_{jk}\Xi_{- i}{}^k{}_{ab},\label{barPsi2}
    \\
    \bar \Psi_0 &= 4\bar A\,S_{-}^2.\label{barPsi0}
\end{align}
These relations follow entirely from the off-shell supersymmetry transformations, and do not require imposing equations of motion.

In four dimensions, a connected component of the fixed-point set of $\xi$ is either an isolated point, a so-called ``nut'', or a two-dimensional surface, a so-called ``bolt'' \cite{Gibbons:1979xm}. We are interested in the class of saddles without bolts. The linearized action of the $U(1)$ isometry generated by $\xi$ on the tangent space at a nut $p$, $T_pM\simeq \mathbb{R}^4$, splits into rotations in two mutually orthogonal two-planes, $T_pM \simeq \mathbb R^2_1\oplus\mathbb R^2_2.$ Introducing polar coordinates $(\rho_i,\varphi_i)$ on each $\mathbb R_i^2$, the Killing vector takes the local form $\xi|_p= b_1^p \partial_{\varphi_1} + b_2^p \partial_{\varphi_2}$, where the coefficients $b_1^p, b_2^p$ are defined as the skew-eigenvalues of the two-form $\frac{1}{2}\dd(\xi_\mu \dd x^\mu)$.

Without bolts, the localization formula of \cite{BenettiGenolini:2026qdm} simplifies to:
\begin{equation}
    \int_M\Psi = I_{\rm nuts} + I_{\partial M},\qquad \int_M\bar{\Psi} = \bar{I}_{\rm nuts} + \bar{I}_{\partial M},\label{eq:loc_formula}
\end{equation}
where the boundary contributions $I_{\partial M}$, $\bar I_{\partial M}$ take the form
\begin{align}
    I_{\partial M} &= -\int_{\partial M}\frac{\xi_\mu \dd x^\mu}{\|\xi\|^2} \wedge \left(\Psi_2 + \Psi_0 \wedge \frac{\dd(\xi_\nu \dd x^\nu)}{\|\xi\|^2}\right),\label{eq:Ibdry}
    \\
    \bar I_{\partial M} &= -\int_{\partial M}\frac{\xi_\mu \dd x^\mu}{\|\xi\|^2} \wedge \left(\bar \Psi_2 + \bar \Psi_0 \wedge \frac{\dd(\xi_\nu \dd x^\nu)}{\|\xi\|^2}\right),\qquad \|\xi\|^2 = \xi_\mu \xi^\mu,\label{eq:barIbdry}
\end{align}
and the fixed-point contributions $I_{\rm nuts}$, $\bar I_{\rm nuts}$ are given by
\begin{align}
    I_{\rm nuts} &= \sum_{\text{nuts }p}\frac{4\pi^2}{b_1^p b_2^p}\,\Psi_0|_p,\qquad \bar I_{\rm nuts} = \sum_{\text{nuts }p}\frac{4\pi^2}{b_1^p b_2^p}\bar{\Psi}_0|_p.\label{eq:Inuts}
\end{align}

We are interested in the chiral multiplets whose lowest components are the prepotentials,
\begin{equation}
    A|_{\CF} = \CF (X^I,A|_{\mathbb W^2}),\qquad \bar A|_{\bar \CF} =  \bar \CF (\bar X^I,\bar A|_{\overline{\mathbb W}^2}).\label{eq:A_prepot}
\end{equation}
Then, to determine the boundary terms \eqref{eq:Ibdry} and \eqref{eq:barIbdry}, we need to explicitly evaluate $\Psi_2$ and $\bar{\Psi}_2$, given by \eqref{Psi2} and \eqref{barPsi2}. To this end, we apply the standard product rules for chiral multiplets (see Sec.~4.1 of \cite{deWit:2017cle}). For the composite multiplet associated with $\CF$, the constituent chiral multiplets are the reduced chiral multiplets $\mathbb V^I$ associated with the vector multiplets and the Weyl-squared multiplet $\mathbb W^2$. These product rules give
\begin{equation}
    B^{ij}|_{\CF}
    =
    \CF_I\,B^{ij}|_{\mathbb V^I}
    +
    \CF_A\,B^{ij}|_{\mathbb W^2},
    \qquad F^+_{ab}|_{\CF} =
    \CF_I\,F^+_{ab}|_{\mathbb V^I}
    +
    \CF_A\,F^+_{ab}|_{\mathbb W^2}.
    \label{eq:F-composite}
\end{equation}
The bosonic components of the reduced vector multiplet are
\begin{equation}
    B^{ij}|_{\mathbb V^I}
    =
    -Y^{I,ij},
    \qquad
    F^+_{ab}|_{\mathbb V^I}
    =
    F^{I+}_{ab}
    -
    \frac14\bar X^I T^+_{ab},
    \label{eq:vector-reduced-chiral}
\end{equation}
while the components of the Weyl-squared multiplet relevant for
$\Psi_2$ are
\begin{equation}
    B^{ij}|_{\mathbb W^2}
    =
    16\,T^{+ab}\,
    \varepsilon^{k(i}
    R(\mathcal V)_{ab}{}^{j)}{}_{k},\qquad 
    F^+_{ab}|_{\mathbb W^2} =
    -16\,\widehat R(M)_{ab}{}^{cd}T^+_{cd}.
    \label{eq:W2-F}
\end{equation}
Consequently,
\begin{align}
    B^{ij}|_{\CF}
    &=
    -\CF_I Y^{I,ij}
    +
    16\CF_A T^{+ab}
    \varepsilon^{k(i}R(\mathcal V)_{ab}{}^{j)}{}_{k},
    \label{eq:B-F-explicit}
    \\
    F^+_{ab}|_{\CF}
    &=
    \CF_I
    \left(
        F^{I+}_{ab}
        -
        \tfrac14\bar X^I T^+_{ab}
    \right)
    -
    16\CF_A\widehat R(M)_{ab}{}^{cd}T^+_{cd}.
    \label{eq:F-F-explicit}
\end{align}
Similarly, for the chiral multiplet $\bar{\Phi}|_{\bar \CF}$ we obtain: %
\begin{align}
    \bar B^{ij}|_{\bar \CF}
    &=
    -\bar \CF_I Y^{I,ij}
    +
    16\bar \CF_A T^{-ab}
    \varepsilon^{k(i}R(\mathcal V)_{ab}{}^{j)}{}_{k},
    \label{eq:barB-barF-explicit}
    \\
    F^-_{ab}\big|_{\bar \CF}
    &=
    \bar \CF_I
    \left(
        F^{I-}_{ab}
        -
        \tfrac14 X^I T^-_{ab}
    \right)
    -
    16\bar \CF_A\widehat R(M)_{ab}{}^{cd}T^-_{cd}.
    \label{eq:barF-barF-explicit}
\end{align}

\paragraph{Boundary term for scalar fields: }The scalar boundary term \eqref{eq:new-scalar-bdry-term} may also be
incorporated into the equivariant description. Writing
\begin{equation}
    I_{\partial M}^X=\int_{\partial M}\Theta_X,\qquad \Theta_X = -\frac{\i}{8\pi}\star (\bar \CF_I \CD_\mu X^I - \CF_I \CD_\mu \bar X^I)\dd x^\mu
\end{equation}
and extending $\Theta_X$ into the bulk, we may associate to it the
equivariantly exact polyform
\begin{equation}
    \Psi_X=\dd_\xi\Theta_X =\dd\Theta_X-\iota_\xi\Theta_X .
\end{equation}
Indeed, on a supersymmetric background $\mathcal L_\xi\Theta_X=0$, and
hence
\begin{equation}
    \dd_\xi\Psi_X = \dd_\xi^2\Theta_X = -\mathcal L_\xi\Theta_X =0.
\end{equation}
Thus any $\xi$-invariant boundary term of this type has an immediate
equivariantly closed completion. In this case the BVAB formula is essentially tautological: $(\Psi_X)_0=0$, so there are no nut contributions, while the boundary contribution reduces precisely to $\int_{\partial M}\Theta_X=I_{\partial M}^X$.

\subsection{The on-shell action in ungauged SUGRA}
\label{sec:on-shell-action-in-flatspace-SUGRA}

To determine the contribution of the multi-center black-hole saddles to the supersymmetric index, we now evaluate their Euclidean on-shell action. In the path integral, we fix the inverse temperature $\beta$, the asymptotic electric potentials $\lim_{r\to\infty}W_\tau^I=2 \i\mu^I$, the total magnetic charges $\Gamma^I$, and the asymptotic scalar values $y^{I(0)}$, $\bar y^{I(0)}$. These boundary data admit the family of shifted saddles described in Subsec.~\ref{sec:shifted_saddles1}. Here we focus on evaluating the on-shell action on a single member of this family, specified by magnetic charges $\Gamma_a^I$ and holonomy-shift vectors $n_a^I$. We will refer to this data as the ``mixed ensemble.'' The corresponding action naturally decomposes as
\begin{equation}
    I_{\rm mixed}
    =
    I_{\mathbb{VW}}
    +
    I_{\partial M}^X
    +
    I_{\mathbb H}
    +
    I_{\rm GHY}^{\rm HD},
    \label{eq:onshell-action-decomposition}
\end{equation}
where $I_{\mathbb{VW}}$ denotes the vector- and Weyl-multiplet bulk action given by \eqref{eq:IVW}, $I_{\partial M}^X$ is the scalar boundary term introduced in \eqref{eq:new-scalar-bdry-term}, $I_{\mathbb H}$ is the hypermultiplet action and $I_{\rm GHY}^{\rm HD}$ denotes a regularized gravitational boundary term appropriate to the higher-derivative theory. We use background subtraction \cite{Gibbons:1976ue} to regularize $I_{\rm GHY}^{\rm HD}$. We stress that \eqref{eq:onshell-action-decomposition} depends on all gauge field parameters $\Gamma_a^I$ and $n_a^I$, not only on the asymptotic holonomy $\mu^I$ and the total magnetic charge $\Gamma^I = \sum_a \Gamma^I_a$.

\paragraph{Hypermultiplet action: }
First, let us note that the hypermultiplet part of the action vanishes. This is easy to see in the gauge where the hypermultiplet scalars are constant, $A_i^\alpha = (-\chi)^{1/2}\delta_i^\alpha$. Since $\mathcal V_\mu{}^i{}_j = 0$ we have $\CD_\mu A_i^\alpha = \partial_\mu A_i^{\alpha} -\mathcal V_\mu{}^j{}_{i}A_j^\alpha =0$. The equations of motion for the scalars $A_i^\alpha$ yield $R = 3D$, and thus,
\begin{equation}
    I_{\mathbb{H}} = -\frac{1}{8\pi}\int\dd^4x\,\sqrt{g}\left\{\chi\left(-\frac16R+\frac12D\right) -\frac12\varepsilon^{ij}\varepsilon_{\alpha\beta}(\CD_\mu A_i{}^\alpha)(\CD^\mu A_j{}^\beta)\right\} = 0.\label{eq:hyper=0}
\end{equation}
This result is in harmony with the analysis in~\cite{Binder:2021euo} where it was shown that hypermultiplets do not affect the on-shell action of supersymmetric asymptotically AdS$_4$ two-derivative Euclidean supergravity backgrounds.

\paragraph{Bulk contribution: }
Using the localization formula \eqref{eq:loc_formula}, we may write the action $I_{\mathbb{VW}}$ as
\begin{equation}
    I_{\mathbb{VW}} = -\frac{\i}{8\pi}\int_M(\Psi_4 - \bar \Psi_4) = -\frac{\i}{8\pi}(I_{\rm nuts} - \bar{I}_{\rm nuts} + I_{\partial M} - \bar{I}_{\partial M}),
\end{equation}
with $I_{\rm nuts}, \bar{I}_{\rm nuts}, I_{\partial M}$ and $\bar{I}_{\partial M}$ given in \cref{eq:Inuts,eq:Ibdry,eq:barIbdry}.

Let us first focus on the fixed-point contributions. We need to calculate the weights $b_1^p$, $b_2^p$ and the Killing-spinor bilinears \eqref{eq:kill_spin_bilinears}. Using the embedding condition \eqref{eq:Killing-spinor-embedding}, we obtain
\begin{equation}\label{eq:SminSplrel}
    S_- = f^{-2}\bar{\epsilon}_{i+}\epsilon^i_{+} = f^{-2}S_+.
\end{equation}
The metric \eqref{12BPS_metric} possesses the Killing vector $\xi^a\,e_a = c_\xi\partial_\tau$, where the constant $c_\xi$ fixes the normalization of the supercharge. Its zeroth component in the orthonormal frame is $\xi^0 = c_\xi\,e^{U}$. On the other hand,
\begin{equation}
    \xi^0 = 2\i\bar{\epsilon}_{i+}\gamma^0 \epsilon_{-}^i = 2\i\bar{\epsilon}_{i+}\gamma^0(-\i f^{-1}\gamma^0 \epsilon_+^i) = 2 f^{-1}S_+.
\end{equation}
Therefore,
\begin{equation}
    S_+ = \frac{c_\xi}{2}e^U f,\qquad S_- = \frac{c_\xi}{2}e^U f^{-1}.\label{eq:SS}
\end{equation}

The fixed points of $\xi^\mu$ are the $2n$ isolated points (``nuts'') $\vec{x} = \vec{x}_{N,a}$ and $\vec{x} = \vec{x}_{S,a}$. Near a generic nut $\alpha$, spacetime is locally $\mathbb{R}^4$, and the metric takes the form
\begin{equation}
    \dd s^2 \sim \dd R_{\alpha}^2 + \frac{1}{4}R_{\alpha}^2\left[(\dd\psi + s_{\alpha} \cos(\theta)\dd\varphi)^2 + \dd\theta^2 + \sin^2(\theta)\dd\varphi^2\right],
\end{equation}
where $s_{N,a} = +1$, $s_{S,a} = -1$ and $\tau = L\psi$. The orthonormal frame fields \eqref{halfBPS_on_frame} simplify to
\begin{equation}
    e^0 = \frac{R_\alpha}{2}(\dd\psi + s_{\alpha} \cos(\theta)\dd\varphi),\qquad e^1 = \dd R_\alpha,\qquad e^2 = \frac{R_\alpha}{2}\dd\theta,\qquad e^3 = \frac{R_\alpha}{2}\sin(\theta)\dd\varphi.
\end{equation}
Hence, to leading order at small $R_\alpha$, we have
\begin{equation}
    \dd(\xi_\mu \dd x^\mu) = \frac{4\pi c_\xi}{\beta}(-e^0 \wedge e^1 - s_{\alpha}\,e^2 \wedge e^3).
\end{equation}
Thus, a consistent choice of the coefficients $b_1,b_2$ is\footnote{In the presence of a $\mathbb{T}$-log multiplet arising from an (anti-)chiral multiplet of weight $w'$, we have \cite{BenettiGenolini:2026qdm}:
\begin{align}
    (b_1^p - b_2^p)^2 &= \frac{2}{w'}S_{+}^2 {A}|_{\bbT} \;\,\qquad (b_1^p + b_2^p)^2 = \frac{1}{16} S_+^2 {A}|_{{\mathbb W}^2}&\text{at a positive-chirality nut }p,\nonumber
    \\
    (b_1^p + b_2^p)^2 &= \frac{2}{w'}S_{-}^2 \bar A|_{\bbT},\qquad (b_1^p - b_2^p)^2 = \frac{1}{16}S_-^2 \bar A|_{\overline{\mathbb W}^2} &\text{at a negative-chirality nut }p. \label{eq:b1b2_eqs}
\end{align}
In our case $b_1^{N,a} - b_2^{N,a} = 0 = b_1^{S,a} + b_2^{S,a}$, consistent with the absence of $\mathbb{T}$-log multiplets. We can also verify the second set of equalities in \eqref{eq:b1b2_eqs}:
\begin{equation}
    \frac{1}{16}S_+^2 {A}|_{{\mathbb W}^2} = \frac{1}{16}(\frac{c_\xi}{2}e^U f)^2 \cdot 64 e^{2U}f^2 T_m^+ T^{+m} = \left(\frac{4\pi c_\xi}{\beta}\right)^2 =(b_1 + b_2)^2
\end{equation}
at the north poles, and similarly, $\frac{1}{16}S_-^2 \bar{A}|_{\overline{\mathbb W}^2} = \frac{1}{16}(\frac{c_\xi}{2}e^U f^{-1})^2 \cdot 64 e^{2U}f^2 T_m^- T^{-m} = \left(\frac{4\pi c_\xi}{\beta}\right)^2 =(b_1 - b_2)^2$ at the south poles. Here, in the second step we used the near-nut expansions $e^{2U} \sim \frac{\rho}{L} + \O(\rho^2)$ and $T^{\pm}_m \dd x^m \sim \mp\frac{\dd \rho}{\rho}$.
}:
\begin{align}
    b_1^{N,a} = b_2^{N,a} = -\frac{2\pi c_\xi}{\beta},\quad b_1^{S,a} = -b_2^{S,a} = -\frac{2\pi c_\xi}{\beta}.\label{eq:nut_bs}
\end{align}

Using \eqref{Psi0} with $A = \CF(X,A|_{\mathbb W^2})$, the homogeneity of the prepotential, and the definitions~\eqref{eq:rescaled_hol_vars}, we can recast \eqref{eq:Inuts} as
\begin{equation}
    I_{\rm nuts} = -\sum_{{\rm nuts}\,p} \frac{16\pi^2}{b_1^p b_2^p}\,\CF(-\tfrac{\i c_\xi}{2} e^{2U}\mathcal{Y}^I,-\tfrac{c_\xi^2}{4}e^{4U}\Upsilon).
\end{equation}
At a negative-chirality nut, $\cY^I$ and $\Upsilon$ are finite, whereas $e^{2U}\to 0$. By homogeneity, when both arguments of $\CF$ approach zero, $\CF$ also approaches zero. Therefore, $I_{\rm nuts}$ receives contributions only from positive-chirality nuts (the north poles), where $e^{2U}\cY^I\to -\frac{4\pi \i}{\beta}\gamma_{N,a}^I$ and $e^{4U}\Upsilon \to -64\cdot (\frac{4\pi}{\beta})^2$. From this and Eq.~\eqref{eq:nut_bs}, we obtain
\begin{equation}
    I_{\rm nuts} = -16\pi^2 \sum_{a=1}^n \CF(\gamma_{N,a}^I,64).
\end{equation}
An analogous argument gives $\bar{I}_{\rm nuts} = -16\pi^2 \sum_{a=1}^n \bar{\CF}(\gamma_{S,a}^I,64)$, and hence the fixed-point contribution to the on-shell action is
\begin{equation}
    -\frac{\i}{8\pi}(I_{\rm nuts} - \bar{I}_{\rm nuts}) =  2\pi \i\sum_{a=1}^n[\CF(\gamma_{N,a}^I,64) - \bar{\CF}(\gamma_{S,a}^I,64)].\label{eq:I_nuts_final}
\end{equation}

We can now evaluate the boundary terms $I_{\partial M}$ and $\bar{I}_{\partial M}$. We take the boundary to be a hypersurface of constant $|\vec{x}|\equiv r=r_c$, with $r_c$ much larger than the characteristic length scale $G_N M_{\rm ADM}$, $\partial M = \{r=\text{const.} = r_c\gg G_N M_{\rm ADM}\} \simeq S^1_\beta \times S^2.$ The outward-pointing unit normal is $n = e^U \partial_r$. We choose the orientation ${\rm vol}_M = e^0 \wedge e^1 \wedge e^2 \wedge e^3$, with $e^a$ defined in \eqref{halfBPS_on_frame}. The volume form induced on $\partial M$ is ${\rm vol}_{\partial M} = \iota_n {\rm vol}_{M} = -e^0\wedge e^2\wedge e^3.$ Since $\xi^\mu \partial_\mu = c_\xi \partial_\tau$, we have $\|\xi\|^2 = c_\xi^2e^{2U}$, $\xi_\mu \dd x^\mu = c_\xi e^{2U}(\dd \tau + \omega)$, and $\dd (\xi_\mu \dd x^\mu) = c_\xi e^{2U}(2\dd U \wedge (\dd\tau + \omega) + \dd\omega).$ Substituting these expressions into \eqref{eq:Ibdry}, we obtain
\begin{align}
    I_{\partial M} = -\frac{1}{c_\xi}\int_{\partial M}(\dd\tau + \omega)\wedge \Psi_2 - \frac{1}{c_\xi^2}\int_{\partial M}\Psi_{0}\,(\dd\tau + \omega)\wedge \dd \omega\label{BVAB_bdry_simplified1}.
\end{align}

A direct calculation shows that $\Psi_0$ remains finite as $r\to \infty$, $\Psi_0 = -c_\xi^2 \CF(\i y^{I(0)}, 0) + \O(r^{-1}).$ At zero NUT charge, $R(\omega) = \O(r^{-3})$, and therefore $\dd \omega = \star R(\omega) = \O(r^{-1})$ vanishes as $r\to \infty$. Consequently, the second contribution to \eqref{BVAB_bdry_simplified1} vanishes in the limit of large $r_c$. We are left with the contribution from $\Psi_2$,
\begin{equation}
    I_{\partial M} = -\frac{1}{c_\xi}\int_{\partial M}(\dd\tau + \omega)\wedge \Psi_2\label{BVAB_bdry_simplified2}
\end{equation}
where
\begin{equation}
    (\Psi_2)_{ab} = 2S_+ (F^+_{ab}|_{\CF} + \tfrac{1}{2}\CF(X,A|_{\mathbb W^2}) T^-_{ab}) - B^{ij}|_{\CF}\varepsilon_{jk}\Xi_{+i}{}^{k}{}_{ab}.
\end{equation}
Here $B^{ij}|_{\CF}$ and $F_{ab}^+|_{\CF}$ are given by \eqref{eq:B-F-explicit} and \eqref{eq:F-F-explicit}, respectively. In particular, $B^{ij}|_{\CF}$ is a linear combination of $Y^{I\,ij}$ and $R(\mathcal{V})$, both of which vanish for the 1/2-BPS configurations of interest. Therefore, $B^{ij}|_{\CF}=0$. In $F_{ab}^+|_{\CF}$, we can neglect the term proportional to the Riemann tensor, since it falls off at infinity faster than $r^{-2}$. Thus, we can write
\begin{equation}
    \frac{1}{c_\xi}\Psi_2 = \frac{2S_+}{c_\xi}[\CF_I (F^{I+} - \tfrac{1}{4}\bar{X}^I T^+) + \tfrac{1}{2}\CF\,T^-] + \O(r^{-1}).\label{eq:Psi2_simplified1}
\end{equation}
Using \eqref{Tplus_sol}, \eqref{Tminus_sol} and expanding $U = \frac{U^{(1)}}{r} + \O(r^{-2})$, we obtain
\begin{equation}
    T^{\pm} = \mp \frac{4f^{\mp 1}U^{(1)}}{r^2}\left(\dd \tau \wedge \dd r \pm r^2 {\rm vol}_{S^2}\right) + \O(r^{-3}).\label{eq:Tpm_expansion}
\end{equation}
Expanding the field-strength tensor at infinity, we similarly find 
\begin{equation}
    F^{I+} = -\frac{\i}{r^2}\left[y^{I(1)}  + U^{(1)}(y^{I(0)} + \bar{y}^{I(0)})\right]\left(\dd\tau \wedge \dd r+r^2 {\rm vol}_{S^2}\right) + \O(r^{-3}).\label{eq:Fminus_expansion}
\end{equation}
Substituting \eqref{eq:Tpm_expansion} and \eqref{eq:Fminus_expansion} into \eqref{eq:Psi2_simplified1}, noting that $\frac{2S_+}{c_\xi}\cdot f^{-1} = e^{U}\xrightarrow{r\to \infty} 1$, and using $X^I=-\i e^Uf^{-1}\cY^I$, $\bar X^I=\i\,e^Uf\,\bar \cY^I$ together with the homogeneity relations $\CF_I(X,0)=-\i e^Uf^{-1}\,\CF_I(\cY,0)$ and $\CF(X,0)=-e^{2U}f^{-2}\CF(\cY,0)$, one finds that all terms proportional to $U^{(1)}$ cancel. Therefore,
\begin{equation}
    \frac{1}{c_\xi}\iota_{S^2_\infty}^*\Psi_2 = - \CF_I(y^{(0)},0)y^{I(1)}{\rm vol}_{S^2} + \O(r^{-1}).
\end{equation}

Substituting this expression into \eqref{BVAB_bdry_simplified2} and taking into account the orientation of $\partial M$, ${\rm vol}_{\partial M} = -e^0 \wedge e^2 \wedge e^3$, we obtain
\begin{equation}
    I_{\partial M} = -4\pi \beta\,\CF_I(y^{(0)},0)\,y^{I(1)}.
\end{equation}
Similarly, the anti-chiral sector gives $\bar{I}_{\partial M} = -4\pi \beta\,\bar \CF_I(\bar y^{(0)},0)\,\bar y^{I(1)}.$ Therefore,
\begin{equation}
    -\frac{\i}{8\pi}(I_{\partial M} - \bar{I}_{\partial M}) = \frac{\i \beta}{2}\left[\CF_I(y^{(0)},0)\,y^{I(1)} - \bar \CF_I(\bar y^{(0)},0)\,\bar y^{I(1)}\right].\label{eq:BVAB_bdry_final}
\end{equation}

\paragraph{Scalar boundary term: }
Let us now evaluate the scalar boundary term $I_{\partial M}^X$, given by \eqref{eq:new-scalar-bdry-term}. To this end, it is convenient to choose an $\SO(1,1)_{\rm R}$ gauge in which $f = 1 + \O(r^{-2})$. Then $A_m = \frac{1}{2}e^{2U}R(\omega)_m = \O(r^{-3})$, so, to the order relevant for the large-$r$ boundary integral, we may replace the covariant derivatives $\CD_\mu$ by ordinary derivatives. Using $X^I=-\i e^U\cY^I$ and $\bar X^I=\i e^U\bar \cY^I$, we can write
\begin{align}
    I_{\partial M}^X = \frac{\i}{8\pi}\int_{\partial M}\dd^3x\sqrt{h}\big\{(\partial_r U)\left[\CF_I(\cY^J,0)\bar \cY^I-\bar\CF_I(\bar \cY^J,0)\cY^I\right]\nonumber 
    \\
    + \CF_I(\cY^J,0)\partial_r\bar \cY^I - \bar\CF_I(\bar \cY^J,0)\partial_r \cY^I\big\}.
\end{align}
The combination $\i[\bar\CF_I(\bar \cY,0)\cY^I-\CF_I(\cY,0)\bar \cY^I]$ approaches $e^{-\CK}$ at asymptotic infinity. Since the right-hand side of \eqref{Deom} vanishes at large $r$, in the gauge $\chi=-2$ we have $e^{-\CK}=1+\O(r^{-1})$. Using Eq.~\eqref{ADM_mass}, we obtain
\begin{equation}
    I_{\partial M}^X = -\frac{\i\beta}{4}\left[\CF_I(y^{(0)},0)\bar y^{I(1)}-\bar\CF_I(\bar y^{(0)},0)y^{I(1)}-\CF_{IJ}(y^{(0)},0)y^{I(1)}\bar y^{J(0)}+\bar\CF_{IJ}(\bar y^{(0)},0)\bar y^{I(1)}y^{J(0)}\right].\label{I_X_bdry}
\end{equation}
Adding this expression to \eqref{eq:BVAB_bdry_final}, we obtain\footnote{To verify this, note that
\begin{align}
    \phi^I &= -y^{I(0)} - \bar{y}^{I(0)},\qquad\;  \phi_I = -\CF_I(y^{(0)},0) - \bar\CF_I(\bar y^{(0)},0),
    \\
    \Gamma^I &= \i(y^{I(1)} - \bar{y}^{I(1)}),\qquad \Gamma_I = \i\left[\CF_{IJ}(y^{(0)},0)y^{J(1)}-\bar\CF_{IJ}(\bar y^{(0)},0)\bar y^{J(1)}\right].
\end{align}}:
\begin{align}
    -\frac{\i}{8\pi}(I_{\partial M} - \bar{I}_{\partial M}) + I_{\partial M}^X = -\frac{\beta}{4}(\phi_I \Gamma^I + \phi^I \Gamma_I).\label{I_bdry_tot}
\end{align}
In the case of the two-derivative theory, this result, combined with \eqref{eq:I_nuts_final}, agrees with Eq.~(3.39) of \cite{Boruch:2025biv}.

\paragraph{Higher-derivative GHY term: }
Finally, higher-derivative curvature interactions require a corresponding generalization of the Gibbons--Hawking--York boundary term \cite{York:1972sj,Gibbons:1976ue}. A perturbative construction of this term for generic curvature-squared gravity coupled to Maxwell fields was given in \cite{Cremonini:2009ih}; see also \cite{Myers:1987yn,Dyer:2008hb,Teimouri:2016ulk}. For asymptotically flat solutions, however, all higher-derivative corrections to this boundary term vanish at spatial infinity. Since the explicit curvature-squared boundary terms contain additional powers of the extrinsic curvature, intrinsic curvature, or field strengths, they decay faster than $r^{-2}$, approaching zero even when multiplied by the $\O(r^2)$ boundary volume element. The additional fields appearing in the supersymmetric completion of Einstein--Maxwell theory do not change this conclusion. Thus, only the ordinary GHY term survives,
\begin{equation}
	I_{\rm GHY}^{\rm HD} = -\frac{1}{8\pi}\int \dd^{3}x\sqrt{h}\,K - I_{\rm ct}.
\end{equation}
Here, the counterterm $I_{\rm ct}$ is defined using the background-subtraction method \cite{Gibbons:1976ue}. This gives us the standard result:
\begin{equation}
	I_{\rm GHY}^{\rm HD} = \frac{\beta M_{\rm ADM}}{2} = \frac{\beta}{4}\symp{\phi}{\Gamma}.\label{eq:GHY_reg}
\end{equation}

\paragraph{The vanishing of D-terms:}
So far, our analysis has focused on the chiral superspace (F-term) contributions to the $\CN = 2$ SUGRA action. In general, however, the action also contains full-superspace integrals (so-called D-terms) \cite{deWit:2010za,Murthy:2013xpa}, which take the form
\begin{equation}
    \int \dd^4 x\,\dd^4 \theta\,\dd^4 \bar{\theta}\,E\,\mathcal{H}(\Phi,\bar{\Phi}),\qquad \mathcal H({\Phi,\bar{\Phi}}) = \sum_{n}f_n(\Phi)\bar{g}_n(\bar{\Phi}),
\end{equation}
where $x,\theta,\bar{\theta}$ are superspace coordinates, $E$ is the superspace volume element, and $f_n$ (resp. $\bar{g}_n$) are functions of the chiral (resp. anti-chiral) multiplets $\Phi$ (resp. $\bar{\Phi}$) with appropriate Weyl and chiral weights. As shown in \cite{deWit:2010za,Murthy:2013xpa}, the full-superspace integral can be rewritten as a chiral-superspace integral, with chiral measure $\mathcal{E}$, involving the kinetic multiplets $\mathbb T(\bar{g}_n(\bar{\Phi}))$,
\begin{equation}
    \int \dd^4 x\,\dd^4\theta\,\dd^4\bar{\theta}\,E\,f_n(\Phi)\bar{g}_n(\bar{\Phi}) = \int\dd^4 x\,\dd^4\theta\,\mathcal{E}\,f_n(\Phi)\mathbb{T}(\bar{g}_n(\bar{\Phi})).
\end{equation}
Any kinetic multiplet $\mathbb{T}(\bar{g}_n)$ can be written as a difference of two $\mathbb{T}$-log multiplets, $\mathbb{T}(\bar{g}_n) = \mathbb{T}(\log(\bar{\Phi}' e^{\bar{g}_n}) ) - \mathbb{T}(\log(\bar{\Phi}'))$. Therefore, we can encode the D-term corrections to the effective action in a $\mathbb T$-log dependence of the prepotential, making the localization formula of \cite{BenettiGenolini:2026qdm} also applicable to the D-term corrections.

We can split the prepotential $\mathcal F$ into $\mathbb{T}$-log-dependent and $\mathbb{T}$-log-independent parts,
\begin{equation}
    \mathcal{F} = \mathcal{F}_0(X,A|_{\mathbb W^2}) + \lambda \mathcal{F}_D(X,A|_{\mathbb W^2}, A|_{\mathbb T}),\label{eq:Dterm-corrected-F}
\end{equation}
where $A|_{\mathbb T}$ denotes the lowest components of the $\mathbb{T}$-log multiplets, $\mathcal{F}_D(X,A|_{\mathbb W^2}, A|_{\mathbb T} = 0) = 0$, and $\lambda$ is a coupling parametrizing the D-term corrections. In the previous sections, we focused on the theory with $\lambda = 0$. Suppose that $\mathcal{S}_{\lambda\neq 0}$ is a supersymmetric configuration of the theory with D-term corrections, satisfying the same boundary conditions as $\mathcal{S}_{\lambda = 0}$. The corresponding on-shell action is
\begin{equation}
    I_\lambda[\mathcal{S}_\lambda] = I_0[\mathcal{S}_\lambda] + \lambda I_D[\mathcal{S}_\lambda],
\end{equation}
where $I_0$ arises from the prepotential $\CF_0$, and $I_D$ arises from the prepotential $\CF_{D}$. Then
\begin{equation}
    \frac{\dd I_{\lambda}}{\dd\lambda} = I_D[\mathcal{S}_\lambda] + \int \dd^4 x\,\frac{\dd \mathcal{S}_\lambda(x)}{\dd \lambda}\frac{\delta I_{\lambda}}{\delta \mathcal{S}_\lambda(x)} = I_D[\mathcal{S}_\lambda],
\end{equation}
since $\mathcal{S}_\lambda$ is a stationary point of $I_{\lambda}$, $\frac{\delta I_\lambda}{\delta \mathcal{S}_\lambda} = 0$.

By the localization formula \eqref{eq:loc_formula}, $I_D$ can only receive contributions from the Killing vector fixed points and asymptotic boundary terms. Since $\mathcal{S}_\lambda$ is asymptotically flat, and $I_D$ consists of higher-derivative corrections, which rapidly vanish in the $1/r$ expansion, we expect that there are no boundary contributions. 

As shown in \cite{BenettiGenolini:2026qdm}, we have
\begin{align}
    &A|_{\mathbb T}\propto (b_1^p - b_2^p)^2,\qquad \text{at a positive-chirality nut }p,\nonumber 
    \\
    &\bar{A}|_{\mathbb T} \propto (b_1^p + b_2^p)^2,\qquad \text{at a negative-chirality nut }p.\label{eq:Tlog-nuts}
\end{align}
For $\lambda = 0$, we know that $b_1^p - b_2^p = 0$ ($b_1^p + b_2^p = 0$) at the north (south) poles. Consequently, the anti-chiral components $A|_{\mathbb T}$ vanish on the north poles and the chiral components $\bar{A}|_{\mathbb T}$ vanish on the south poles, implying that the fixed-point contributions to $I_D$ vanish at $\lambda = 0$. We can always set the Killing vector normalization so that $b_1, b_2$ take discrete values $\pm 1$, even when $\lambda\neq 0$. Thus, $b_1,b_2$ do not change under continuous deformations of $\lambda$, which, by \eqref{eq:Tlog-nuts}, implies that $A|_{\mathbb T}$ vanishes on the north poles (and $\bar{A}|_{\mathbb T}$ vanishes on the south poles). Therefore, $I_D = 0$ for every $\lambda,$ so that
\begin{equation}
    \frac{\dd I_\lambda}{\dd\lambda} = 0,\qquad I_{\lambda}[\mathcal{S}_\lambda] = I_0[\mathcal{S}_0].
\end{equation}
This proves that D-terms do not contribute to the on-shell action and that they do not affect the values of the fields at the fixed-point (even though they can in principle deform the saddle away from the fixed points), extending the non-renormalization results of \cite{deWit:2010za,Murthy:2013xpa} to the rotating, multi-center index saddles.

\subsection{Computing the on-shell action in the canonical and mixed ensembles}
\label{sec:on-shell-action-in-three-ensembles}

We are now ready to put all the results above together to compute the on-shell action. Before computing this action for the OSV ensemble, it is useful to first understand the action in two ensembles whose interpretation in terms of boundary conditions imposed in the GPI is more straightforward.

\paragraph{Mixed ensemble:} To start, let us consider the ensemble for the index obtained from considering Dirichlet boundary conditions for all fields in the Weyl, vector and hypermultiplets. In terms of a trace over the black hole Hilbert space, such an index is computed by\footnote{Due to the insertion of $(J_3)^2$, $Z_\text{BH}(\Gamma^I, \mu^I; \Omega_\infty)$ is technically the helicity supertrace in the mixed ensemble. At the saddle-point level, the insertion of $(J_3)^2$ does not affect the value of the on-shell action and thus will not play an important role in our analysis. Nevertheless, for such saddle-points to contribute to the path integral and for $Z_\text{BH}(\Gamma^I, \mu^I; \Omega_\infty)$ to be non-zero, the insertion of $(J_3)^2$ is necessary in order to give a non-zero weight to a set of fermionic modes coming from the gravitino that would otherwise be zero modes and cause $Z_\text{BH}(\Gamma^I, \mu^I; \Omega_\infty)$ to vanish \cite{H:2023qko, Anupam:2023yns}.    } 
\be 
Z_\text{BH}(\Gamma^I, \mu^I; \Omega_\infty) = \Tr_{\substack{\text{Fixed }\Gamma^I }} (-1)^F (J_3)^2 e^{-\beta H + \beta \mu^I \Gamma_I}\,,
\label{eq:mixed-ensemble}
\ee
which is thus a mixed ensemble where electric chemical potentials and magnetic charges are held fixed. This is, however, not yet the relevant ensemble for the OSV conjecture since the Hamiltonian is $H=\{Q,Q^{\dagger}\}+|\mathcal{Z}|$ and since BPS states with different electric charges have different central charges $|\mathcal{Z}|$ they are weighted differently by their BPS mass. Nevertheless, the gravitational action in this ensemble is the easiest to compute: with such boundary conditions no additional boundary terms are needed beyond those computed above. Collecting the results \eqref{eq:hyper=0}, \eqref{eq:I_nuts_final}, \eqref{I_bdry_tot} and \eqref{eq:GHY_reg}, we find that the on-shell action in the mixed ensemble for each saddle is given by:
\begin{equation}
    -I_\text{mixed}(\{\Gamma^I_a\}, \mu^I, \{n^I_a\}; \Omega_\infty) %
    =  {+} \frac{\beta \phi_I \Gamma^I}{2} - 2\pi \i\sum_{a=1}^n\,[\CF(\gamma_{N,a}^I,64) - \bar{\CF}(\gamma_{S,a}^I,64)]\,,\label{eq:Imix_res}
\end{equation}
where $\gamma_{S,a}^I$ and $\gamma_{N,a}^I$ are the residues given by \cref{eq:shifted_gammaN,eq:sum_upper_gamma} in terms of the magnetic charge and electric chemical potential, and $\phi_I$ can be interpreted as a magnetic chemical potential that is given by \eqref{eq:el_pot}. Here and below, each saddle is labeled by the distribution of magnetic charges among its $n$ black holes, $\{\Gamma^I_a\}\equiv\{\Gamma^I_1,\dots,\Gamma^I_n\}$ with $\sum_{a=1}^n\Gamma^I_a=\Gamma^I$, and by the collection of dual-lattice vectors $\{n^I_a\}\equiv\{n^I_1,\dots,n^I_n\}$, $n^I_a\in\Lambda^*$, which specify the gauge-field configuration around each black hole. Quantities that depend only on the total charges, such as $Z_\text{BH}(\Gamma^I, \mu^I; \Omega_\infty)$, include the sum over all such saddles.

For single-center black holes there is still an infinite number of saddles, each labeled by the integer shifts $n^I$ in \eqref{eq:shifted_gammaN},\footnote{For single-center black holes we drop the subscript $a$ in $n^I_a$ that is associated with each black hole. } 
\begin{equation}
Z_\text{BH}(\Gamma^I, \mu^I; \Omega_\infty) \supset \sum_{n^I \in \Lambda^*} Z_\text{1-loop}\left(\Gamma^I, \mu^I + \frac{2\pi \i n^I}{\beta}; \Omega_\infty\right) e^{+ \tfrac{1}{2}{\beta \phi_I \Gamma^I} - 2\pi \i \,[\CF(\gamma_{N}^I,64) - \bar{\CF}(\gamma_{S}^I,64)]} \left[1 + O(G_N)\right],
\label{eq:ZBH-single-center-canonical}
\end{equation}
where the dependence on $n^I$ of the action is implicit through \eqref{eq:el_pot}, \eqref{eq:shifted_gammaN}, and \eqref{eq:sum_upper_gamma}. Just like the action, the one-loop determinant depends only on the combination $\mu^I + 2\pi \i \frac{n^I}{\beta}$ and computing it goes beyond the scope of this paper. Additionally, we have to include multi-center saddles, in which the total magnetic charge is divided among the black holes and the dipole charge of each black hole depends on the integer shift $n^I_a$, 
\begin{align}
Z_\text{BH}(\Gamma^I, \mu^I; \Omega_\infty) &\supset \sum_{\substack{\Gamma^I_1, \dots, \Gamma^I_n \in \Lambda^*\\ \sum_a \Gamma^I_a = \Gamma^I}} \,\,\sum_{n^I_1, \dots, n^I_n \in \Lambda^*} \text{Vol}\left(\{\Gamma^I_a\}, \mu^I, \{n^I_a\}; \Omega_\infty\right) Z_\text{1-loop}\left(\{\Gamma^I_a\}, \mu^I, \{n^I_a\}; \Omega_\infty\right) \nonumber \\ &\qquad\qquad\qquad \qquad  \times \,  e^{\tfrac{1}{2}{\beta \phi_I \Gamma^I} - 2\pi \i \,\sum_{a=1}^n [\CF(\gamma_{N, a}^I,64) - \bar{\CF}(\gamma_{S,a}^I,64)]} \left[1 + O(G_N)\right]\,,
\label{eq:ZBH-multi-center-canonical}
\end{align} 
where, again, $\gamma_{N,a}^I$ and $\gamma_{S,a}^I$ depend on $n^I_a$. The leading-order quantum fluctuations now include both a one-loop determinant and a volume factor for the non-trivial moduli of saddles present for multi-center black hole configurations \cite{Boruch:2025biv}; once again, computing such terms goes beyond the scope of the paper. We will also not be concerned in this paper with which of the saddles in \eqref{eq:ZBH-single-center-canonical} and \eqref{eq:ZBH-multi-center-canonical} are actually picked up by the GPI; instead, we will focus on the on-shell action of each saddle.

The action in each exponent is computed to all orders in the topological string coupling fixed at the north and south poles of the black hole. For the single-center black hole, the second term in the action \eqref{eq:Imix_res} is expected from the OSV conjecture since the prepotential $\CF(\gamma_{N,a}^I,64) $ can be identified with the %
free energy of the topological string \cite{Antoniadis:1993ze,Bershadsky:1993cx}. The first term, however, is due to the non-trivial weighting of BPS states in \eqref{eq:mixed-ensemble} by their BPS mass and therefore has no interpretation within the OSV conjecture: for instance, it has a non-trivial dependence on $\beta$ coming from $\phi_I$, making it difficult to verify that \eqref{eq:Imix_res} is indeed the free energy associated with a protected index.

\paragraph{Canonical ensemble:}
To perform the calculation in the ensemble considered in the OSV conjecture, we have to first go from the mixed ensemble to the canonical ensemble. This is not only a useful intermediate step in our proof of OSV but also serves as an intermediate check of our results since we can compare the on-shell action in the two-derivative approximation to the results obtained in \cite{Boruch:2023gfn, Boruch:2025biv}. The canonical ensemble is defined as 
\begin{equation}
Z_\text{BH}(\Gamma^I, \Gamma_I; \Omega_\infty) = \Tr_{\substack{\text{Fixed }(\Gamma^I, \Gamma_I)}} (-1)^F (J_3)^2 e^{-\beta H } = \Omega(\Gamma^I, \Gamma_I; \Omega_\infty) e^{-\beta M_{\rm BPS}(\Gamma; \Omega_\infty)}\,,
\label{eq:canonical-ensemble}
\end{equation}
where in the second equality we have used the fact that only BPS states contribute to the index whose energy is fixed by \eqref{eq:ADM-mass} and have defined the canonical helicity supertrace $\Omega(\Gamma^I, \Gamma_I; \Omega_\infty)$ as the difference between the number of bosonic and fermionic BPS microstates weighted by $(J_3)^2$ in the sector with fixed $\Gamma$.
 As before, this ensemble receives contributions from both single-center and multi-center configurations. The on-shell action of such configurations can be computed from the action of the analogous configurations in the mixed ensemble, by relating the two ensembles as\footnote{We also provide a direct gravitational calculation of the on-shell action by including an electromagnetic boundary term in App.~\ref{app:KerrNewman_puzzle}. What makes the derivation in the appendix particularly subtle is the fact that the boundary term depends on how the gauge connection is specified in patches, a fact that we are forced to deal with for dyonic rotating black holes.}
 \begin{align}
      Z_\text{BH}(\Gamma^I, \mu^I; \Omega_\infty) &= \sum_{\substack{\Gamma^I_1, \dots, \Gamma^I_n \in \Lambda^* \\ \sum_a \Gamma^I_a = \Gamma^I}}\sum_{n^I_1, \dots, n^I_n \in  \Lambda^*} Z_\text{BH}(\{\Gamma^I_a\}, \mu^I, \{n^I_a\}; \Omega_\infty) \nonumber \\ &= \sum_{\substack{\Gamma^I_1, \dots, \Gamma^I_n \in \Lambda^* \\ \sum_a \Gamma^I_a = \Gamma^I}} \sum_{\Gamma_{I,1}, \dots, \Gamma_{I,n} \in \Lambda} Z_\text{BH}(\{\Gamma^I_a\}, \{\Gamma_{I,a}\}; \Omega_\infty) e^{+\beta \mu^I \sum_a \Gamma_{I,a}}
      \label{eq:from-canonical-to-grand-canonical}
 \end{align}
 where $Z_\text{BH}(\{\Gamma^I_a\}, \mu^I, \{n^I_a\}; \Omega_\infty) $ represents the contribution of each saddle in \eqref{eq:ZBH-multi-center-canonical} and~$Z_\text{BH}(\{\Gamma^I_a\}, \{\Gamma_{I,a}\}; \Omega_\infty)$~represents~the contribution of a black hole configuration with charges $(\Gamma^I_a, \Gamma_{I,a})$ in the canonical ensemble, 
 \be 
 Z_\text{BH}(\Gamma^I, \Gamma_I; \Omega_\infty) = \sum_{\substack{\Gamma^I_1, \dots, \Gamma^I_n \in \Lambda^* \\ \sum_a \Gamma^I_a = \Gamma^I}} \sum_{\substack{\Gamma_{I,1}, \dots, \Gamma_{I,n} \in \Lambda \\ \sum_a \Gamma_{I,a} = \Gamma_I}}Z_\text{BH}(\{\Gamma^I_a\}, \{\Gamma_{I,a}\}; \Omega_\infty)\,.
 \ee
 The two summands in the first and second lines of \eqref{eq:from-canonical-to-grand-canonical} are related through the Poisson summation formula,
 \be 
 Z_\text{BH}(\{\Gamma^I_a\}, \{\Gamma_{I,a}\}; \Omega_\infty) = (\text{Vol}(\Lambda))^n \int \left(\prod_a \prod_I dn_a^I \right)  Z_\text{BH}(\{\Gamma^I_a\}, \mu^I, \{n^I_a\}; \Omega_\infty) e^{- \sum_a (2\pi \i n^I_a + \beta \mu^I) \Gamma_{I,a}} \,,
 \label{eq:Poisson-summation-mixed-to-canonical}
 \ee
 where the volume of the lattice $\Lambda$ is $\text{Vol}(\Lambda)=  \prod_{I=0}^{n_v} q_I$. 
By taking the saddle point value of the integrals over $n^I_a$, we see that the on-shell action in the ``canonical ensemble,'' with fixed electric and magnetic charges, $\Gamma_{I,a}$ and $\Gamma^I_a$, is:
\begin{align}
    -I_{\rm can}(\beta,\{\Gamma_{I,a}\},\{\Gamma^I_a\}; \Omega_\infty) = - \beta M_{\rm BPS}-2\pi \i \sum_{a=1}^n\,[\CF(\gamma_{N,a}^I,64) - \bar{\CF}(\gamma_{S,a}^I,64)]\nonumber 
    \\
    + \beta (\tfrac{1}{2}\phi^I - \mu^I  )\Gamma_I - 2\pi \i \sum_a n_{a\,*}^I \Gamma_{I,a},\label{eq:Ican_res1}
\end{align}
where $\gamma_{N,a}^I$ and $\gamma_{S,a}^I$ are determined as a function of the charge vectors $(\Gamma^{I}_a,\Gamma_{I,a})$ through \eqref{eq:sum_upper_gamma} and \eqref{eq:sum_lower_gamma}, $\phi^I$ is given by \eqref{eq:el_pot}, the BPS mass is given by \eqref{eq:ADM-mass}, and $n_a^I$ is fixed to a saddle point value $n_{a\,*}^I$ as a function of $\gamma_{N,a}^I$ and $\gamma_{S,a}^I$ through equations~\eqref{eq:shifted_gammaN} and \eqref{eq:shifted_gammaS}. It follows that
\begin{align}
    \beta(\mu^I - \tfrac{1}{2}\phi^I)\Gamma_I + 2\pi\i \sum_{a=1}^n n_{a\,*}^I\Gamma_{I,a} &= \sum_{a=1}^n (\beta\mu ^I - \tfrac{\beta}{2}\phi^I + 2\pi \i n^I_{a\,*}) \Gamma_{I,a} \nonumber 
    \\
    &= \i\pi \sum_{a=1}^n(\gamma^I_{N,a} - \gamma_{S,a}^I)\Gamma_{I,a} + i\pi \sum_{a=1}^n\Gamma^{I}_a\Gamma_{I,a}.
\end{align}
Substituting this expression in \eqref{eq:Ican_res1}, and using the homogeneity of the prepotentials to write
\begin{align}
    2\CF(\gamma_{N,a}^I,64) &= \CF_I(\gamma_{N,a}^J,64) \gamma_{N,a}^I + 128\CF_A(\gamma_{N,a}^J,64),
    \\
    2\bar \CF(\gamma_{S,a}^I,64) &= \bar \CF_I(\gamma_{S,a}^J,64) \gamma_{S,a}^I + 128\bar \CF_A(\gamma_{S,a}^J,64),
\end{align}
we can recast \eqref{eq:Ican_res1} as:\footnote{When going from \eqref{eq:Ican_res1} to \eqref{eq:canonical-action-final-result}, we have dropped a factor of $i \pi \sum_a \Gamma^{I}_a  \Gamma_{I,a}\in i \pi \mathbb Z $ in the action. This can change the phase of the contribution of each saddle, and it would be interesting to understand whether, when also accounting for the phase of the one-loop determinant, this correctly reproduces the phases of a microscopic index.  }
\begin{equation}
\log Z_\text{BH}(\{\Gamma^I_a\}, \{\Gamma_{I,a}\}; \Omega_\infty) \approx  - I_{\rm can}(\beta,\{\Gamma_{I,a}\},\{\Gamma^I_a\}; \Omega_\infty) = -\beta M_{\rm BPS} + \sum_{a=1}^n S^{\rm Wald}_a,
 \label{eq:canonical-action-final-result}
\end{equation}
where
\begin{equation}
    S^{\rm Wald}_a = \i\pi\left[\symp{\gamma_{N,a}}{\gamma_{S,a}} - 128\CF_A(\gamma_{N,a}^J,64) + 128\bar{\CF}_A(\gamma_{S,a}^J,64)\right]
    \label{eq:Wald-entropy}
\end{equation}
is precisely the Wald entropy \cite{Wald:1993nt,Iyer:1994ys,LopesCardoso:1998tkj,Sen:2005wa,Hegde:2024bmb} of the $a$-th black hole that accounts for all higher-derivative corrections, derived in \cite{Adhikari:2025eok} in the AdS$_2\times S^2$ near-horizon region of an extremal black hole. $\log Z_\text{BH}(\{\Gamma^I_a\}, \{\Gamma_{I,a}\}; \Omega_\infty)$ takes precisely the form expected for the logarithm of the index: it is linear in $\beta$ with the mass term $M_{\rm BPS}$ containing the only dependence on $\Omega_\infty$, and the term independent of $\beta$ can be identified as the logarithm of the difference between bosonic and fermionic black hole degeneracies.  Since the Wald entropy is obtained from a sum of bosonic and fermionic degeneracies rather than from the difference, this shows that, at fixed charge, the ensemble is dominated by black holes with a single type of spin statistics;\footnote{This makes precise the statement that the index equals the degeneracy for such black holes, made in \cite{Sen:2008vm,Sen:2009vz,Dabholkar:2010rm}.} the degeneracy of black holes with opposite spin statistics is not suppressed in the string coupling but is rather suppressed in $G_N$. %
The result in~\eqref{eq:canonical-action-final-result} generalizes the single black hole analysis of \cite{Hegde:2024bmb} to multicentered saddles, where the standard form of the quantum statistical relation fails and one is required to compute the on-shell action directly \cite{Boruch:2025biv}. 
Our result is also consistent with the recent findings of \cite{Mahajan:2026fhc}, which showed that in any higher derivative theory, the on-shell action for any saddle admitting a Killing spinor is dependent on temperature and scalar moduli only through the  $\beta M_{\rm BPS}$ in \eqref{eq:canonical-action-final-result}.

\section{A proof of OSV to all orders in the topological string coupling}
\label{sec:gravitational-index-to-all-orders}

We can now put the above results together to compute the action in the OSV ensemble,
\begin{equation}
Z_\text{BH}^{\text{OSV}}( \Gamma^I, \mu^I; \Omega_\infty) = \Tr_{\substack{\text{Fixed }\Gamma^I }} (-1)^F (J_3)^2 e^{-\beta \{Q, Q^\dagger \} + \beta \mu^I \Gamma_I}\,.
\label{eq:OSV-ensemble}
\end{equation}
In contrast to the canonical and mixed ensembles discussed above, $Z_\text{BH}^{\text{OSV}}( \Gamma^I, \mu^I; \Omega_\infty)$ is not associated with a natural set of  boundary conditions that we can directly impose in the GPI. Instead, we compute it by relating it to the canonical ensemble,
 \begin{align}
      Z_\text{BH}^{\text{OSV}}(\Gamma^I, \mu^I; \Omega_\infty) &= \sum_{\substack{\Gamma^I_1, \dots, \Gamma^I_n \in \Lambda^* \\ \sum_a \Gamma^I_a = \Gamma^I}}\sum_{n^I_1, \dots, n^I_n \in \Lambda^*} Z_\text{BH}^{\text{OSV}}(\{\Gamma^I_a\}, \mu^I, \{n^I_a\}; \Omega_\infty) \nonumber \\ &= \sum_{\substack{\Gamma^I_1, \dots, \Gamma^I_n \in \Lambda^* \\ \sum_a \Gamma^I_a = \Gamma^I}} \sum_{\Gamma_{I,1}, \dots, \Gamma_{I,n} \in \Lambda} Z_\text{BH}(\{\Gamma^I_a\}, \{\Gamma_{I,a}\}; \Omega_\infty) e^{\beta|\mathcal{Z}(\Gamma, \Omega_\infty)| + \beta \mu^I \sum_a \Gamma_{I,a}}
      \label{eq:from-canonical-to-OSV}
 \end{align}
 where $Z_\text{BH}^{\text{OSV}}(\{\Gamma^I_a\}, \mu^I, \{n^I_a\}; \Omega_\infty)$ represents the contribution of the saddles in the OSV ensemble analogous to those in the mixed ensemble in \eqref{eq:ZBH-multi-center-canonical} and $Z_\text{BH}(\{\Gamma^I_a\}, \{\Gamma_{I,a}\}; \Omega_\infty) $ is the contribution of each saddle in the canonical ensemble as in \eqref{eq:from-canonical-to-grand-canonical}. The two quantities are once again related through the Poisson summation formula as in \eqref{eq:Poisson-summation-mixed-to-canonical}, this time with the Fourier transform being with respect to the electric charges
\begin{align}
Z_\text{BH}^{\text{OSV}}(\{\Gamma^I_a\}, \mu^I, \{n^I_a\}; \Omega_\infty) &= \frac{1}{(\text{Vol}(\Lambda))^n } \int  \left(\prod_a \prod_I d\Gamma_{I, a} \right)  \underbrace{Z_\text{BH}(\{\Gamma^I_a\}, \{\Gamma_{I,a}\}; \Omega_\infty) e^{\beta |\mathcal{Z}(\Gamma, \Omega_\infty)|}}_{\approx \exp\left[{-  \tilde{I}_{\rm can}(\beta,\{\Gamma_{I,a}\},\{\Gamma^I_a\})}\right] } \nonumber\\ &\hspace{5.1cm}\times e^{ \sum_a (2\pi \i n^I_a + \beta \mu^I) \Gamma_{I,a}}\,.
\label{eq:Poisson-summation-canonical-to-OSV}
\end{align}
Our goal is  to now evaluate the integral over the electric charges by the saddle-point method.  For this, we can define the canonical action, with the ADM mass Boltzmann term subtracted,\footnote{Such on-shell actions can be used to directly compute $\Omega(\Gamma^I, \Gamma_I; \Omega_\infty)$.} as 
\begin{equation}
    \tilde{I}_{\rm can}(\beta,\{\Gamma_{I,a}\},\{\Gamma^I_a\}) = I_{\rm can} - \beta M_{\rm ADM} = 2 \pi \i\sum_{a=1}^n[\CF(\gamma^I_{N,a},64) - \bar{\CF}(\gamma_{S,a}^I,64)] + 2\pi \sum_{a=1}^n \delta^I_a \Gamma_{I,a}.\label{eq:tilde-Ican}
\end{equation}

To find the saddle-point approximation for the integrals \eqref{eq:Poisson-summation-canonical-to-OSV}, we differentiate the first term in \eqref{eq:tilde-Ican} at fixed magnetic charges $\Gamma_a^I$. We get $\dd \gamma_{N,a}^I = \i\dd \delta^I_a$, $\dd \gamma_{S,a}^I = -\i\,\dd \delta^I_a$ and consequently,
\begin{equation}
    \dd[\CF(\gamma^I_{N,a},64) - \bar{\CF}(\gamma_{S,a}^I,64)] =\i (\CF_I(\gamma_{N,a}^J,64) + \bar{\CF}_I(\gamma_{S,a}^J,64))\,\dd\delta_a^I = \i\,\Gamma_{I,a}\dd\delta_a^I,
\end{equation}
and therefore the saddle-point equation becomes
\begin{equation}
    \frac{\partial \tilde{I}_{\rm can}}{\partial \Gamma_{I,a}}\dd\Gamma_{I,a} %
    = (2\pi \i\,n_a^I + \beta \mu^I)\dd\Gamma_{I,a}\,,
\end{equation}
which fixes the dipole charges to %
\begin{equation}
2\pi \delta^I_a = 2\pi \i n_a^I + \beta \mu^I\,.
\label{eq:delta-in-OSV}
\end{equation}
Plugging this back into \eqref{eq:tilde-Ican} and then further into \eqref{eq:Poisson-summation-canonical-to-OSV}, we find that in the OSV ensemble the saddle-point action is given by\footnote{The on-shell actions in \eqref{eq:tilde-Ican} and \eqref{eq:on-shell-action-OSV} are both independent of the scalar moduli at infinity, $\Omega_\infty$. Nevertheless, whether a given saddle contributes or not to the path integral can depend on $\Omega_\infty$. For this reason, we leave the dependence of $Z_\text{BH}^{\text{OSV}}(\{\Gamma^I_a\}, \mu^I, \{n^I_a\}; \Omega_\infty) $ on $\Omega_\infty$ explicit.}
\begin{align}
   \log Z_\text{BH}^{\text{OSV}}(\{\Gamma^I_a\}, \mu^I, \{n^I_a\}; \Omega_\infty)  &= - {I}_{\rm OSV}(\{\Gamma^I_a\}, \mu^I, \{n^I_a\}) \nonumber
   \\ &= -\tilde{I}_{\rm can}(\beta,\{\Gamma_{I,a}\},\{\Gamma^I_a\}) + 2\pi\sum_{a=1}^n \delta_a^I\, \Gamma_{I,a} \nonumber 
    \\
    &= -2 \pi \i\sum_{a=1}^n[\CF(\gamma^I_{N,a},64) - \bar{\CF}(\gamma_{S,a}^I,64)],
    \label{eq:on-shell-action-OSV}
\end{align}
where the residues $\gamma^I_{N,a}$ and $\gamma_{S,a}^I$ are determined from \eqref{eq:sum_upper_gamma} and \eqref{eq:delta-in-OSV} to be 
\begin{align}
    \gamma_{N,a}^I&=\frac{\Gamma_a^I}{2}
                       +\frac{\i \beta \mu^I - 2\pi n_a^I}{2\pi},
    &\gamma_{S,a}^I&=\frac{\Gamma_a^I}{2}
                       -\frac{\i \beta \mu^I - 2\pi n_a^I}{2\pi}.
    \label{eq:OSV-pole-parametrization}
\end{align}
Note that the saddle-point evaluations of~\eqref{eq:Poisson-summation-mixed-to-canonical} and~\eqref{eq:Poisson-summation-canonical-to-OSV} are, at leading order in $G_N$, Legendre transforms of the classical action. Since the action retains the full prepotential $\mathcal F(X,A)$, the result \eqref{eq:on-shell-action-OSV} holds to all orders in $g_{\rm top}$.

To compare with the original proposal of OSV \cite{Ooguri:2004zv}\footnote{Note that our prepotential equals minus the prepotential from \cite{Ooguri:2004zv}. Their scalar attractor values are $C X^\Lambda$ equal to $2\gamma_N^I$ in our conventions, and their electromagnetic charges and potentials $p^{\Lambda}$, $q_{\Lambda}$, $\phi^\Lambda$ correspond to ours $\Gamma^I$, $-\Gamma_I$ and $\beta \mu^I$, respectively.}, let us focus on the saddle with $n_a^I  = 0$ whose action we denote by ${I}_{\rm OSV}(\{\Gamma^I_a\}, \mu^I)$. For such a saddle, the $\tau$-component of the gauge field is purely imaginary, and in the zero-temperature limit ($\beta \to \infty$ with $\alpha^I = \beta \mu^I = \text{fixed}$) such a Euclidean saddle can be analytically continued to the real Lorentzian extremal geometry initially considered by OSV.  To make this explicit, we can rewrite the prepotential in terms of the topological string free energy as explained in the introduction,%
\begin{align}
    F_{\rm top}(t^A,g_{\rm top})
    &= - 2\pi \i\,
    \mathcal F\!\left(
        \frac{2\pi \i}{g_{\rm top}}\,(1,t^A),
        64
    \right)\,,\nonumber  \\ \bar F_{\rm top}(t^A,g_{\rm top})
    &= +2\pi\i\,
    \bar{\mathcal F}\!\left(
        \frac{2\pi \i}{g_{\rm top}}\,(1,t^A),
        64
    \right).
    \label{eq:Ftop-from-prepotential}
\end{align} 

For the single black-hole saddle with $n_a^I = 0$, to evaluate the action in \eqref{eq:on-shell-action-OSV} we must thus use 
\begin{align}
    t_N^A&=\frac{\Gamma^A + \frac{\i \beta \mu^A}{\pi}}{\Gamma^0 + \frac{\i \beta \mu^0}{\pi}}, &g_{{\rm top},N}&=\frac{4\pi\i}{{\Gamma^0} +\frac{\i \beta \mu^0}{\pi}},\nonumber 
    \\
    t_S^A&=\frac{\Gamma^A - \frac{\i \beta \mu^A}{\pi}}{\Gamma^0- \frac{\i \beta \mu^0}{\pi}}, &g_{{\rm top},S}&=\frac{-4\pi\i}{\Gamma^0 -\frac{\i \beta \mu^0}\pi},
    \label{eq:OSV-topological-moduli}
\end{align}
where the sign of $g_{\rm top}$ is arbitrary and does not affect the closed-string genus expansion. We choose it such that  $g_{{\rm top},S} = (g_{{\rm top},N})^*$ for real $\mu^I$.

Combining \eqref{eq:on-shell-action-OSV}, \eqref{eq:Ftop-from-prepotential}, and \eqref{eq:OSV-topological-moduli} proves the factorization of the single-center saddle weight,
\begin{equation}
    e^{-{I}_{\rm OSV}(\{\Gamma^I_a\}, \mu^I)}
    =Z_{{\rm top},N}\,\bar Z_{{\rm top},S},
    \quad \text{where }\quad Z_{{\rm top}}=e^{F_{\rm top}}.
    \label{eq:OSV-factorized-saddle-weight}
\end{equation}
For the unshifted saddle with $n_a^I = 0$ and real $\mu^I$, the two prepotentials are related by complex conjugation since $\gamma_N^I=(\gamma_S^I)^*$, $t_N^A=(t_S^A)^*$, and $g_{{\rm top},N}=g_{{\rm top},S}^*$. Hence\footnote{For complex chemical potentials $\mu^I$, the product in \eqref{eq:OSV-factorized-saddle-weight} is the appropriate analytic continuation.  }
\begin{equation}
    e^{-{I}_{\rm OSV}(\{\Gamma^I_a\}, \mu^I)}
    =\left|Z_{\rm top}(t,g_{{\rm top}})\right|^2,
    \label{eq:OSV-modulus-square}
\end{equation}
with the shorthand notation, $t\equiv t_N$ and $g_{\text{top}} \equiv g_{\text{top},N}$.
Since we derived the action to arbitrary order in derivatives, \eqref{eq:OSV-modulus-square} holds to all orders in the genus expansion of $ F_{\rm top}(t,g_{\rm top})$. To be explicit, this expansion is given by
\begin{equation}
    F_{\rm top}(t,g_{\rm top})
    =\sum_{g=0}^{\infty}g_{\rm top}^{2g-2}F_{{\rm top},g}(t)\,,
    \label{eq:OSV-topological-genus-series}
\end{equation}
which determines the derivative expansion of the prepotential
\begin{equation}
    \CF(X,A)=\sum_{g=0}^{\infty}\CF^{(g)}(X)A^g,
    \qquad
    \CF^{(g)}(\lambda X)=\lambda^{2-2g}\CF^{(g)}(X)\,,
    \label{eq:OSV-genus-prepotential}
\end{equation}
with $A = 64$ and $X = \frac{2\pi \i}{g_{\rm top}}\,(1,t^A) $.
The first few terms in the expansion are given by\footnote{Here, we use the standard Calabi--Yau normalization $C_{ABC}=\int_{\mathcal Y}\omega_A\wedge\omega_B\wedge\omega_C = -6 D_{ABC}$ and $c_{2A}=\int_{\mathcal Y}c_2(\mathcal Y)\wedge\omega_A$.} 
\begin{equation}
    F_{\rm top}(t,g_{\rm top})
    = \frac{(2\pi \i)^3}{g_{\rm top}^2}
      D_{ABC}t^At^Bt^C+\frac{\chi(\mathcal Y)\zeta(3)}{2g_{\rm top}^2}
      -\frac{\pi\i}{12}c_{2A}t^A+ O(g_{\rm top}^2),
    \label{eq:OSV-large-volume-check}
\end{equation}
where $\chi(\mathcal Y)$ is the Euler characteristic of the Calabi--Yau threefold, as in \cite{Bershadsky:1993cx,Ooguri:2004zv}. At every genus, $F_{{\rm top},g}(t)$ also receives worldsheet instanton contributions, and for $g\geq 2$ it additionally contains constant map contributions; such terms are also included in our computation of the on-shell action. The constant map contributions are given by \cite{Marino:1998pg,Faber:1998gsw,Gopakumar:1998ii} 
\begin{equation}
    F_{{\rm top},g}^{\rm const}
    =
    -\frac{\chi(\mathcal Y)}{2}\,
    \frac{\lvert B_{2g}B_{2g-2}\rvert}
    {2g(2g-2)(2g-2)!},
    \qquad g\geq2,
    \label{eq:constant-map-genus-g}
\end{equation}
where $B_{2g}$ are the Bernoulli numbers.
The worldsheet instanton contributions can be organized as \cite{Gopakumar:1998ii,Gopakumar:1998jq,Denef:2007vg}
\begin{equation}
    F_{\rm top}^{\rm inst}(t,g_{\rm top})
    =\sum_{g\geq0}\ \sum_{\substack{\beta_{\rm ws}\ {\rm effective}\\\beta_{\rm ws}\neq0}}
      N_{g,\beta_{\rm ws}}\,
      (-g_{\rm top}^2)^{g-1}
      e^{2\pi\i\beta_{{\rm ws},A}t^A},
    \label{eq:OSV-worldsheet-instantons}
\end{equation}
where $N_{g,\beta_{\rm ws}}$ are the connected Gromov--Witten invariants, including their multiple-cover and degenerate contributions. Here, $\beta_{\rm ws}$ denotes the homology class of the curve wrapped by the world sheet. More explicitly, take $\{\omega_A\}_{A=1}^{n_v}$, $n_v=h^{1,1}(\mathcal{Y})$ to be an integral basis of $H^2(\mathcal{Y},\mathbb Z)$,
chosen to generate the Kähler cone, and $\{C^A\}$ as the dual basis of 2-cycles in
$H_2(\mathcal{Y},\mathbb Z)$, $\int_{C^A}\omega_B=\delta^A{}_B$. The complexified Kähler moduli are the periods ,
\begin{equation}
  t^A=\frac{1}{4\pi^2\alpha'}\int_{C^A}\big(B_2+\i J\big)\equiv b^A+\i\,v^A ,
  \qquad
  B_2+\i J=4\pi^2\alpha'\,t^A\omega_A ,
\end{equation}
where $B_2$ is the NS-NS two-form and $J$ the string-frame Kähler form.
In supergravity variables, $t^A=X^A/X^0$, and the CY volume is
${\rm vol}(\mathcal{Y})=\tfrac{(4\pi^2\alpha')^3}{6}\,C_{ABC}v^Av^Bv^C$.
A worldsheet wrapping a curve of class $\beta_{\rm ws}=\beta_{{\rm ws},A}C^A$, with
$\beta_{{\rm ws},A}=\int_{\beta_{\rm ws}}\omega_A\in\mathbb Z_{\geq0}$ for effective classes, then has action
\begin{equation}
  e^{-\frac{1}{2\pi\alpha'}\int_{\beta_{\rm ws}}J\,+\,\frac{\i}{2\pi\alpha'}\int_{\beta_{\rm ws}}B_2}
  = e^{2\pi\i\,\beta_{{\rm ws},A}t^A},
\end{equation}
which is the weight appearing in \eqref{eq:OSV-worldsheet-instantons}.

To summarize, we have proven the OSV conjecture to all orders in the topological string coupling by using the index saddle that, in the zero-temperature limit, can be analytically continued to the extremal Lorentzian black hole geometry initially considered in the conjecture. Nevertheless, there are a multitude of other saddles that contribute to the helicity supertrace in $Z_\text{BH}^{\text{OSV}}( \Gamma^I, \mu^I; \Omega_\infty)$. Each of these saddles has an on-shell action \eqref{eq:on-shell-action-OSV} that localizes to contributions of the prepotential evaluated at the north and south poles of the geometry, a factorization property that we will now use to provide a non-perturbative refinement of the conjecture.

\section{A non-perturbative refinement of OSV}
\label{sec:gravitational-index-to-all-refinement}
Our proposed refinement of OSV is that the factorization of the helicity supertrace \eqref{eq:OSV-ensemble} continues to hold saddle by saddle,
\begin{equation}
Z_\text{BH}^{\text{OSV}}(\{\Gamma^I_a\}, \mu^I, \{n^I_a\}; \Omega_\infty) \underset{\substack{\text{saddle-point}\\{\text{approx.}}}}{=} \prod_{a=1}^n Z_\text{top}(t_{N,a}, g_{\text{top}, N, a}) \bar{Z}_\text{top}(t_{S,a}, g_{\text{top}, S, a})\,, 
\end{equation}
where, as before, each saddle is labeled by the number of black holes $n$, by the distribution of magnetic charges among the black holes $\{\Gamma_a^I\}$, and by the dual-lattice vectors $\{n_a^I\}$ that specify the possible smooth gauge connections for a given value of the magnetic charges.
As in the introduction, the equality holds in the saddle-point approximation, i.e., at least at leading order in the $G_N \to 0$ limit but to all orders in the topological string coupling.  The values of the moduli at which each topological string partition function is evaluated, as well as the values of the topological string couplings, are independently given by the attractor values at the north and south poles of each black hole,
\begin{align}
    t_{N, a}^A&=\frac{\Gamma^A_a + \frac{\i \beta \mu^A}{\pi} - 2n^A_a}{\Gamma^0_a + \frac{\i \beta \mu^0}{\pi} - 2n^0_a},&g_{{\rm top},N,a}&=\frac{4\pi\i}{\Gamma^0_a +\frac{\i \beta \mu^0}{\pi} - 2n^0_a},\nonumber
    \\
    t_{S,a}^A&=\frac{\Gamma^A_a -\frac{\i \beta \mu^A}{\pi} + 2n^A_a}{\Gamma^0_a -\frac{\i \beta \mu^0}{\pi} + 2n^0_a} , &g_{{\rm top},S,a}&=\frac{-4\pi\i}{\Gamma^0_a -\frac{\i \beta \mu^0}{\pi} + 2n^0_a}.
    \label{eq:OSV-topological-moduli-general}
\end{align}
This refinement allows us to extend the regime of applicability of OSV to cases where the dominant saddle is not the single-center black hole with $n^I = 0$ but instead is a configuration with a higher number of black holes and/or with $n^I_a \neq 0$. This applies, for instance, to cases where the charges are such that enigmatic black hole configurations dominate over the single-center black hole \cite{Denef:2007vg, Boruch:2025biv}.  Below, we comment on the different kinds of saddles that can contribute to $Z_\text{BH}^{\text{OSV}}(\Gamma^I, \mu^I; \Omega_\infty) $. 

\subsection{Shifted saddles and the quantization of electric charges}
\label{sec:shifted-saddle}

The calculation in Sec.~\ref{sec:gravitational-index-to-all-orders} does not yield a helicity supertrace that is periodic in the electric chemical potential. This is required in order for the electric charges to be quantized. The reason we do not see this periodicity is that we are not including the full sum over single black hole saddles. For this we simply need to take the result in \eqref{eq:on-shell-action-OSV} and focus on the saddles with one black hole, $n=1$, but with $n^I \neq 0$. 

For such saddles $g_{\text{top}, N} \neq (g_{\text{top}, S})^*$ and  $t^A_N \neq (t^A_S)^*$. Thus, the partition function does not factorize into a holomorphic and anti-holomorphic contribution; nevertheless, it does factorize into a product of two topological string partition functions, one evaluated at the attractor values at the north pole and the other at the south pole of the saddle. The electric charges for each individual saddle are not quantized. This is because  \eqref{eq:sum_lower_gamma} dictates that the electric charge is
\begin{equation}
    \Gamma_{I} =  \CF_I\left(\frac{\Gamma^J}{2} +\frac{\i \beta \mu^J - 2\pi n^J }{2\pi}, 64\right) +  \bar\CF_I\left( \frac{\Gamma^J}{2} -\frac{\i \beta \mu^J - 2\pi n^J }{2\pi}, 64\right)\,,
\end{equation}
which need not be quantized for $n^J \in \Lambda^*$.
However, the sum over saddles automatically ensures that the overall partition function can be re-expressed in terms of a sum over electric charges in $\Lambda$ as in \eqref{eq:from-canonical-to-OSV} since the partition function is periodic in the chemical potential $\beta \mu^I \sim \beta\mu^I + 2\pi \i\,n^I$. This is because under  $\beta \mu^I \to \beta\mu^I + 2\pi \i\,n^I$ the contribution of one saddle to $Z_\text{BH}^{\text{OSV}}(\Gamma^I, \mu^I; \Omega_\infty)$ is mapped to that of another saddle.  

The sum over such saddles is somewhat formal in this case since the explicit sum over charges in \eqref{eq:from-canonical-to-OSV} need not converge; this is because unless there are large cancellations between configurations with different charges the fast growth of the Wald entropy in \eqref{eq:Wald-entropy} with the charge causes the sum to diverge \cite{Dabholkar:2005by,Dabholkar:2005dt,Pioline:2006ni,Denef:2007vg}. Consequently, the sum over $n^I$ is also not expected to converge. This can be seen even at the two-derivative level by expanding the leading order in~\eqref{eq:OSV-large-volume-check} for large values of $n_a^A$ using \eqref{eq:OSV-topological-moduli-general}. The problem is worsened by the worldsheet instanton contributions in \eqref{eq:OSV-worldsheet-instantons} where at large $n_a^A$ for real $\mu^0$ there is an exponential divergence in  $n_a^A$. This is not as bad as it looks -- if one is instead interested solely in computing the canonical helicity supertrace $\Omega(\Gamma^I, \Gamma_I; \Omega_\infty)$ in~\eqref{eq:canonical-ensemble}, which for a single black hole is clearly finite, through Poisson resummation, one simply needs to perform a Fourier transform in $n^I$ as in \eqref{eq:Poisson-summation-mixed-to-canonical} with a defining contour for  $n^I$ where the result converges.

\subsection{Multi-center black holes}
\label{sec:multi-center}

In addition to the shifted single-center saddles discussed above, one should consider multi-center saddles. Such saddles have a rich moduli space previously discussed in \cite{Boruch:2025biv}, determined by the finite-temperature regularity conditions on the pole locations $\vec x_{N,a}$ and $\vec x_{S,a}$. 

Similarly to the unshifted single-center saddle discussed in Sec.~\ref{sec:gravitational-index-to-all-orders}, the saddles with $n^I_a = 0$ can be analytically continued to the Lorentzian multi-center black hole geometry considered in the past, see for instance \cite{Bates:2003vx,Denef:2007vg}. Given a set of magnetic charges and an electric chemical potential, we can compare the action of the unshifted single-center saddle to that of an unshifted two-center saddle for a given distribution of the magnetic charges. The configurations where the action of the two-center saddle is smaller than that of the single-center saddle are analogous to the enigmatic configurations discussed in \cite{Denef:2007vg}. While the existence of such saddles was established in the canonical ensemble in \cite{Denef:2007vg}, finding these configurations in the OSV ensemble as contributions from unshifted saddles is less straightforward since the electric chemical potential $\mu^I$ determines the electric charges $\Gamma_{I,a}$ of both black holes. To illustrate the difficulty of constructing enigmatic configurations in the OSV ensemble as compared to the canonical ensemble, note that the total electric charges of the single- and two-center configurations need not even be equal in the OSV ensemble, whereas in the canonical ensemble the configurations whose action we are comparing are defined to have the same electric charge. This is because, in the OSV ensemble, the total charge is given by \eqref{eq:sum_lower_gamma},
\begin{equation}
\sum_{a=1}^n\Gamma_{I,a} =  \sum_{a=1}^n \left[  \CF_I\left(\frac{\Gamma^J_a}{2} +\frac{\i \beta \mu^J - 2\pi n^J_a }{2\pi}, 64\right) +  \bar\CF_I\left( \frac{\Gamma^J_a}{2} -\frac{\i \beta \mu^J - 2\pi n^J_a }{2\pi}, 64\right)\right]\,,
\end{equation}
which can clearly depend on how we divide the magnetic charges among the black holes. Nevertheless, it is possible to find enigmatic unshifted two-black-hole saddles and we describe an example in App.~\ref{sec:enigmas}. 

As for the single-center saddles, the saddles with $n^I_a \neq 0$ ensure the quantization of electric charges for each black hole; the electric charges for the individual saddles are not necessarily quantized and are also different amongst the saddles. The problems that we encountered with divergences in the single-center case are still there, but if we are once again solely interested in recovering the contribution of multi-center configurations to $\Omega(\Gamma^I, \Gamma_I; \Omega_\infty)$ one needs to perform a Fourier transform in $n^I_a$ as in \eqref{eq:Poisson-summation-mixed-to-canonical}, again with a defining contour where the result converges.

\section{Discussion}
\label{sec:discussion}

In this paper we used the gravitational path integral to derive the OSV relation to all orders in the topological string coupling and at leading order in $G_N$. We generalized the finite-temperature supersymmetric black hole saddles to include the higher-derivative corrections encoded by the prepotential and used equivariant localization to compute their on-shell action. The action localizes to contributions at the north and south poles of each black hole, which are identified with the holomorphic and anti-holomorphic topological string free energies, respectively. For the unshifted single-center saddle, this gives the squared modulus of the topological string partition function, as conjectured by OSV. The same calculation applies to shifted gauge-field configurations and multi-center black holes, yielding a non-perturbative refinement in $G_N$ that holds saddle by saddle. This refinement accounts for the quantization of electric charges and remains relevant when the single-center saddle is absent or subdominant. 

Our results for the conformal supergravity action of the supersymmetric Euclidean saddles pass important structural checks compatible with expectations from topological string theory. We find that D-terms in 4d $\mathcal{N}=2$ conformal supergravity do not contribute to the action and explicitly compute the contribution of F-terms. The result shows that the on-shell action coming from such terms is independent of the moduli of the multicentered non-extremal saddles associated with the positions and relative orientation of the centers \cite{Boruch:2025biv} and, up to wall-crossing phenomena where a saddle could stop contributing to the GPI, also independent of the choice of scalar moduli at infinity.

The calculation of the 4d Euclidean conformal supergravity action facilitated by equivariant localization is naturally performed in the mixed ensemble. This is also in harmony with expectations from the underlying type IIA string theory. The type IIA supergravity action is formulated in terms of the 10d metric, dilaton and $B_2$ field, together with the RR forms $C_{(1)}$ and $C_{(3)}$. Therefore, it is natural to fix the values of $C_{(1)}$ and $C_{(3)}$ at asymptotic infinity. This corresponds to fixing the number of D6- and D4-branes since they couple magnetically to $C_{(1)}$ and $C_{(3)}$ and the chemical potentials for D0- and D2-branes, since they couple electrically to the RR forms. This perspective is in the spirit of the Fradkin-Tseytlin approach to the string theory effective action, see~\cite{Fradkin:1985fq,Fradkin:1985ys}, and was recently emphasized in \cite{Gautason:2025per,vanMuiden:2026nsp,Bobev:2026gir} in the context of AdS/CFT. To be more explicit, note that the 9d boundary of the type IIA uplift of our saddles is of the form $S^1_{\beta}\times S^2 \times \mathcal{Y}$. For a general CY manifold $\mathcal{Y}$, there are no 1-cycles and therefore we need to specify the holonomy of $C_{(1)}$ on $S^1_{\beta}$. Similarly, we should specify the period of $C_{(3)}$ on the $h^{1,1}(\mathcal{Y})=n_v$ $S^1_{\beta}\times C^A$ 3-cycles. The periods of $C_{(3)}$ along the $2(h^{2,1}(\mathcal{Y})+1)$ 3-cycles of $\mathcal{Y}$ describe the boundary conditions for the D2-brane instantons which do not contribute to the perturbative path integral series. We therefore conclude that the type IIA string theory path integral for the boundary conditions of interest to us should be specified in terms of the electric chemical potentials for D0- and D2-branes and the number, or charge, of D4- and D6-branes. This corresponds precisely to the 4d supergravity mixed ensemble employed in our analysis.

We now proceed with a discussion of several open problems and possible generalizations of our results.

\paragraph{On-shell and off-shell in the presence of a Killing spinor.} The method of equivariant localization is not limited to computing the action of on-shell geometries but in principle can be used to compute the action of any geometry that admits a well-defined Killing spinor. Nevertheless, the action of a generic geometry that admits a Killing spinor need not agree with that of the saddles, because the localization formula depends on the values of the fields at the fixed points of $\xi$, and whether these values are fixed by the boundary conditions depends on the ensemble. The results of App.~\ref{app:off-shell-attractor} make this precise. With Dirichlet boundary conditions for all fields, as in the mixed ensemble, the magnetic charge of each black hole is topological, see~\eqref{eq:magnetic-mapping-to-infinity}, and the gauge-field regularity condition \eqref{eq:equivariant-shifted-saddles} relates the values of the scalars at each pole to the fixed chemical potential $\mu^I$ using only the Killing spinor equations. The attractor values at the fixed points therefore hold off shell. In the canonical ensemble, by contrast, the values at the fixed points are determined by the electric charges through \eqref{eq:sum_lower_gamma}, which holds off shell only for the horizon charges $\Gamma_{a,I}^{\rm horizon}$; these coincide with the asymptotic charges held fixed in this ensemble only once Maxwell's equations are imposed, see \eqref{eq:electric-mapping-to-infinity}. In the canonical ensemble, the attractor values at the fixed points are thus only determined on shell. Since the action in the OSV ensemble, \eqref{eq:on-shell-action-OSV}, depends only on the residues $\gamma^I_{S,a}$ and $\gamma^I_{N,a}$, which are fixed by $\mu^I$, $n^I_a$ and $\Gamma^I_a$ through \eqref{eq:OSV-pole-parametrization}, we expect the action to stay fixed under any off-shell deformation of each saddle that preserves the Killing spinor $\epsilon$. This suggests that, in the OSV ensemble, the contribution of each saddle to the GPI could be one-loop exact. There is already an indication that this is the case: in $\mathcal N=4$ and $\mathcal N=8$ 4d supergravity, where the OSV partition function can be evaluated from the exact microscopic degeneracies, the answer in the OSV ensemble appears to be one-loop exact \cite{Shih:2005he}. 

\paragraph{One-loop corrections and supersymmetric localization.} Our calculation was performed at leading order in $G_N$ and to all orders in $g_{\rm top}$, but did not include loop corrections around the Euclidean saddles that contribute to the indexed partition function. If the factorization underlying OSV extends to these quantum corrections, the one-loop determinant should likewise factorize into contributions from the north and south poles of each black hole. Since our backgrounds preserve supersymmetry, the Atiyah--Bott fixed-point theorem suggests a route to this calculation through the equivariant index of the fluctuation complex \cite{Pestun:2007rz,Jeon:2018kec,Iliesiu:2022kny,H:2023qko}. The relevant fixed points are those of the Killing vector $\partial_\tau$ constructed from the Killing spinor bilinear that we have also used above in the BVAB fixed-point formula to compute the on-shell action. As with the  BVAB fixed-point formula, applying this method to our complex saddle-point geometries with an asymptotic boundary requires a careful treatment of boundary conditions, boundary contributions, and possible zero modes. This analysis will also elucidate the measure one needs to use when performing Laplace-type transformations when changing ensembles. Recently, a suggestive expression inspired by twisted holography was proposed in~\cite{BenettiGenolini:2026kjy} for similar one-loop contributions to the GPI in the context of AdS$_4$/CFT$_3$ and it will be interesting to understand whether it can be adapted to the OSV context. More ambitiously, it is desirable to develop world-sheet supersymmetric localization methods, perhaps along the lines of~\cite{Beasley:2005iu}, that can be used to evaluate the GPI exactly.

\paragraph{Relation to microstates.} Our results establish the validity of the OSV conjecture to all orders in the string coupling, together with a non-perturbative refinement, and should be useful to make precise connections with the microscopic evaluation of the black hole helicity supertrace in string theory. As mentioned above, type IIA reductions on $T^6$ and $T^2\times K3$ that preserve extended supersymmetry, are the simplest examples to understand \cite{Shih:2005he}. Extending this to general CY compactifications that preserve 4d $\mathcal{N}=2$ supersymmetry is more challenging and perhaps the quintic CY 3-fold is a natural explicit example to consider, see~\cite{Huang:2006hq}. More generally, it will be interesting to connect our results to the work of Denef-Moore~\cite{Denef:2007vg} and understand better the enigmatic multi-center configurations we briefly discussed as well as wall-crossing phenomena.

\paragraph{Application to small black holes.} Our all-orders expression for the on-shell action in terms of the topological string coupling also allows us to determine the contribution of small black holes, for which higher-derivative corrections are essential even at leading order \cite{Dabholkar:2004yr,Dabholkar:2004dq,Sen:2004dp}. As shown in \cite{Chowdhury:2024ngg}, one can use scaling arguments to determine the charge dependence of this contribution, leaving an overall numerical coefficient undetermined. Chen, Murthy, and Turiaci conjectured a formula for the higher-derivative on-shell action and, assuming its temperature independence, evaluated it in the zero-temperature limit to obtain the precise leading microscopic index \cite{Chen:2024gmc}. Our conformal supergravity and equivariant localization analysis supplies a direct finite-temperature derivation of this action including both F-terms and D-terms. It thus confirms the conjectured result for the index in the canonical ensemble in~\cite{Chen:2024gmc} and explains how to include the small-black-hole contribution within the OSV relation in the mixed ensemble. 

\paragraph{Uplift to 5d flat space and AdS$_3\times S^2$.} The contributions of 5d black holes, black rings, and black lenses to the index have recently been discussed in the context of two-derivative supergravity \cite{Adhikari:2024zif,Cassani:2025iix,Bandyopadhyay:2025jbc,Boruch:2025qdq,Boruch:2025sie}. These saddles can be obtained by uplifting the 4d index saddles discussed in this paper on the M-theory circle $S^1_M$ and then decompactifying this circle. One needs to find an uplift of the 4d higher-derivative saddles we studied here that solves the higher-derivative 5d equations of motion. Presumably, using the conformal supergravity formalism can streamline this analysis and would extend our derivation of the OSV relation and its non-perturbative refinement to five-dimensional black objects. Furthermore, for $\Gamma^0=0$, taking the low-temperature limit while scaling the size of $S^1_M$ with the thermal circle gives a decoupled, asymptotically AdS$_3\times S^2$ geometry at finite temperature. The resulting gravitational saddles directly compute contributions to the elliptic genus of the dual MSW CFT~\cite{Maldacena:1997de}. At the two-derivative level, they include rotating BTZ$\times S^2$ saddles with $S^1\times S^2$ horizons, BMPV-type black holes with $S^3$ horizons, black lenses with $S^3/\mathbb Z_k$ horizons, and their allowed multi-center bound states. Extending this construction to the higher-derivative theory would place OSV and its saddle-by-saddle refinement directly in the framework of AdS/CFT, making contact with the AdS$_3$/CFT$_2$ derivation of OSV in \cite{Gaiotto:2006ns}, see \cite{Castro:2008ne} for a review. This analysis will likely be facilitated by the recent results on equivariant localization of the on-shell action of 5d supersymmetric Euclidean supergravity solutions, \cite{Cassani:2024kjn,Cassani:2025iix,Cassani:2026teb,Colombo:2025ihp,Colombo:2025yqy}.

\paragraph{OSV in AdS.} It is natural to wonder whether there is a version of the OSV relation for asymptotically AdS$_4$ black holes. There are indeed suggestive results from various perspectives in the literature that point to a version of the OSV formula. Higher-derivative gauged supergravity combined with results from supersymmetric localization strongly suggests that the supergravity prepotential is determined by the dual 3d SCFT free energy on $S^3$ in the presence of appropriate real mass deformations, see~\cite{Bobev:2020egg,Bobev:2021oku,Genolini:2021urf,BenettiGenolini:2026qdm,Bobev:2022eus,Bobev:2025ltz,Hristov:2024cgj,Hristov:2022lcw}. This is backed up by an OSV-type relation between the superconformal and topologically twisted indices dual to the Kerr-Newman and Reissner-Nordstr\"om AdS$_4$ black holes and the square of the $S^3$ partition function, see~\cite{Bobev:2026lvl} and references therein. Finally, it was recently emphasized that these QFT supersymmetric localization calculations and the agreement with the holographically dual AdS$_4$ backgrounds in the context of M-theory are most naturally formulated in the grand canonical ensemble with fixed chemical potential for M2-branes~\cite{Bobev:2026gir}, with further support from equivariant localization~\cite{BenettiGenolini:2026cyc,BenettiGenolini:2026kjy}. Despite these similarities between the OSV conjecture for asymptotically flat and AdS black holes, there is an important difference, namely the relation to topological strings, which is more tenuous in the AdS setup. It will certainly be very interesting to put the OSV conjecture in AdS$_4$ on a more solid footing and to clarify the underlying connection to topological strings.

\subsection*{Acknowledgments}

\noindent We are grateful to Mina Aganagic, Roberto Emparan, Fri\dh rik Freyr Gautason, Junho Hong, Raghu Mahajan, Jesse van Muiden, Sameer Murthy, Valentin Reys, Ashoke Sen, James Sparks and Gustavo Joaquin Turiaci for useful discussions. NB is supported in part by FWO projects G003523N, G094523N, and G0E2723N, as well as by the KU Leuven C1 project C16/25/01. JB is supported by the Leinweber Foundation and the Leinweber Institute for Theoretical Physics at Stanford University. LVI was supported in part by the Leinweber Institute for Theoretical Physics at UC Berkeley, by
the Department of Energy, Office of Science, Office of
High Energy Physics through the award DE-SC0025522,
and by the Department of Energy through QuantISED
award DE-SC0019380.  NB, JB, and LVI would like to acknowledge the hospitality of the Galileo Galilei Institute for Theoretical Physics during the “Pathways to Quantum Black Holes” workshop where this project was initiated. This work was performed in part
at the Aspen Center for Physics, which is supported by
National Science Foundation grant PHY-2210452.

\subsection*{Statement on AI use}

The manuscript was written by the authors, who take full responsibility for any errors. AI helped with typographical errors, matching conventions to other papers, and in the understanding of the role of D-terms in the higher-derivative theory. 

\appendix 

\section{Conventions}\label{app:conventions}

In this appendix, we provide a dictionary between the conventions used in this paper and the Euclidean supergravity conventions of \cite{deWit:2017cle,Adhikari:2025eok,BenettiGenolini:2026qdm,Gutowski:2012yb}. The latter are naturally adapted to the timelike dimensional reduction of five-dimensional Lorentzian $\mathcal N=2$ supergravity. In contrast, the conventions used in the main text are adapted to the Euclidean theory obtained by Wick rotation of four-dimensional Lorentzian $\mathcal N=2$ supergravity, as in \cite{Boruch:2023gfn,Boruch:2025biv}. These conventions are directly related to the standard Lorentzian special-K\"ahler structure, with vector multiplet scalars $X^I$, $\bar X^I$ being complex conjugates upon imposing the four-dimensional Lorentzian reality contour. By contrast, in the theory arising from timelike dimensional reduction, vector multiplet scalars $X_+^I$, $X_-^I$ are independent real projective coordinates of Euclidean special para-K\"ahler geometry \cite{Cortes:2005uq}. 

A convenient way to obtain a map between the two sets of conventions is to use their common five-dimensional origin: timelike reduction of five-dimensional Lorentzian supergravity gives the conventions of \cite{Gutowski:2012yb,deWit:2017cle}, while Wick rotating the five-dimensional theory first and subsequently reducing to four dimensions gives the conventions used here (see \cite{Ciceri:2023mjl} for a detailed discussion of the five-dimensional uplift\footnote{Note however that the orientation used in \cite{Ciceri:2023mjl} is ${\rm vol}_M = e^1\wedge e^2\wedge e^3 \wedge e^4$, where $e^1$ is along the radial Euclidean AdS${}_2$ direction and $e^2$ is along Euclidean time. This orientation is opposite to ours.}).  As in the main text, here we restrict ourselves to bosonic fields. We do not specify the corresponding continuation of the fermions. Then, the five-dimensional Wick rotation amounts to multiplying the fields in the $0$th vector multiplet by $\i$, while the remaining multiplets are identified directly, with the understanding that their reality conditions may differ in the two Euclidean formulations. 

Denoting the quantities arising from timelike KK reduction with a superscript $(\mathsf{t})$, we have:
\begin{equation}
\mathbb{V}^0 = \i \mathbb{V}^{0(\mathsf{t})},
\qquad
\bigl(X^0, \bar{X}^0 ,\, W^0_\mu,\, Y^{0,ij}\bigr)
= 
\bigl(\i X_-^{0(\mathsf{t})} , \i X_+^{0(\mathsf{t})} ,\, \i W^{0(\mathsf{t})}_\mu,\, \i Y^{0,ij(\mathsf{t})}\bigr).\label{timelike_KKvect_multiplet}
\end{equation}
The remaining vector multiplets are identified,
\begin{equation}
\mathbb{V}^A = \mathbb{V}^{A(\mathsf{t})},
\qquad
\bigl(X^A, \bar{X}^A ,\, W^A_\mu,\, Y^{A,ij}\bigr)
= 
\bigl(X_-^{A(\mathsf{t})} , X_+^{A(\mathsf{t})} ,\, W^{A(\mathsf{t})}_\mu,\, Y^{A,ij(\mathsf{t})}\bigr).\label{timelike_vect_multiplet}
\end{equation}
Consequently, the special coordinates are identified as 
\begin{equation}
    t_-^{A(\mathsf t)} = \frac{X_-^{A(\mathsf t)}}{X_-^{0(\mathsf t)}} = \i\frac{X^A}{X^0} = \i t^A,\qquad t_+^{{A}(\mathsf t)} = \frac{X_+^{{A}(\mathsf t)}}{X_+^{0(\mathsf t)}} = \i\frac{\bar{X}^{A}}{\bar{X}^0} = \i\bar{t}^{A}.
\end{equation}

The two independent real prepotentials $\CF^{-(\mathsf t)}$ and $\CF^{+(\mathsf t)}$ of the timelike-reduced theory are related to the holomorphic and anti-holomorphic prepotentials used in our conventions by
\begin{equation}
    \CF(X,A|_{\mathbb W^2}) = \i \CF^{-(\mathsf{t})}(X_-^{(\mathsf t)},A_-^{(\mathsf t)}|_{\mathbb W^2}),\qquad \bar \CF(\bar X,\bar A|_{\overline{\mathbb W}^2}) = -\i \CF^{+(\mathsf{t})}(X_+^{(\mathsf{t})},A_+^{(\mathsf{t})}|_{\mathbb W^2}).
\end{equation}
Together with the continuation of the vector-multiplet scalars given above, these relations map the special para-K\"ahler potential of the timelike-reduced theory to the standard special-K\"ahler potential from \cite{Boruch:2023gfn},
\begin{equation}
    (e^{-\CK})^{(\mathsf{t})} = \CF_I^{+(\mathsf{t})}X_-^{I(\mathsf{t})} + \CF^{-(\mathsf{t})}_IX_+^{I(\mathsf{t})} = -\i(\CF_I\bar X^I - \bar \CF_I X^I) = \i\la\Omega,\bar{\Omega}\ra = e^{-\mathcal{K}}.
\end{equation}
Here $\Omega = (X^I,\CF_I)$, $\bar{\Omega} = (\bar X^I, \bar{\CF}_I)$ are the unnormalized period vectors.

From \eqref{timelike_KKvect_multiplet} and \eqref{timelike_vect_multiplet}, we see that the vector multiplet field strengths are related as:
\begin{equation}
    F_{\mu\nu}^{0} = \i F_{\mu\nu}^{0(\mathsf t)},\qquad F_{\mu\nu}^A = F_{\mu\nu}^{A(\mathsf t)}.
\end{equation}
We next determine the corresponding continuation of the dual field strengths $G_I$. In our conventions, they are defined by (see e.g. Eq.~(2.13) of \cite{LopesCardoso:2000qm})
\begin{align}
    G_{\mu\nu\, I}^{+} &= \CF_{IJ}(F_{\mu\nu}^{J+} - \frac{1}{4}\bar X^{J}T_{\mu\nu}^{+}) + \CF_{IA} F_{\mu\nu}^{+}|_{{\mathbb W}^2} + \frac{1}{4}\bar \CF_IT_{\mu\nu}^{+},
    \\
    G_{\mu\nu\, I}^{-} &= \bar \CF_{IJ}(F_{\mu\nu}^{J-} - \frac{1}{4}X^{J}T_{\mu\nu}^{-}) + \bar \CF_{IA} F_{\mu\nu}^{-}|_{\overline{\mathbb W}^2} + \frac{1}{4}\CF_IT_{\mu\nu}^{-},
\end{align}
where, as in the main text, the subscript $A$ on $\CF_{IA}$ and $\bar\CF_{IA}$ denotes the derivative with respect to $A|_{\mathbb W^2}$ and $\bar A|_{\overline{\mathbb W}^2}$, respectively (not a vector-multiplet index), and, up to fermionic fields, $F_{ab}^{+}|_{\mathbb W^2} = - 16 R(M)^{cd}{}_{ab}T^{+}_{cd}$ and $F_{ab}^{-}|_{\overline{\mathbb W}^2} = - 16 R(M)^{cd}{}_{ab}T^{-}_{cd}$. In the timelike-reduced theory, the dual field strengths are instead defined as
\begin{equation}
    G_{\mu\nu\, I}^{\pm (\mathsf{t})} = \pm \left[\CF_{IJ}^{\mp(\mathsf{t})}(F_{\mu\nu}^{J\pm(\mathsf{t})} - \frac{1}{4}X_{\pm}^{J(\mathsf{t})}T_{\mu\nu}^{\pm(\mathsf{t})}) + \CF_{IA}^{\mp(\mathsf{t})} F_{\mu\nu}^{\pm (\mathsf{t})}|_{\mathbb W^2_{\mp}} - \frac{1}{4}\CF_I^{\pm(\mathsf{t})}T_{\mu\nu}^{\pm(\mathsf{t})}\right].
\end{equation}
Using the identification of the vector multiplets and prepotentials described above, together with the direct identification of the Weyl multiplet, $T^{\pm(\mathsf t)}_{\mu\nu}=T^{\pm}_{\mu\nu}$, we find
\begin{equation}
    G_{\mu\nu,0}^{\pm}= G_{\mu\nu,0}^{\pm(\mathsf{t})}, \qquad G_{\mu\nu,A}^{\pm} = \i G_{\mu\nu,A}^{\pm(\mathsf{t})},
\end{equation}
which gives us $F^{I(\mathsf t)}\wedge G_I^{(\mathsf t)} = -\i F^I\wedge G_I$. 

We also need to be careful about the overall sign of the action. The Lagrangian $\CL^{(\mathsf t)}$ used in
\cite{Adhikari:2025eok,deWit:2017cle,Gutowski:2012yb} enters the Euclidean path integral with a positive sign, $Z \sim \exp [+\int \dd^4x\,\CL^{(\mathsf t)}]$. To see this, consider the five-dimensional Wick rotation of the Lorentzian path integral,
\begin{equation}
    \i S_{\rm 5d}^{\rm L}
    =
    \frac{\i}{\kappa_5^2}
    \int \dd t_L\,\dd^4x\,\CL_{\rm 5d}^{\rm L}= 
    \frac{1}{\kappa_5^2}
    \int \dd t_E\,\dd^4x\,\CL^{(\mathsf t)}
    =\frac{1}{\kappa_4^2}
    \int \dd^4x\,\CL^{(\mathsf t)},
\end{equation}
where $t_L=-\i t_E$ and $\kappa_5^2=\kappa_4^2\int\dd t_E$. We use units where the four-dimensional Newton's constant is equal to one, $\kappa_4^2 = 8\pi.$ In the main text, we instead use the standard convention $Z\sim e^{-I}$. Accordingly,
\begin{equation}
    I = -\frac{1}{\kappa_4^2} \int \dd^4x\,\CL^{(\mathsf t)} .
\end{equation}
The same convention for the Euclidean path-integral weight is used in \cite{BenettiGenolini:2026qdm}.

Finally, in \cite{Adhikari:2025eok,deWit:2017cle,Gutowski:2012yb} the Riemann tensor is defined such that the Ricci scalar $R^{(\mathsf t)}$ is negative on a round sphere, the Poincar\'e supergravity spin connection is given by the Cartan equations $\dd e^a - \omega^{a}{}_b \wedge e^b = 0$, and the curvature tensor is $R^a{}_b = \dd \omega^{a}{}_b - \omega^{a}{}_c\wedge \omega^c{}_b$. We use the opposite sign convention for the spin connection, $\dd e^a+\omega^a{}_b\wedge e^b=0$, together with $R^a{}_b=\dd\omega^a{}_b+\omega^a{}_c\wedge\omega^c{}_b$. The curvature two-form consequently changes sign, and the two Ricci scalars are related by $R=-R^{(\mathsf t)}$.

\section{Deriving Poincar\'e supergravity from conformal supergravity} 
\label{app:poincare-reduction}

In this appendix, we explicitly derive two-derivative Poincar\'e $\mathcal N=2$ supergravity from the conformal-supergravity formulation used in the main text. We specialize the prepotential to the two-derivative theory, fix the additional superconformal gauge symmetries, and eliminate the auxiliary fields. The resulting action contains the Einstein--Hilbert term, the electromagnetic kinetic terms, and a nonlinear sigma model with special-K\"ahler target space. Throughout the calculation we retain the boundary term \eqref{eq:new-scalar-bdry-term} imposing Dirichlet boundary conditions for vector multiplet scalars.

\paragraph{Two-derivative conformal supergravity: }
The prepotentials depend only on the vector-multiplet scalars and are homogeneous of degree two,
\begin{equation}
    \CF(\lambda X)=\lambda^2\CF(X),\qquad
    \bar\CF(\lambda\bar X)=\lambda^2\bar\CF(\bar X).
    \label{eq:poincare-homogeneity}
\end{equation}
Barred and unbarred fields are treated as independent in the complexified Euclidean theory. The bosonic action is
\begin{equation}
    I^{(2)}=I_{\mathbb{VW}}^{(2)}+I_{\mathbb H} +I_{\partial M}^{X}. \label{eq:poincare-action-decomposition}
\end{equation}
Throughout this appendix we omit the gravitational boundary terms. Translating Eq.~(6.3) of~\cite{deWit:2017cle} with the dictionary in App.~\ref{app:conventions}, we obtain:
\begin{equation}
    I_{\mathbb{VW}}^{(2)}
    =-\frac{\i}{8\pi}\int_M\dd^4x\sqrt g\,
        (\CL-\bar\CL),
    \label{eq:poincare-vector-action}
\end{equation}
where
\begin{align}
    \CL={}&\CF_I\Box_c\bar X^I
       +\frac18\CF_I
          \left(F_{ab}^{I-}-\frac14X^IT_{ab}^{-}\right)T^{ab-}
       +\frac1{32}\CF\,(T_{ab}^{-})^2
       \nonumber\\
       &-\frac14\CF_{IJ}
          \left(F_{ab}^{I+}-\frac14\bar X^IT_{ab}^{+}\right)
          \left(F^{J+ab}-\frac14\bar X^JT^{+ab}\right)
       -\frac18\CF_{IJ}Y_{ij}^{I}Y^{Jij},
       \label{eq:2derivL}\\
    \bar\CL={}&\bar\CF_I\Box_cX^I
       +\frac18\bar\CF_I
          \left(F_{ab}^{I+}-\frac14\bar X^IT_{ab}^{+}\right)T^{ab+}
       +\frac1{32}\bar\CF\,(T_{ab}^{+})^2
       \nonumber\\
       &-\frac14\bar\CF_{IJ}
          \left(F_{ab}^{I-}-\frac14X^IT_{ab}^{-}\right)
          \left(F^{J-ab}-\frac14X^JT^{-ab}\right)
       -\frac18\bar\CF_{IJ}Y_{ij}^{I}Y^{Jij}.
       \label{eq:2derivLbar}
\end{align}
The superconformal d'Alembertians appearing above are
\begin{equation}
    \Box_cX^I=\left(\CD_\mu\CD^\mu-\frac16R-D\right)X^I,
    \qquad
    \Box_c\bar X^I=
       \left(\CD_\mu\CD^\mu-\frac16R-D\right)\bar X^I.
    \label{eq:conformal-laplacian}
\end{equation}
Here the special conformal transformations have been fixed by $b_\mu=0$, and
\begin{equation}
    \CD_\mu X^I=(\nabla_\mu -A_\mu)X^I,\qquad
    \CD_\mu\bar X^I=(\nabla_\mu + A_\mu)\bar X^I.
    \label{eq:scalar-derivatives}
\end{equation}

The action of an ungauged hypermultiplet is given by Eq.~(2.8) of \cite{Bobev:2021oku} with the R-symmetry gauge coupling set to zero, namely,
\begin{align}
    I_{\mathbb H}
    &=-\frac{1}{8\pi}\int_M\dd^4x\sqrt g\,
       \left[\chi\left(-\frac16R+\frac12D\right)
       -\frac12\varepsilon^{ij}\varepsilon_{\alpha\beta}
          (\CD_\mu A_i{}^\alpha)(\CD^\mu A_j{}^\beta)\right],
    \label{eq:2deriv-hyper-action}
\end{align}
where $\chi=\frac12\varepsilon^{ij}\varepsilon_{\alpha\beta} A_i{}^\alpha A_j{}^\beta.$ The surface contribution relating this first-derivative representative to the hypermultiplet Laplacian action vanishes in the covariantly constant compensator gauge used below. The vector-multiplet scalar boundary term is given by Eq.~\eqref{eq:new-scalar-bdry-term}.

To obtain the Poincar\'e supergravity action, we now eliminate the auxiliary fields and fix the additional superconformal gauge symmetries. We begin by integrating out the auxiliary vector-multiplet triplets $Y^{I,ij}$. We then fix the compensating hypermultiplet to $A_i{}^\alpha=\sqrt{-\chi}\,\delta_i{}^\alpha$ and integrate out the $\SU(2)_{\rm R}$ connection $\mathcal V_\mu{}^i{}_j$. Local dilatations are fixed by setting $\chi=-2$, while special conformal transformations have already been gauge-fixed by $b_\mu=0$; the corresponding $S$-supersymmetry gauge condition is irrelevant for the bosonic truncation considered here. We subsequently integrate out the auxiliary two-form $T_{ab}$, which produces the standard vector-multiplet period matrix and electromagnetic kinetic terms. The equation of motion of the auxiliary scalar $D$ then fixes the K\"ahler potential to $e^{-\CK}=1$. Finally, we integrate out the $\SO(1,1)_{\rm R}$ connection $A_\mu$. In the two-derivative theory, the remaining $\SO(1,1)_{\rm R}$ gauge freedom may be used to impose $X^0=\bar X^0$, leaving the projective coordinates $t^A=X^A/X^0$ and $\bar t^{\bar A}=\bar X^{\bar A}/\bar X^0$ as the physical vector-multiplet scalars.

\paragraph{The compensator and the Einstein--Hilbert term: }
The auxiliary triplets $Y_{ij}^I$ appear only quadratically in the Lagrangians \eqref{eq:2derivL}, \eqref{eq:2derivLbar}. Their equations of motion therefore set
\begin{equation}
    Y_{ij}^I=0.
    \label{eq:poincare-Y}
\end{equation}
We next integrate out the $\SU(2)_{\rm R}$ gauge field
$\mathcal V_\mu{}^i{}_j$. It appears through the covariant derivatives in the kinetic term for the compensating hypermultiplet scalars,
\begin{equation}
    \CD_\mu A_i{}^\alpha
    =
    \partial_\mu A_i{}^\alpha
    -
    \mathcal V_\mu{}^j{}_i A_j{}^\alpha.
\end{equation}
We gauge-fix the $\SU(2)_{\rm R}$ symmetry by requiring that $A_i{}^\alpha=\sqrt{-\chi}\,\delta_i{}^\alpha$. Using local dilatation symmetry, we can set $\chi = -2.$ The equation of motion of the $\SU(2)_{\rm R}$ gauge field $\mathcal V_\mu{}^i{}_j$ then implies
\begin{equation}
    \mathcal V_\mu{}^i{}_j=0,
\end{equation}
and hence $\CD_\mu A_i{}^\alpha=0$. The hypermultiplet scalar equations of motion consequently reduce to
\begin{equation}
    3D=R.\label{eq:D=R/3}
\end{equation}
With this relation, the compensating-hypermultiplet action vanishes on shell, $I_{\mathbb H}|_{\rm on\,shell}=0.$ In the two-derivative theory, the prepotentials are independent of the Weyl multiplet field $A|_{\mathbb W^2}$, and thus the equation of motion for the $D$-field, \eqref{Deom}, reduces to:
\begin{equation}
    e^{-\CK} = -\frac{\chi}{2} = 1.
\end{equation}

Substituting \eqref{eq:D=R/3} in \eqref{eq:conformal-laplacian}, we can rewrite the kinetic term for the scalars in the action $I^{(2)}_{\mathbb{VW}}$ as:
\begin{equation}
    I^{(2)}_{\mathbb{VW}}\supset -\frac{\i}{8\pi}\int \dd^4x\sqrt{g}\,\left[\CF_{I}(\CD_\mu \CD^\mu - \tfrac{1}{2}R)\bar{X}^I - \bar{\CF}_I(\CD_\mu \CD^\mu - \tfrac{1}{2}R)X^I\right].
\end{equation}
Using $-\i(\CF_I \bar{X}^I - \bar{\CF}_I {X}^I) = e^{-\CK} = 1,$ we recover the standard Einstein--Hilbert term,
\begin{equation}
    I^{(2)}_{\mathbb{VW}}\supset  \frac{1}{8\pi}\int \dd^4x\sqrt{g}\,\left[-\i(\CF_{I}\CD_\mu \CD^\mu \bar{X}^I - \bar{\CF}_I \CD_\mu \CD^\mu X^I) - \tfrac{1}{2}R\right].
\end{equation}
Adding to this expression the scalar boundary term $I_{\partial M}^X$, we get:
\begin{equation}
    I^{(2)}_{\mathbb{VW}} + I_{\partial M}^X \supset -\frac{1}{8\pi}\int \dd^4x\sqrt{g}\,\left[N_{IJ}(\CD_\mu X^I)(\CD^\mu \bar{X}^J) + \tfrac{1}{2}R\right],\label{eq:2deriv-scalar&EH}
\end{equation}
where
\begin{equation}
    N_{IJ} = -\i (\CF_{IJ} - \bar{\CF}_{IJ}).
\end{equation}

\paragraph{Electromagnetic kinetic terms: }We now integrate out the auxiliary two-form $T_{ab}$. Using the homogeneity relations $\CF_I=\CF_{IJ}X^J$ and $\bar\CF_I=\bar\CF_{IJ}\bar X^J$, the terms involving the SD parts of $F^I_{ab}$ and $T_{ab}$ can be organized as
\begin{align}
    -\frac{\i}{64}(N_{MN}\bar X^M \bar X^N)\,(T_{ab}^{+})^2 +\frac{\i}{8}(N_{IJ}\bar X^J) F^{I+ab}T_{ab}^{+} -\frac14\CF_{IJ}F_{ab}^{I+}F^{J+ab}\subset \CL-\bar\CL. \label{eq:TF-lag-SD}
\end{align}
Similarly, the terms involving the ASD fields $F_{ab}^-$ and $T_{ab}^-$ are
\begin{equation}
    -\frac{\i}{64}(N_{MN}X^M X^N)\,(T_{ab}^{-})^2
       +\frac{\i}{8}(N_{IJ}X^J) F^{I-ab}T_{ab}^{-}
       +\frac14\bar\CF_{IJ}F_{ab}^{I-}F^{J-ab} \subset \CL - \bar{\CL}.\label{eq:TF-lag-ASD}
\end{equation}
Integrating out $T_{ab}^{\pm}$ gives
\begin{equation}
    T_{ab}^{+}=\frac{4}{N_{MN}\bar X^M \bar X^N} N_{IJ}\bar X^I F_{ab}^{J+},\qquad
    T_{ab}^{-}=\frac4{N_{MN}X^M X^N}N_{IJ}X^I F_{ab}^{J-}.
    \label{eq:T-solutions}
\end{equation}

We define
\begin{equation}
    \CN_{IJ}=-\bar\CF_{IJ} - \i\frac{(N_{IK}X^K)(N_{JL}X^L)}{N_{MN}X^M X^N},\qquad \bar\CN_{IJ}=-\CF_{IJ} + \i\frac{(N_{IK}\bar X^K)(N_{JL}\bar X^L)}{N_{MN}\bar X^M \bar X^N}. \label{eq:period-matrices}
\end{equation}
Using \eqref{eq:T-solutions} and \eqref{eq:period-matrices} in \eqref{eq:TF-lag-SD}, we obtain:
\begin{equation}
    \frac{1}{4}\bar\CN_{IJ} F_{ab}^{I+}F^{J+ab}\subset \CL - \bar{\CL}.
\end{equation}
Similarly, the Lagrangian for the ASD sector \eqref{eq:TF-lag-ASD} becomes
\begin{equation}
    -\frac14\CN_{IJ}F_{ab}^{I-}F^{J-ab}\subset \CL - \bar{\CL}.
\end{equation}
Moreover, the dual field strength tensors, defined in \eqref{eq:Gminus} and \eqref{eq:Gplus}, reduce to
\begin{equation}
    G_{I\,ab}^+ = -\bar \CN_{IJ} F^{J+}_{ab},\qquad G_{I\,ab}^- = -{\CN}_{IJ} F^{J-}_{ab}.
\end{equation}
Equivalently,
\begin{equation}
    G_I = G_I^+ + G_I^- = \i\,{\rm Im}(\CN_{IJ})\star F^J - {\rm Re}(\CN_{IJ})F^J,
\end{equation}
where ${\rm Re}(\CN_{IJ}) = \frac{1}{2}(\CN_{IJ} + \bar{\CN}_{IJ})$ and ${\rm Im}(\CN_{IJ}) = \frac{1}{2\i}(\CN_{IJ}-\bar{\CN}_{IJ}).$ Since $\star F^{I\pm}=\pm F^{I\pm}$, the orthogonality of SD and ASD two-forms, $A^+_{ab}B^{-ab}=0$ for any SD $A^+_{ab}$ and ASD $B^-_{ab}$, implies
\begin{equation}
    \frac{1}{2}F^I\wedge G_I =\frac14\left(F_{ab}^{I+}G_I^{+ab} -F_{ab}^{I-}G_I^{-ab}\right) \sqrt g\,\dd^4x = -\frac{1}{4}(\bar \CN_{IJ}F_{ab}^{I+}F^{J+ab} - {\CN}_{IJ}F_{ab}^{I-}F^{J-ab})\sqrt g\,\dd^4x.
\end{equation}
The action for vector multiplet gauge fields is therefore
\begin{align}
    I_{\rm EM}^{(2)} =\frac{\i}{16\pi}\int_M F^I\wedge G_I. \label{eq:2deriv-EM-action}
\end{align}

\paragraph{Scalar kinetic terms: }
Combining the results \eqref{eq:2deriv-scalar&EH} and \eqref{eq:2deriv-EM-action}, we find:
\begin{align}
    I^{(2)} &=\frac1{8\pi}\int_M\dd^4x\sqrt g\, \left[-\frac12R-N_{IJ}\CD_\mu X^I\CD^\mu\bar X^J\right] +\frac{\i}{16\pi}\int_M F^I\wedge G_I. \label{eq:poincare-before-A-elimination}
\end{align}
We now focus on the kinetic term for the vector-multiplet scalars,
\begin{equation}
    I_X = -\frac{1}{8\pi}\int \dd^4 x\sqrt{g}\,N_{IJ}(\CD_\mu X^I)(\CD^\mu \bar X^J).
\end{equation}
Importantly, not all scalar fields $X^I$ and $\bar X^I$ are dynamical. Fixing the $\SO(1,1)_{\rm R}$ symmetry allows us to set $X^0=\bar X^0$. Moreover, the condition $e^{-\CK}=1$ constrains the field $x\equiv X^0=\bar X^0$. We can therefore express $I_X$ entirely in terms of the special coordinates
\begin{equation}
    t^A = \frac{X^A}{X^0},\qquad \bar{t}^{\bar A} = \frac{\bar X^A}{\bar{X}^0}.
\end{equation}
To show this explicitly, let us define:
\begin{equation}
    Z^I=(1,t^A), \qquad \bar Z^I=(1,\bar t^{\bar A}).
\end{equation}
Then, in the gauge where $X^0 = \bar{X}^0$ we have
\begin{equation}
    X^I = x Z^I(t),\qquad \bar{X}^I = x\bar{Z}^I(\bar{t}).
\end{equation}
Writing
\begin{equation}
    \CF(X)=x^2 f(t),\qquad
    \bar\CF(\bar X)=x^2\bar f(\bar t),
    \label{eq:poincare-projective-prepotential}
\end{equation}
the derivatives of the prepotential are
\begin{align}
    \CF_A&=x f_A,&
    \CF_0&=x(2f-t^Af_A),\nonumber\\
    \CF_{AB}&=f_{AB},&
    \CF_{0A}&=f_A-t^Bf_{AB},\nonumber\\
    \CF_{00}&=2f-2t^Af_A+t^At^Bf_{AB}.
    \label{eq:projective-derivatives}
\end{align}
Here $\CF_A = \frac{\partial}{\partial X^A}\CF$ and
$f_A = \frac{\partial}{\partial t^A}f$ should not be confused with derivatives of the prepotential with respect to the Weyl-squared multiplet in the higher-derivative theory. The corresponding formulae for the anti-holomorphic prepotential are obtained by replacing each quantity by its barred counterpart. Since $N_{IJ}$ depends only on $t$ and $\bar t$, we can define the K\"ahler potential $k(t,\bar{t})$ as
\begin{align}
    h(t,\bar t) \equiv \i\left[Z^I\bar\CF_I(\bar Z) -\bar Z^I\CF_I(Z)\right] =N_{IJ}Z^I\bar Z^J,\qquad k(t,\bar t)=-\log h(t,\bar t).
    \label{eq:projective-Kahler-potential}
\end{align}
Since the K\"ahler potential is constant,
\begin{equation}
    1 = e^{-\CK(X,\bar{X})} = x^2 h(t,\bar t),
\end{equation}
the field $x$ depends on the special coordinates $t^A,\bar{t}^{\bar{A}}.$ 

To determine an explicit form of the action for $t^A,\bar{t}^{\bar A}$, it is convenient to introduce
\begin{equation}
    E_A^I= \frac{\partial Z^I}{\partial t^A},\qquad \bar E_{\bar B}^I=\frac{\partial \bar Z^I}{\partial\bar t^{\bar B}}.
\end{equation}
Homogeneity gives $\CF_{IJK}X^K=0$ and $\bar{\CF}_{IJK}\bar{X}^K = 0$. Consequently, differentiating $e^{-\CK} = N_{IJ}X^I \bar{X}^J$ gives
\begin{equation}
    X^0\bar{X}^0\partial_Ah =\frac{\partial X^I}{\partial t^A}\frac{\partial}{\partial X^I} e^{-\mathcal K} =X^0E_A^IN_{IJ}\bar{X}^J =X^0\bar{X}^0E_A^IN_{IJ}\bar{Z}^J,
\end{equation}
and consequently,
\begin{align}
    h_A = E_A^I N_{IJ}\bar{Z}^J = -hk_A,\qquad 
    h_{\bar{A}} = Z^I N_{IJ}\bar{E}^J_{\bar A} = -h k_{\bar{A}}.\label{eq:hA-eqs}
\end{align}
Differentiating $e^{-\CK}$ twice, we obtain:
\begin{equation}
    X^0\bar{X}^0 h_{A\bar B} =\frac{\partial X^I}{\partial t^A}\frac{\partial \bar{X}^J}{\partial \bar t^{\bar B}} \frac{\partial^2}{\partial X^I\partial \bar X^J}\left( e^{-\mathcal K}\right) =X^0\bar X^0 E_A^I \bar E_{\bar B}^J N_{IJ},
\end{equation}
and therefore,
\begin{equation}
    h_{A\bar B} = E_A^IN_{IJ}\bar E_{\bar B}^J.\label{eq:hAB-eqs}
\end{equation}
The special-K\"ahler metric is:
\begin{equation}
    k_{A\bar B} =\partial_A\partial_{\bar B}k =-\frac{h_{A\bar B}}h+\frac{h_Ah_{\bar B}}{h^2} = -\frac{1}{h}N_{IJ}P_A^I \bar P_{\bar B}^J, \label{eq:poincare-projective-metric}
\end{equation}
where
\begin{equation}
    P_A^I=E_A^I+k_AZ^I,\qquad \bar P_{\bar B}^J=\bar E_{\bar B}^J+k_{\bar B}\bar Z^J,
    \label{eq:poincare-horizontal-tangents}
\end{equation}
are the horizontal (K\"ahler-covariant) tangent vectors along the directions $t^A$, $\bar{t}^{\bar A}$. They satisfy 
\begin{equation}
    N_{IJ}P_A^I\bar Z^J=0,\qquad N_{IJ}Z^I\bar P_{\bar B}^J=0.
\end{equation}
The above relations also determine all components of $N_{IJ}$:
\begin{align}
    N_{AB}
      &=-h\left(k_{A\bar B}-k_Ak_{\bar B}\right),
         \nonumber\\
    N_{A0}
      &=h\left[-k_A+
          \left(k_{A\bar B}-k_Ak_{\bar B}\right)\bar t^{\bar B}\right],
         \nonumber\\
    N_{0B}
      &=h\left[-k_{\bar B}+
          t^A\left(k_{A\bar B}-k_Ak_{\bar B}\right)\right],
         \nonumber\\
    N_{00}
      &=h\left[1+k_At^A+k_{\bar B}\bar t^{\bar B}
         -\left(k_{A\bar B}-k_Ak_{\bar B}\right) t^A\bar t^{\bar B}\right].
    \label{eq:N-components}
\end{align}

We are now ready to eliminate the $\SO(1,1)_{\rm R}$ gauge field and evaluate the covariant derivatives $\CD_\mu X^I$ and $\CD_\mu\bar X^I$. Define
\begin{equation}
    \alpha_\mu=\partial_\mu\log x-A_\mu,\qquad
    \bar\alpha_\mu=\partial_\mu\log x+A_\mu.
    \label{eq:2deriv-alpha}
\end{equation}
Then
\begin{align}
    \CD_\mu X^I =x\left(Z^I\alpha_\mu+E_A^I\partial_\mu t^A\right),\qquad  \CD_\mu\bar X^I =x\left(\bar Z^I\bar\alpha_\mu +\bar E_{\bar B}^I \partial_\mu\bar t^{\bar B}\right).
    \label{eq:projective-covariant-derivatives}
\end{align}
Using \eqref{eq:hA-eqs} and \eqref{eq:hAB-eqs}, their contraction becomes
\begin{align}
    N_{IJ}\CD_\mu X^I\CD^\mu\bar X^J
      =e^{-\CK}\Bigl[\left(\alpha_\mu-k_A\partial_\mu t^A\right) \left(\bar\alpha^\mu -k_{\bar B}\partial^\mu\bar t^{\bar B}\right) -k_{A\bar B}\partial_\mu t^A \partial^\mu\bar t^{\bar B}\Bigr].
    \label{eq:poincare-scalar-factorization}
\end{align}
Differentiating $e^{-\CK}=x^2h=1$ gives
\begin{equation}
    \left(\alpha_\mu-k_A\partial_\mu t^A\right) + \left(\bar\alpha_\mu - k_{\bar B}\partial_\mu\bar t^{\bar B}\right) = - \partial_\mu \CK=0.
    \label{eq:poincare-alpha-sum}
\end{equation}
On the other hand, the variation of \eqref{eq:poincare-scalar-factorization} with respect to $A_\mu$ yields
\begin{equation}
    \left(\alpha_\mu-k_A\partial_\mu t^A\right) -\left(\bar\alpha_\mu - k_{\bar B}\partial_\mu\bar t^{\bar B}\right) =0.
    \label{eq:poincare-alpha-difference}
\end{equation}
Together, these two relations imply
\begin{align}
    \alpha_\mu&=k_A\partial_\mu t^A,\qquad
    \bar\alpha_\mu=k_{\bar B}\partial_\mu\bar t^{\bar B},
    \nonumber
    \\
    \CD_\mu X^I&=xP_A^I\partial_\mu t^A,\qquad
    \CD_\mu\bar X^I=x\bar P_{\bar B}^I \partial_\mu\bar t^{\bar B}.
    \label{eq:poincare-horizontal-derivatives}
\end{align}
It then follows directly from \eqref{eq:poincare-scalar-factorization} that
\begin{align}
    -N_{IJ}\CD_\mu X^I\CD^\mu\bar X^J = k_{A\bar B}\partial_\mu t^A \partial^\mu\bar t^{\bar B}.
    \label{eq:special-sigma-model}
\end{align}

\paragraph{Final result: }
Substituting \eqref{eq:special-sigma-model} into \eqref{eq:poincare-before-A-elimination}, we obtain
\begin{align}
    I^{(2)} &=\frac1{8\pi}\int_M\dd^4x\sqrt g\, \left[-\frac12R + k_{A\bar B}\partial_\mu t^A \partial^\mu\bar t^{\bar B}\right] + \frac{\i}{16\pi}\int_M F^I\wedge G_I.
    \label{eq:poincare-final-action}
\end{align}
This agrees with Eq.~(3.25) of \cite{Boruch:2025biv}.

\section{Defining electromagnetic boundary terms}
\label{app:KerrNewman_puzzle}

In this appendix, we provide a direct gravitational derivation of the electromagnetic boundary term needed to pass from the mixed ensemble to the canonical ensemble. The main subtlety is that, in the presence of magnetic charge, the gauge connection cannot be represented by a single globally defined one-form on the asymptotic boundary $\partial M$, so the naive boundary term is patch dependent.\footnote{Related subtleties in the definition of conserved charges were discussed in \cite{Hanaki:2007mb}. The analysis in this appendix may also be relevant to asymptotically AdS$_4$ solutions and their holographic applications.} In the first subsection we formulate a patch-independent prescription in $\CN=2$ supergravity with an arbitrary number of vector multiplets. In the second subsection, we illustrate why this prescription is necessary by evaluating the naive boundary term for the dyonic Kerr--Newman solution.

\subsection{Electromagnetic boundary terms in $\CN = 2$ SUGRA}
To determine an appropriate boundary term, let us first explicitly verify that \eqref{eq:onshell-action-decomposition} imposes Dirichlet boundary conditions for the gauge fields. Varying the action $I$ over $W^I$, and using \cref{eq:Gplus,eq:Gminus}, we obtain:
\begin{align}
    \delta I_{\rm mix} = \delta I_{\mathbb{VW}} &= \frac{\i}{16\pi}\int \dd^4 x\sqrt{g}\,\left(G_I^{+\mu\nu}\delta F^{I+}_{\mu\nu} - G^{-\mu\nu}_{I}\delta F_{\mu\nu}^{I-}\right)\nonumber 
    \\
    &= \frac{\i}{16\pi}\int \dd^4 x\sqrt{g}\,G_I^{\mu\nu}\left(\delta F^{I+}_{\mu\nu} - \delta F_{\mu\nu}^{I-}\right).
\end{align}
In the second line, we used the orthogonality of SD and ASD two-forms. Since $\frac{1}{2}\,\dd^4 x\sqrt{g}\,A^{\mu\nu} B_{\mu\nu} = A\wedge \star B$ and $\star \delta F^{I\pm} = \pm \delta F^{I\pm}$, we have:
\begin{equation}
    \delta I_{\rm mix} = \frac{\i}{8\pi}\int_M G_I \wedge \delta F^I.
\end{equation}
Assuming that $\delta W^I$ is a globally defined 1-form variation, we can write $\delta F^I = \dd \delta W^I,$ and apply Stokes' theorem to recast $\delta I_{\rm mix}$ as:
\begin{equation}
    \delta I_{\rm mix} = -\frac{\i}{8\pi}\int_M \dd G_I \wedge \delta W^I + \frac{\i}{8\pi}\int_{\partial M}G_I\wedge \delta W^I.
\end{equation}
Thus, $I_{\rm mix}$ defines the Dirichlet variational problem. To impose $\delta G_I|_{\partial M} = 0$, one would naively add $I_N^{\rm naive} = -\frac{\i}{8\pi}\int_{\partial M}G_I \wedge W^I$ \cite{Braden:1990hw,Hawking:1995ap}. However, in the presence of magnetic charges, the gauge field at infinity cannot be specified globally in a single coordinate patch, and thus, $I_{\rm N}^{\rm naive}$ is not well defined and depends on the choice of patches covering the asymptotic sphere $S^2_{\infty}$. As shown explicitly below for the dyonic Kerr--Newman saddle, this ambiguity is removed by subtracting from $W^I$ appearing in $I_{\rm N}^{\rm naive}$ a reference gauge field $W^I_{\rm ref}$, so that $W^I - W_{\rm ref}^I$ is a regular 1-form on the entire $\partial M$,
\begin{equation}
    I_{\rm N} = -\frac{\i}{8\pi}\int_{\partial M}(W^I - W^I_{\rm ref})\wedge G_I.
\end{equation}
We can take $W^I_{\rm ref}$ that depends only on the magnetic charges $\Gamma^I$. For variations preserving the magnetic charges, $\delta \Gamma^I = 0$, it follows that $\delta W_{\rm ref}^I = 0$, and 
\begin{equation}
    \delta I_{\rm mix} + \delta I_{\rm N} = -\frac{\i}{8\pi }\int_{M}\dd G_I\wedge \delta W^I - \frac{\i}{8\pi }\int_{\partial M}(W^I - W^I_{\rm ref})\wedge \delta G_I.
\end{equation}
This clearly imposes Neumann boundary conditions for gauge fields, $\delta G_I|_{\partial M} = 0.$

In the case of the Bates--Denef solution with total magnetic charge $\Gamma^I$, the asymptotic form of $W^I$ is given by \eqref{eq:Wasymptotic}, and we can choose $W^I_{\rm ref} = \Gamma^I(\cos\theta \mp 1)\,\dd\varphi$. Then, at large $r$ we have $W^I - W^I_{\rm ref} = 2\i \mu^I\dd\tau + \O(r^{-1})$. Consequently,
\begin{equation}
    I_{\rm N} = \frac{1}{4\pi}\int_{\partial M}\mu^I \dd\tau \wedge G_I = +\beta \mu^I \Gamma_I.\label{eq:inn}
\end{equation}

\subsection{Subtleties about electromagnetic boundary terms for dyonic Kerr--Newman black holes} 
We now illustrate why the subtraction by a reference connection is necessary. To this end, we consider ordinary Einstein--Maxwell theory and the dyonic Kerr--Newman black hole,
\begin{align}
    &\dd s^2 = \frac{\Delta}{\Sigma}(\dd \tau - \i a \sin^2(\theta) \,\dd\phi)^2 + \frac{\Sigma}{\Delta}\dd r^2 + \Sigma\,\dd\theta^2 + \frac{\sin^2\theta}{\Sigma}((r^2 + a^2)\dd\phi + \i a\dd \tau)^2,\label{eq:kerr_newman_metric}
    \\
    &A = -\frac{Qr}{\Sigma}(\i\dd \tau + a \sin^2(\theta)\dd\phi) - \frac{P \cos(\theta)}{\Sigma}(\i a\,\dd\tau + (r^2 + a^2)\dd\phi) + \begin{pmatrix}
        \text{gauge}
    \end{pmatrix}\label{eq:A0_kerr_newman}
\end{align}
where $\Sigma = r^2 + a^2 \cos^2(\theta)$, $\Delta = r^2 + a^2 - 2M r + 4\pi (Q^2 + P^2)$. $Q$ and $P$ are the electric and magnetic charges of the black hole, $M$ is its mass and $J = Ma$ is its angular momentum. We denote the outer horizon radius by $r_H$. The corresponding corotating coordinates are
\begin{equation}
    \tau'=\tau,
    \qquad
    \varphi'=\varphi+\i\Omega_H\tau,
\end{equation}
where $\Omega_H = \frac{a}{r_H^2 + a^2}$ is the angular velocity of the black hole. To remove the conical singularity at the horizon, we identify $\tau'\sim\tau'+\beta$ at fixed $\varphi'$. The cycle $(\tau',\varphi')\sim(\tau'+\beta,\varphi')$ is contractible at the horizon, and hence the gauge field $A$ must have trivial holonomy along this cycle there. This condition fixes the asymptotic $\tau'$ component of the gauge field to $A_{\tau'}=\i\Phi_e+\O(r^{-1})$, where $\Phi_e=\frac{Qr_H}{r_H^2+a^2}$.

The standard electromagnetic boundary term \cite{Braden:1990hw,Hawking:1995ap} that would naively impose Neumann boundary conditions is
\begin{equation}
    I_{\rm N}^{\rm naive}
    =
    -\int\dd^3 x\sqrt{h}\,n^\mu A^\nu F_{\mu\nu}
    =
    -\int_{\partial M}A\wedge \star F.
    \label{eq:naive_inn}
\end{equation}

At asymptotic infinity, the spacetime boundary is an $S^1_\beta$-fibration over $S^2$. We cover the $S^2$ by northern and southern patches, $U_N$ and $U_S$, containing the north pole $\theta=0$ and south pole $\theta=\pi$, respectively. The asymptotic gauge field in the two patches is
\begin{align}
    A_{{\rm N}/{\rm S}}
    &=
    \i(\Phi_e+\Omega_H P\cos\theta)\dd\tau'
    -P(\cos\theta\mp1)\,\dd\varphi'
    +\O(r^{-1})
    \nonumber\\
    &=
    \i(\Phi_e\pm\Omega_H P)\dd\tau
    -P(\cos\theta\mp1)\dd\varphi
    +\O(r^{-1}).
\end{align}
Here the upper sign refers to the northern patch, and
$A_{\rm N}-A_{\rm S}=2P\,\dd\varphi'$ on the overlap $U_N\cap U_S$. The asymptotic form of the dual field strength is
\begin{equation}
    \star F = -\i Q\sin\theta\,\dd\theta\wedge\dd\varphi.
\end{equation}

Let $E_N\subset U_N$ and $E_S\subset U_S$ be non-overlapping regions such that $E_N\cup E_S=S^2$. Since $E_N$ and $E_S$ have disk topology, the restriction of the $S^1$-fibration to either region is trivial, and we can write the corresponding parts of the boundary as $S^1\times E_N$ and $S^1\times E_S$. We can then interpret \eqref{eq:naive_inn} as
\begin{align}
    I_{\rm N}^{\rm naive}
    =
    -\int_{S^1\times E_N}A_{\rm N}\wedge\star F
    -\int_{S^1\times E_S}A_{\rm S}\wedge\star F.
\end{align}
With orientation
${\rm vol}_{\partial M}=-\dd\tau\wedge\sin\theta\,\dd\theta\wedge\dd\varphi$,
we obtain
\begin{equation}
    I_{\rm N}^{\rm naive}
    =
    4\pi\beta\Phi_e Q
    +
    \beta\Omega_HPQ
    \bigl(
        {\rm Area}(E_N)-{\rm Area}(E_S)
    \bigr),
    \label{eq:naive_KerrNewman_inn}
\end{equation}
where ${\rm Area}(E_{N/S})$ denote the areas of the regions $E_{N/S}$, with
${\rm Area}(E_N)+{\rm Area}(E_S)=4\pi$. Clearly, for $P\Omega_H\neq0$, Eq.~\eqref{eq:naive_KerrNewman_inn} depends on the choice of $E_N$ and $E_S$. Thus, the definition \eqref{eq:naive_inn} is ambiguous.

To resolve this ambiguity, we introduce a reference gauge field $A_{\rm ref}$ in the boundary term $I_{\rm N}$,
\begin{equation}
    I_{\rm N}
    =
    -\int_{\partial M}(A-A_{\rm ref})\wedge\star F.
\end{equation}
As the reference connection $A_{\rm ref}$, we take the boundary pullback of the gauge field of a purely magnetic Kerr--Newman black hole with the same $\beta$, $\Omega_H$ and $P$. In an ensemble with fixed $P$ and $\Omega_H$, $A_{\rm ref}$ does not vary, while $A-A_{\rm ref}$ is a globally defined one-form,
\begin{equation}
    A-A_{\rm ref}
    =
    \i\Phi_e\dd\tau+\O(r^{-1}),
    \qquad
    \star F
    =
    -\i Q\sin\theta\,\dd\theta\wedge\dd\varphi.
\end{equation}
The resulting boundary term
\begin{equation}
    I_{\rm N}
    =
    4\pi\beta\Phi_e Q,
\end{equation}
is independent of the choice of patches covering $S^2$.

\section{Deriving the generalized attractor mechanism from equivariant localization} 
\label{app:off-shell-attractor}
Equivariant localization is not limited to computing the on-shell action: it can also be used to derive the attractor values of the scalar moduli at the north and south poles of the black hole. In this appendix, we use the equivariant formalism reviewed in Subsec.~\ref{sec:equivariant-closed-forms} to reformulate the analysis of the attractor mechanism and the gauge-field regularity conditions from Subsecs.~\ref{sec:new-attractor} and \ref{sec:shifted_saddles1}. Besides providing an independent derivation of these conditions, this reformulation clarifies which of them hold off shell.\footnote{This extends the off-shell gluing and UV/IR relations in Eqs.~(14) and (15) of \cite{BenettiGenolini:2026qdm}.}

Throughout this appendix, we normalize the Killing vector to $\xi = \partial_\tau$, setting $c_\xi = 1$, see~\eqref{eq:SminSplrel} and the discussion below.

We begin by defining a polyform-valued symplectic vector
\begin{equation}
\label{eq:EM-equivariant-form}
    (\mathcal{X}^I,\mathcal{X}_I) = \mathcal{X} = \mathbb{F} - \i\Phi,
\end{equation}
with $\mathbb{F}$ determined by Eqs.~\eqref{eq:symp-field-strength-m} and \eqref{eq:symp-field-strength-p}, and $\Phi$ given by \eqref{eq:el_pot}. As shown in \cite{Adhikari:2025eok}, \cref{eq:symp-field-strength-m,eq:symp-field-strength-p,eq:el_pot} follow solely from the Killing spinor equations and remain valid even when the equations of motion are not satisfied. It follows that $-\iota_\xi \mathbb{F} = \i \dd\Phi$ off shell, and therefore,
\begin{equation}
    (\dd - \iota_\xi)\mathcal{X} = \dd \mathbb{F} = (\dd F^I,\dd G_I).
\end{equation}

Following \cite{BenettiGenolini:2026qdm}, we aim to derive the regularity conditions by integrating the polyform over a $\xi$-invariant two-surface $\Sigma \subset M$. The pullback of any three-form to $\Sigma$ vanishes, and therefore the pullback of $\mathcal{X}$ to $\Sigma$, $\mathcal{X}|_{\Sigma}$, is an equivariantly closed form,%
\begin{equation}
    (\dd - \iota_\xi)\mathcal{X}|_{\Sigma} = (\dd F^I, \dd G_I)|_{\Sigma} = 0.
\end{equation}
Applying the BVAB formula to the two-dimensional manifold $\Sigma$, we obtain
\begin{equation}
    \int_{\Sigma}\mathbb{F} = -2\pi \i\sum_{\text{fixed points }\alpha}\frac{\Phi_{\alpha}}{b_{\alpha}} + \i\int_{\partial\Sigma}\Phi \,\frac{\xi_\mu \dd x^\mu}{\|\xi\|^2},\label{eq:2d-BVAB}
\end{equation}
where $\Phi_{\alpha}$ is the value of the electric potential $\Phi$ at the fixed point $\alpha$, and $b_{\alpha}$ is the skew-eigenvalue of the pullback of $\frac{1}{2}\dd (\xi_\mu \dd x^\mu)$ to $\Sigma$ at the point $\alpha.$ As is the case for the saddle-point geometries studied in this paper, we assume that $\xi$ has no bolts, i.e., $\xi$ can vanish only at isolated points $\alpha$ of $\Sigma$, and not on all of $\Sigma.$

\paragraph{Compact $\Sigma$: The attractor equations in the canonical ensemble. }Let $\Sigma = \Sigma_a$ be the horizon of the $a$-th black hole. Then $\xi$ has two fixed points, namely the north and south poles of the horizon. The horizon $\Sigma_a$ is a closed surface with no boundary, $\partial \Sigma_a = \emptyset$, so the boundary integral on the right-hand side of \eqref{eq:2d-BVAB} vanishes. We define the horizon charge vector $\Gamma_a^{\rm horizon}$ of the $a$-th black hole by the magnetic and electric fluxes through its horizon, 
\begin{equation}
    4\pi \Gamma_a^{\rm horizon} \equiv \int_{\Sigma_a}\mathbb F =  -2\pi \i \left(\frac{\Phi_{N,a}}{b_{N,a}} + \frac{\Phi_{S,a}}{b_{S,a}}\right).\label{eq:equivariant-attractor}
\end{equation}
Since the pullback of \eqref{eq:EM-equivariant-form} to any $\xi$-invariant surface is equivariantly closed whenever the geometry admits a Killing spinor, even off shell, Eq.~\eqref{eq:equivariant-attractor} also holds off shell. However, we stress that the charges $\Gamma_{a}^{\rm horizon}$ are defined by integrals over the individual horizons, and they can in principle be different from the charges $\Gamma_a$ defined by the fluxes through surfaces that enclose only the $a$-th black hole but lie far from its horizon. The Bianchi identity $\dd F^I = 0$ implies that the magnetic fluxes $\Gamma_a^I$ depend only on the homology class of the integration surface and therefore, 
\begin{equation}
    \Gamma_a^I = \Gamma_a^{I, {\rm horizon}},
    \label{eq:magnetic-mapping-to-infinity}
\end{equation}
holds even off shell. In contrast, $\dd G_I$ does not vanish off shell, so the electric fluxes can change under continuous deformations of the integration surface. Only upon imposing Maxwell's equations do the charges $\Gamma_{a, I}$ become topological. Since the sphere at infinity is homologous to the sum of the horizon surfaces through a source-free region, the asymptotic electric charge can be identified with the sum of the horizon electric charges, 
\begin{equation}
   \Gamma_{a,I} = \Gamma_{a,I}^{\rm horizon}, \qquad \text{ and }\qquad  \Gamma_I= \sum_a \Gamma_{a, I}^{\rm horizon}, 
     \label{eq:electric-mapping-to-infinity}
\end{equation}
hold only on shell.

The potential $\Phi$ is uniquely determined by the function $U$ and the $\SO(1,1)_{\rm R}$-invariant scalars $\cY^I$, $\bar \cY^I$, and the behavior of $U$ is determined solely by the Killing spinor equations and regularity conditions, without using the equations of motion. Consequently, \eqref{eq:equivariant-attractor} uniquely determines the values of the scalars $\cY^I$, $\bar{\cY}^I$ at the north and south poles of the $a$-th black hole in terms of the horizon charge vector $\Gamma_a^{\rm horizon}$. Using the pole values of $\Phi$ from Subsec.~\ref{sec:shifted_saddles1}, together with $b_{N,a} = -b_{S,a} = 2\pi/\beta$,\footnote{The weights at the two fixed points of a rotation of a closed surface necessarily have opposite signs.} these conditions are equivalent to the attractor equations relevant for the canonical ensemble, Eqs.~\eqref{eq:sum_upper_gamma} and \eqref{eq:sum_lower_gamma}, derived in Subsec.~\ref{sec:new-attractor}. Equivariant localization nevertheless teaches us something new: because of \eqref{eq:magnetic-mapping-to-infinity} and \eqref{eq:electric-mapping-to-infinity}, we find that Eq.~\eqref{eq:sum_upper_gamma} holds off shell, whereas Eq.~\eqref{eq:sum_lower_gamma} holds off shell only for the horizon charges $\Gamma_{a,I}^{\rm horizon}$, and for the charges $\Gamma_{a,I}$ only on shell. 

This extends the analysis of the gluing conditions in \cite{BenettiGenolini:2026qdm,Hosseini:2019iad}, which considered only the conditions obtained by integrating $\mathcal{X}^I$ and not $\mathcal{X}_I$.

\paragraph{Non-compact $\Sigma$: The gauge-field regularity condition in the mixed ensemble. } Next, we can derive the gauge-field regularity condition relevant for the mixed ensemble, where the electric chemical potentials are fixed. For this, we consider a non-compact surface $\Sigma_\alpha$ passing through a fixed point $\alpha$ of $\xi$ and extending to asymptotic infinity. More precisely, we assume that $\partial \Sigma_\alpha$ is the temporal circle at asymptotic infinity, $\partial\Sigma_\alpha \simeq S^1_{\beta}.$ Integrating $\mathcal{X}$ over $\Sigma_\alpha$ and applying the BVAB formula then relates the local parameters at the pole $\alpha$ to the asymptotic data for the gauge field $W^I$.\footnote{In the context of AdS/CFT, this gives rise to the so-called UV/IR relations of \cite{BenettiGenolini:2026qdm,Bobev:2020pjk}.}

Let $W_{\alpha}^I$ be a gauge field that is regular over all of $\Sigma_\alpha$. Such a regular $W_{\alpha}^I$ exists only if $\Sigma_\alpha$ contains no nuts other than $\alpha$. Using \eqref{eq:2d-BVAB}, we obtain
\begin{equation}
    \int_{\Sigma_{\alpha}}{F}^I  = -\frac{2\pi\i}{b_{\alpha}}\Phi^I_\alpha + \i\beta\phi^I,
\end{equation}
and therefore the gauge-field holonomy along the temporal circle at infinity is
\begin{equation}
    \exp[-\frac{\i}{2}q_I\oint_{\partial \Sigma_\alpha}W_{\alpha}^I] = \exp[-\frac{\i}{2}q_I\int_{\Sigma_{\alpha}}F^I] = \exp[-\frac{\pi}{b_{\alpha}} \Phi^I_{\alpha} q_I + \frac{\beta}{2}\phi^I q_I].
\end{equation}
This expression must equal the holonomy $\exp(\beta \mu^I q_I)$ fixed by the asymptotic boundary conditions, and therefore,
\begin{equation}
    \beta\mu^I = \frac{\beta}{2}\phi^I - \frac{\pi}{b_{\alpha}}\Phi^I_{\alpha} + 2\pi \i n_{\alpha}^I,\label{eq:equivariant-shifted-saddles}
\end{equation}
where $n_{\alpha}^I$ belongs to the dual lattice $\Lambda^*$. One can show that, with the orientation of $\Sigma_\alpha$ induced by $\partial\Sigma_\alpha \simeq S^1_\beta$, $b_{\alpha} = 2\pi /\beta$ for both the south and north poles (in contrast to the closed horizon, where $b_{N,a} = -b_{S,a}$), so \eqref{eq:equivariant-shifted-saddles} is equivalent to Eq.~\eqref{eq:gauge-field-reg-cond}. We stress again that the condition~\eqref{eq:equivariant-shifted-saddles} follows solely from the Killing spinor equations and is therefore valid off shell.

\section{An ``enigmatic'' two-center saddle} %
\label{sec:enigmas}

The goal of this appendix is to show, through an explicit example, that enigmatic saddles do arise in the OSV ensemble. Specifically, we construct an unshifted two-center saddle that dominates over the unshifted single-center saddle with the same total magnetic charge and electric chemical potential, in a regime where the large-volume approximation is under control. To this end, we focus on a two-derivative theory with one vector multiplet, with a prepotential given by
\begin{equation}
    \CF^{(0)}(X)=\frac{(X^1)^3}{X^0}.
\end{equation}
Consider a parameter $L\in\mathbb Z_{>0}$. We shall look at saddles with a common dipole moment $\delta^I$ and the magnetic charges given by
\begin{align}
 \delta^I&=(-L,0),\\
 \Gamma_1^I&=(0,10L^2),
 \nonumber\\
 \Gamma_2^I&=(6L,10L^2),\\
 \Gamma_{\rm total}^I&=(6L,20L^2).
 \label{eq:common-potential-paper-variables}
\end{align}
The two-black-hole and one-black-hole on-shell actions are determined by
\begin{equation}
 -2\pi \i(\CF - \bar{\CF})|_{\rm 2BH}
 =\underbrace{500\pi L^5}_{\text{BH}_1}+\underbrace{50 \pi L^5}_{\text{BH}_2}= 550\pi L^5,
 \qquad
 -2\pi \i(\mathcal F - \bar{\CF})|_{\rm 1BH}=400\pi L^5.
 \label{eq:common-potential-dominance}
\end{equation}
Consequently,
\begin{equation}\label{eq:AppEineq}
 \log\frac{Z_{\rm 2BH}^{(0)}}{Z_{\rm 1BH}^{(0)}}
 = 150\pi  L^5>0,
\end{equation}
so the unshifted two-black-hole saddle is exponentially dominant at large $L$. The stationary electric charges are different, as they should be in the mixed ensemble.  In the ordering
$(\Gamma^0,\Gamma^1;\Gamma_0,\Gamma_1)$, they are
\begin{align}
 \Gamma_1
 &=
   \left(0,10L^2;250L^4,0\right),
 \nonumber\\
 \Gamma_2
 &=
   \left(6L,10L^2;-20L^4,45L^3\right),
 \nonumber\\
 \Gamma_{\rm 1BH}
 &=
    \left(6L,20L^2;-160L^4,180L^3\right).
 \label{eq:common-potential-full-charges}
\end{align}

For completeness, we also display the corresponding pole residues, given by
\begin{equation}
    \gamma_{N,a}^I = \frac{\Gamma_a^I}{2}+\i\delta^I, \qquad \gamma_{S,a}^I =\frac{\Gamma_a^I}{2} - \i\delta^I, \label{eq:common-potential-upper-residues}
\end{equation}
while the lower components follow from
$\gamma_{N,a,I}=\CF_I(\gamma_{N,a})$ and
$\gamma_{S,a,I}=\bar\CF_I(\gamma_{S,a})$.  Thus, again ordering the full
symplectic vectors as $(\gamma^0,\gamma^1;\gamma_0,\gamma_1)$, the two-center
configuration has
\begin{align}
 \gamma_{S,1}
 &=\left(
   \i L,5L^2;125L^4,-75\i L^3\right),
 \nonumber\\
 \gamma_{N,1}
 &=\left(
   -\i L,5L^2;125L^4,75\i L^3\right),
 \nonumber\\
 \gamma_{S,2}
 &=\left(
   (3+\i)L,5L^2;
   -\frac{5}{2}(4-3\i)L^4,\frac{15}{2}(3-\i)L^3\right),
 \nonumber\\
 \gamma_{N,2}
 &=\left(
   (3-\i)L,5L^2;
   -\frac{5}{2}(4+3\i)L^4,\frac{15}{2}(3+\i)L^3\right).
 \label{eq:common-potential-two-center-residues}
\end{align}
For the competing one-black-hole saddle one instead finds
\begin{align}
 \gamma_{S,{\rm 1BH}}
 &=\left(
   (3+\i)L,10L^2;
   -20(4-3\i)L^4,30(3-\i)L^3\right),
 \nonumber\\
 \gamma_{N,{\rm 1BH}}
 &=\left(
   (3-\i)L,10L^2;
   -20(4+3\i)L^4,30(3+\i)L^3\right).
 \label{eq:common-potential-one-center-residues}
\end{align}
These expressions obey
$\gamma_{N,a}+\gamma_{S,a}=\Gamma_a$ separately for every black hole.
They give the south-pole moduli
\begin{equation}
 t_{S,1}=-5\i L,
 \qquad
 t_{S,2}=\frac{3-\i}{2}L,
 \qquad
 t_{S,{\rm 1BH}}=(3-\i)L,
 \qquad
 t_N=t_S^*,
\end{equation}
so the entire comparison is parametrically controlled at large CY volume. The two-center solution lies in a regular stability chamber \cite{Denef:2000nb,Denef:2007vg}; for example, one may take $B_\infty=0$ and $J_\infty^1=2L$. Moreover, $g_{\rm top}=O(L^{-1})$, whereas the difference among the on-shell actions in \eqref{eq:common-potential-dominance} is $O(L^5)$, so the inequality~\eqref{eq:AppEineq} is insensitive to subleading large-volume corrections.

\bibliographystyle{utphys2}
{\small \bibliography{Biblio}{}}

@article{Bandyopadhyay:2025jbc,
    author = "Bandyopadhyay, Subhodip and Punia, Gurmeet Singh and Srivastava, Yogesh K. and Virmani, Amitabh",
    title = "{The gravitational index of a small black ring}",
    eprint = "2504.09982",
    archivePrefix = "arXiv",
    primaryClass = "hep-th",
    doi = "10.1007/JHEP07(2025)200",
    journal = "JHEP",
    volume = "07",
    pages = "200",
    year = "2025"
}

@article{Adhikari:2024zif,
    author = "Adhikari, Soumya and Dharanipragada, Pavan and Goswami, Kaberi and Virmani, Amitabh",
    title = "{Attractor saddle for 5D black hole index}",
    eprint = "2411.12413",
    archivePrefix = "arXiv",
    primaryClass = "hep-th",
    doi = "10.1007/JHEP03(2025)180",
    journal = "JHEP",
    volume = "03",
    pages = "180",
    year = "2025"
}

@article{Hegde:2024bmb,
    author = "Hegde, Subramanya and Sen, Ashoke and Shanmugapriya, P. and Virmani, Amitabh",
    title = "{Supersymmetric index for half BPS black holes in N=2 supergravity with higher curvature corrections}",
    eprint = "2411.08260",
    archivePrefix = "arXiv",
    primaryClass = "hep-th",
    doi = "10.1007/JHEP02(2025)131",
    journal = "JHEP",
    volume = "02",
    pages = "131",
    year = "2025"
}

@article{Huang:2006hq,
    author = "Huang, Min-xin and Klemm, Albrecht and Quackenbush, Seth",
    title = "{Topological string theory on compact Calabi-Yau: Modularity and boundary conditions}",
    eprint = "hep-th/0612125",
    archivePrefix = "arXiv",
    reportNumber = "MAD-TH-06-12",
    doi = "10.1007/978-3-540-68030-7_3",
    journal = "Lect. Notes Phys.",
    volume = "757",
    pages = "45--102",
    year = "2009"
}

@inproceedings{Hristov:2024cgj,
    author = "Hristov, Kiril",
    title = "{Equivariant localization and gluing rules in 4d N = 2 higher derivative supergravity}",
    eprint = "2406.18648",
    archivePrefix = "arXiv",
    primaryClass = "hep-th",
    doi = "10.1007/978-3-032-16202-1_4",
    month = "6",
    year = "2024"
}

@article{Hristov:2022lcw,
    author = "Hristov, Kiril",
    title = "{ABJM at finite N via 4d supergravity}",
    eprint = "2204.02992",
    archivePrefix = "arXiv",
    primaryClass = "hep-th",
    doi = "10.1007/JHEP10(2022)190",
    journal = "JHEP",
    volume = "10",
    pages = "190",
    year = "2022"
}

@article{Bobev:2022eus,
    author = "Bobev, Nikolay and Hong, Junho and Reys, Valentin",
    title = "{Large N partition functions of the ABJM theory}",
    eprint = "2210.09318",
    archivePrefix = "arXiv",
    primaryClass = "hep-th",
    doi = "10.1007/JHEP02(2023)020",
    journal = "JHEP",
    volume = "02",
    pages = "020",
    year = "2023"
}

@article{Bobev:2025ltz,
    author = "Bobev, Nikolay and De Smet, Pieter-Jan and Hong, Junho and Reys, Valentin and Zhang, Xuao",
    title = "{An Airy tale at large N}",
    eprint = "2502.04606",
    archivePrefix = "arXiv",
    primaryClass = "hep-th",
    doi = "10.1007/JHEP07(2025)123",
    journal = "JHEP",
    volume = "07",
    pages = "123",
    year = "2025"
}

@article{Colombo:2025yqy,
    author = "Colombo, Edoardo and Dimitrov, Vasil and Martelli, Dario and Zaffaroni, Alberto",
    title = "{Patch-wise localization with Chern-Simons forms in five dimensional supergravity}",
    eprint = "2511.13824",
    archivePrefix = "arXiv",
    primaryClass = "hep-th",
    month = "11",
    year = "2025"
}

@article{Colombo:2025ihp,
    author = "Colombo, Edoardo and Dimitrov, Vasil and Martelli, Dario and Zaffaroni, Alberto",
    title = "{Equivariant localization in supergravity in odd dimensions}",
    eprint = "2502.15624",
    archivePrefix = "arXiv",
    primaryClass = "hep-th",
    month = "2",
    year = "2025"
}

@article{Cassani:2026teb,
    author = "Cassani, Davide and Turetta, Enrico",
    title = "{The black hole at the end of the cone: localizing the anomaly polynomial on toric geometries}",
    eprint = "2606.16986",
    archivePrefix = "arXiv",
    primaryClass = "hep-th",
    month = "6",
    year = "2026"
}

@article{Cassani:2025iix,
    author = "Cassani, Davide and Ruip{\'e}rez, Alejandro and Turetta, Enrico",
    title = "{Bubbling saddles of the gravitational index}",
    eprint = "2507.12650",
    archivePrefix = "arXiv",
    primaryClass = "hep-th",
    doi = "10.21468/SciPostPhys.19.5.134",
    journal = "SciPost Phys.",
    volume = "19",
    number = "5",
    pages = "134",
    year = "2025"
}

@article{Cassani:2024kjn,
    author = "Cassani, Davide and Ruip{\'e}rez, Alejandro and Turetta, Enrico",
    title = "{Localization of the 5D supergravity action and Euclidean saddles for the black hole index}",
    eprint = "2409.01332",
    archivePrefix = "arXiv",
    primaryClass = "hep-th",
    doi = "10.1007/JHEP12(2024)086",
    journal = "JHEP",
    volume = "12",
    pages = "086",
    year = "2024"
}

@inproceedings{Mahajan:2026fhc,
    author = "Mahajan, Raghu and Sen, Ashoke",
    title = "{Temperature and Moduli Independence of the Supersymmetric Index from Gravity}",
    eprint = "2610.02306",
    archivePrefix = "arXiv",
    primaryClass = "hep-th",
    month = "10",
    year = "2026"
}

@article{Binder:2021euo,
    author = "Binder, Damon J. and Freedman, Daniel Z. and Pufu, Silviu S. and Zan, Bernardo",
    title = "{The holographic contributions to the sphere free energy}",
    eprint = "2107.12382",
    archivePrefix = "arXiv",
    primaryClass = "hep-th",
    reportNumber = "PUPT-2626, MIT-CTP/5317",
    doi = "10.1007/JHEP01(2022)171",
    journal = "JHEP",
    volume = "01",
    pages = "171",
    year = "2022"
}

@article{Castro:2008ne,
    author = "Castro, Alejandra and Davis, Joshua L. and Kraus, Per and Larsen, Finn",
    title = "{String Theory Effects on Five-Dimensional Black Hole Physics}",
    eprint = "0801.1863",
    archivePrefix = "arXiv",
    primaryClass = "hep-th",
    doi = "10.1142/S0217751X08039724",
    journal = "Int. J. Mod. Phys. A",
    volume = "23",
    pages = "613--691",
    year = "2008"
}

@article{BenettiGenolini:2026kjy,
    author = "Benetti Genolini, Pietro and Gaar, Florian and Gauntlett, Jerome P. and Park, Jaeha and Sparks, James",
    title = "{Localizing the M-Theory Path Integral}",
    eprint = "2609.31829",
    archivePrefix = "arXiv",
    primaryClass = "hep-th",
    month = "9",
    year = "2026"
}

@article{Gaiotto:2006ns,
    author = "Gaiotto, Davide and Strominger, Andrew and Yin, Xi",
    title = "{From AdS(3)/CFT(2) to black holes/topological strings}",
    eprint = "hep-th/0602046",
    archivePrefix = "arXiv",
    doi = "10.1088/1126-6708/2007/09/050",
    journal = "JHEP",
    volume = "09",
    pages = "050",
    year = "2007"
}

@article{Aganagic:2012si,
    author = "Aganagic, Mina and Schaeffer, Kevin",
    title = "{Refined Black Hole Ensembles and Topological Strings}",
    eprint = "1210.1865",
    archivePrefix = "arXiv",
    primaryClass = "hep-th",
    doi = "10.1007/JHEP01(2013)060",
    journal = "JHEP",
    volume = "01",
    pages = "060",
    year = "2013"
}

@article{deWit:2010za,
    author = "de Wit, Bernard and Katmadas, Stefanos and van Zalk, Maaike",
    title = "{New supersymmetric higher-derivative couplings: Full N=2 superspace does not count!}",
    eprint = "1010.2150",
    archivePrefix = "arXiv",
    primaryClass = "hep-th",
    reportNumber = "ITP-UU-10-26, NIKHEF-2010-024",
    doi = "10.1007/JHEP01(2011)007",
    journal = "JHEP",
    volume = "01",
    pages = "007",
    year = "2011"
}

@article{Murthy:2013xpa,
    author = "Murthy, Sameer and Reys, Valentin",
    title = "{Quantum black hole entropy and the holomorphic prepotential of N=2 supergravity}",
    eprint = "1306.3796",
    archivePrefix = "arXiv",
    primaryClass = "hep-th",
    doi = "10.1007/JHEP10(2013)099",
    journal = "JHEP",
    volume = "10",
    pages = "099",
    year = "2013"
}

@article{Shih:2005he,
    author = "Shih, David and Yin, Xi",
    title = "{Exact black hole degeneracies and the topological string}",
    eprint = "hep-th/0508174",
    archivePrefix = "arXiv",
    doi = "10.1088/1126-6708/2006/04/034",
    journal = "JHEP",
    volume = "04",
    pages = "034",
    year = "2006"
}

@article{Chowdhury:2024ngg,
    author = "Chowdhury, Chandramouli and Sen, Ashoke and Shanmugapriya, P. and Virmani, Amitabh",
    title = "{Supersymmetric index for small black holes}",
    eprint = "2401.13730",
    archivePrefix = "arXiv",
    primaryClass = "hep-th",
    doi = "10.1007/JHEP04(2024)136",
    journal = "JHEP",
    volume = "04",
    pages = "136",
    year = "2024"
}

@article{Hosseini:2019iad,
    author = "Hosseini, Seyed Morteza and Hristov, Kiril and Zaffaroni, Alberto",
    title = "{Gluing gravitational blocks for AdS black holes}",
    eprint = "1909.10550",
    archivePrefix = "arXiv",
    primaryClass = "hep-th",
    reportNumber = "IPMU19-0132",
    doi = "10.1007/JHEP12(2019)168",
    journal = "JHEP",
    volume = "12",
    pages = "168",
    year = "2019"
}

@article{Bobev:2020pjk,
    author = "Bobev, Nikolay and Charles, Anthony M. and Min, Vincent S.",
    title = "{Euclidean black saddles and AdS$_{4}$ black holes}",
    eprint = "2006.01148",
    archivePrefix = "arXiv",
    primaryClass = "hep-th",
    doi = "10.1007/JHEP10(2020)073",
    journal = "JHEP",
    volume = "10",
    pages = "073",
    year = "2020"
}

@article{Chen:2024gmc,
    author = "Chen, Yiming and Murthy, Sameer and Turiaci, Gustavo J.",
    title = "{Gravitational index of the heterotic string}",
    eprint = "2402.03297",
    archivePrefix = "arXiv",
    primaryClass = "hep-th",
    doi = "10.1007/JHEP09(2024)041",
    journal = "JHEP",
    volume = "09",
    pages = "041",
    year = "2024"
}

@article{Boruch:2025sie,
    author = "Boruch, Jan and Emparan, Roberto and Iliesiu, Luca V. and Murthy, Sameer",
    title = "{Novel black saddles for 5d gravitational indices and the index enigma}",
    eprint = "2510.23699",
    archivePrefix = "arXiv",
    primaryClass = "hep-th",
    month = "10",
    year = "2025"
}

@article{Boruch:2025qdq,
    author = "Boruch, Jan and Emparan, Roberto and Iliesiu, Luca V. and Murthy, Sameer",
    title = "{The gravitational index of 5d black holes and black strings}",
    eprint = "2501.17909",
    archivePrefix = "arXiv",
    primaryClass = "hep-th",
    doi = "10.1007/JHEP06(2025)145",
    journal = "JHEP",
    volume = "06",
    pages = "145",
    year = "2025"
}

@article{Dijkgraaf:2005bp,
    author = "Dijkgraaf, Robbert and Gopakumar, Rajesh and Ooguri, Hirosi and Vafa, Cumrun",
    title = "{Baby universes in string theory}",
    eprint = "hep-th/0504221",
    archivePrefix = "arXiv",
    reportNumber = "CALT-68-2557, HUTP-05-A019, ITFA-2005-14",
    doi = "10.1103/PhysRevD.73.066002",
    journal = "Phys. Rev. D",
    volume = "73",
    pages = "066002",
    year = "2006"
}

@article{Kolanowski:2026gii,
    author = "Kolanowski, Maciej and Marolf, Donald",
    title = "{How to tame your (black hole) saddles: Lessons from the Lorentzian Gravitational Path Integral}",
    eprint = "2603.24681",
    archivePrefix = "arXiv",
    primaryClass = "hep-th",
    month = "3",
    year = "2026"
}

@article{Teimouri:2016ulk,
    author = "Teimouri, Ali and Talaganis, Spyridon and Edholm, James and Mazumdar, Anupam",
    title = "{Generalised Boundary Terms for Higher Derivative Theories of Gravity}",
    eprint = "1606.01911",
    archivePrefix = "arXiv",
    primaryClass = "gr-qc",
    doi = "10.1007/JHEP08(2016)144",
    journal = "JHEP",
    volume = "08",
    pages = "144",
    year = "2016"
}

@article{Myers:1987yn,
    author = "Myers, Robert C.",
    title = "{Higher Derivative Gravity, Surface Terms and String Theory}",
    reportNumber = "NSF-ITP-87-16",
    doi = "10.1103/PhysRevD.36.392",
    journal = "Phys. Rev. D",
    volume = "36",
    pages = "392",
    year = "1987"
}

@article{Cremonini:2009ih,
    author = "Cremonini, Sera and Liu, James T. and Szepietowski, Phillip",
    title = "{Higher Derivative Corrections to R-charged Black Holes: Boundary Counterterms and the Mass-Charge Relation}",
    eprint = "0910.5159",
    archivePrefix = "arXiv",
    primaryClass = "hep-th",
    reportNumber = "MCTP-09-51, MIFP-09-44, DAMTP-2009-72",
    doi = "10.1007/JHEP03(2010)042",
    journal = "JHEP",
    volume = "03",
    pages = "042",
    year = "2010"
}

@article{Kolanowski:2026ylt,
    author = "Kolanowski, Maciej and Marolf, Donald and Wang, Zi-Yue and Zheng, Wenwen",
    title = "{Superconformal indices and black hole saddles}",
    eprint = "2608.07660",
    archivePrefix = "arXiv",
    primaryClass = "hep-th",
    month = "8",
    year = "2026"
}

@article{Chemissany:2012pf,
    author = "Chemissany, W. and Ferrara, S. and Kallosh, R. and Shahbazi, C. S.",
    title = "{N=2 Supergravity Counterterms, Off and On Shell}",
    eprint = "1208.4801",
    archivePrefix = "arXiv",
    primaryClass = "hep-th",
    reportNumber = "CERN-PH-TH-2012-272",
    doi = "10.1007/JHEP12(2012)089",
    journal = "JHEP",
    volume = "12",
    pages = "089",
    year = "2012"
}

@article{LopesCardoso:2000qm,
    author = "Lopes Cardoso, Gabriel and de Wit, Bernard and Kappeli, Jurg and Mohaupt, Thomas",
    title = "{Stationary BPS solutions in N=2 supergravity with R**2 interactions}",
    eprint = "hep-th/0009234",
    archivePrefix = "arXiv",
    reportNumber = "SPIN-2000-21, SPIN-00-21, ITP-UU-00-24, SU-ITP-00-19",
    doi = "10.1088/1126-6708/2000/12/019",
    journal = "JHEP",
    volume = "12",
    pages = "019",
    year = "2000"
}

@article{Cortes:2005uq,
    author = "Cortes, Vicente and Mayer, Christoph and Mohaupt, Thomas and Saueressig, Frank",
    title = "{Special geometry of euclidean supersymmetry. II. Hypermultiplets and the c-map}",
    eprint = "hep-th/0503094",
    archivePrefix = "arXiv",
    reportNumber = "ITP-05-06, SPIN-05-04, FSU-TPI-02-05",
    doi = "10.1088/1126-6708/2005/06/025",
    journal = "JHEP",
    volume = "06",
    pages = "025",
    year = "2005"
}

@article{Guica:2007wd,
    author = "Guica, Monica and Strominger, Andrew",
    editor = "Baulieu, Laurent and de Boer, Jan and Douglas, Michael R. and Rabinovici, Eliezer and Vanhove, Pierre and Windey, Paul",
    title = "{Cargese lectures on string theory with eight supercharges}",
    eprint = "0704.3295",
    archivePrefix = "arXiv",
    primaryClass = "hep-th",
    doi = "10.1016/j.nuclphysbps.2007.06.007",
    journal = "Nucl. Phys. B Proc. Suppl.",
    volume = "171",
    pages = "39--68",
    year = "2007"
}

@article{Bobev:2026gir,
    author = "Bobev, Nikolay and Gautason, Fridrik Freyr and van Muiden, Jesse",
    title = "{Holographic Tests of the $\mu$ Ensemble}",
    eprint = "2607.06493",
    archivePrefix = "arXiv",
    primaryClass = "hep-th",
    month = "7",
    year = "2026"
}

@article{BenettiGenolini:2026cyc,
    author = "Benetti Genolini, Pietro and Gaar, Florian and Gauntlett, Jerome P. and Park, Jaeha and Sparks, James",
    title = "{Airy functions from quantum M-theory}",
    eprint = "2607.07255",
    archivePrefix = "arXiv",
    primaryClass = "hep-th",
    month = "7",
    year = "2026"
}

@article{Bobev:2026lvl,
    author = "Bobev, Nikolay and Choi, Sunjin and Hong, Junho and Reys, Valentin",
    title = "{Towards OSV in AdS}",
    eprint = "2606.23893",
    archivePrefix = "arXiv",
    primaryClass = "hep-th",
    month = "6",
    year = "2026"
}

@article{Pioline:2006ni,
    author = "Pioline, Boris",
    title = "{Lectures on black holes, topological strings and quantum attractors}",
    eprint = "hep-th/0607227",
    archivePrefix = "arXiv",
    reportNumber = "LPTENS-06-27",
    doi = "10.1088/0264-9381/23/21/S05",
    journal = "Class. Quant. Grav.",
    volume = "23",
    pages = "S981",
    year = "2006"
}

@book{Lauria:2020rhc,
    author = "Lauria, Edoardo and Van Proeyen, Antoine",
    title = "{${\cal N}=2$ Supergravity in $D=4,5,6$ Dimensions}",
    eprint = "2004.11433",
    archivePrefix = "arXiv",
    primaryClass = "hep-th",
    doi = "10.1007/978-3-030-33757-5",
    isbn = "978-3-030-33755-1, 978-3-030-33757-5",
    volume = "966",
    month = "3",
    year = "2020"
}

@article{Bobev:2020egg,
    author = "Bobev, Nikolay and Charles, Anthony M. and Hristov, Kiril and Reys, Valentin",
    title = "{The Unreasonable Effectiveness of Higher-Derivative Supergravity in AdS$_4$ Holography}",
    eprint = "2006.09390",
    archivePrefix = "arXiv",
    primaryClass = "hep-th",
    doi = "10.1103/PhysRevLett.125.131601",
    journal = "Phys. Rev. Lett.",
    volume = "125",
    number = "13",
    pages = "131601",
    year = "2020"
}

@article{Antoniadis:1993ze,
    author = "Antoniadis, Ignatios and Gava, E. and Narain, K. S. and Taylor, T. R.",
    title = "{Topological amplitudes in string theory}",
    eprint = "hep-th/9307158",
    archivePrefix = "arXiv",
    reportNumber = "NUB-3071, IC-93-202, CPTH-A258-0793",
    doi = "10.1016/0550-3213(94)90617-3",
    journal = "Nucl. Phys. B",
    volume = "413",
    pages = "162--184",
    year = "1994"
}

@article{Bershadsky:1993cx,
    author = "Bershadsky, M. and Cecotti, S. and Ooguri, H. and Vafa, C.",
    title = "{Kodaira-Spencer theory of gravity and exact results for quantum string amplitudes}",
    eprint = "hep-th/9309140",
    archivePrefix = "arXiv",
    reportNumber = "HUTP-93-A025, RIMS-946, SISSA-142-93-EP",
    doi = "10.1007/BF02099774",
    journal = "Commun. Math. Phys.",
    volume = "165",
    pages = "311--428",
    year = "1994"
}

@article{Dyer:2008hb,
    author = "Dyer, Ethan and Hinterbichler, Kurt",
    title = "{Boundary Terms, Variational Principles and Higher Derivative Modified Gravity}",
    eprint = "0809.4033",
    archivePrefix = "arXiv",
    primaryClass = "gr-qc",
    doi = "10.1103/PhysRevD.79.024028",
    journal = "Phys. Rev. D",
    volume = "79",
    pages = "024028",
    year = "2009"
}

@article{Gutowski:2012yb,
    author = "Gutowski, Jan B. and Sabra, W. A.",
    title = "{Euclidean N=2 Supergravity}",
    eprint = "1209.2029",
    archivePrefix = "arXiv",
    primaryClass = "hep-th",
    doi = "10.1016/j.physletb.2012.10.050",
    journal = "Phys. Lett. B",
    volume = "718",
    pages = "610--614",
    year = "2012"
}

@article{Ciceri:2023mjl,
    author = "Ciceri, Axel and Jeon, Imtak and Murthy, Sameer",
    title = "{Localization on AdS$_{3}${\texttimes} S$^{2}$. Part I. The 4d/5d connection in off-shell Euclidean supergravity}",
    eprint = "2301.08084",
    archivePrefix = "arXiv",
    primaryClass = "hep-th",
    doi = "10.1007/JHEP07(2023)218",
    journal = "JHEP",
    volume = "07",
    pages = "218",
    year = "2023"
}

@article{Hegde:2023jmp,
    author = "Hegde, Subramanya and Virmani, Amitabh",
    title = "{Killing spinors for finite temperature Euclidean solutions at the BPS bound}",
    eprint = "2311.09427",
    archivePrefix = "arXiv",
    primaryClass = "hep-th",
    doi = "10.1007/JHEP02(2024)203",
    journal = "JHEP",
    volume = "02",
    pages = "203",
    year = "2024"
}

@article{Hanaki:2007mb,
    author = "Hanaki, Kentaro and Ohashi, Keisuke and Tachikawa, Yuji",
    title = "{Comments on charges and near-horizon data of black rings}",
    eprint = "0704.1819",
    archivePrefix = "arXiv",
    primaryClass = "hep-th",
    reportNumber = "TIT-HEP-570",
    doi = "10.1088/1126-6708/2007/12/057",
    journal = "JHEP",
    volume = "12",
    pages = "057",
    year = "2007"
}

@article{deWit:2017cle,
    author = "de Wit, Bernard and Reys, Valentin",
    title = "{Euclidean supergravity}",
    eprint = "1706.04973",
    archivePrefix = "arXiv",
    primaryClass = "hep-th",
    doi = "10.1007/JHEP12(2017)011",
    journal = "JHEP",
    volume = "12",
    pages = "011",
    year = "2017"
}

@article{Mohaupt:2000mj,
    author = "Mohaupt, Thomas",
    title = "{Black hole entropy, special geometry and strings}",
    eprint = "hep-th/0007195",
    archivePrefix = "arXiv",
    doi = "10.1002/1521-3978(200102)49:1/3<3::AID-PROP3>3.0.CO;2-#",
    journal = "Fortsch. Phys.",
    volume = "49",
    pages = "3--161",
    year = "2001"
}

@article{Boruch:2023gfn,
    author = "Boruch, Jan and Iliesiu, Luca V. and Murthy, Sameer and Turiaci, Gustavo J.",
    title = "{New forms of attraction: attractor saddles for the black hole index}",
    eprint = "2310.07763",
    archivePrefix = "arXiv",
    primaryClass = "hep-th",
    doi = "10.1007/JHEP04(2025)087",
    journal = "JHEP",
    volume = "04",
    pages = "087",
    year = "2025"
}

@article{Bobev:2021oku,
    author = "Bobev, Nikolay and Charles, Anthony M. and Hristov, Kiril and Reys, Valentin",
    title = "{Higher-derivative supergravity, AdS$_{4}$ holography, and black holes}",
    eprint = "2106.04581",
    archivePrefix = "arXiv",
    primaryClass = "hep-th",
    doi = "10.1007/JHEP08(2021)173",
    journal = "JHEP",
    volume = "08",
    pages = "173",
    year = "2021"
}

@article{Hawking:1995ap,
    author = "Hawking, S. W. and Ross, Simon F.",
    title = "{Duality between electric and magnetic black holes}",
    eprint = "hep-th/9504019",
    archivePrefix = "arXiv",
    reportNumber = "DAMTP-R-95-8",
    doi = "10.1103/PhysRevD.52.5865",
    journal = "Phys. Rev. D",
    volume = "52",
    pages = "5865--5876",
    year = "1995"
}

@article{Adhikari:2025eok,
    author = "Adhikari, Soumya and Bhattacharjee, Abhinava and Virmani, Amitabh",
    title = "{BPS solutions of 4d Euclidean $ \mathcal{N}=2 $ supergravity with higher derivative interactions}",
    eprint = "2511.18771",
    archivePrefix = "arXiv",
    primaryClass = "hep-th",
    doi = "10.1007/JHEP03(2026)136",
    journal = "JHEP",
    volume = "03",
    pages = "136",
    year = "2026"
}

@article{Beasley:2006us,
    author = "Beasley, Chris and Gaiotto, Davide and Guica, Monica and Huang, Lisa and Strominger, Andrew and Yin, Xi",
    title = "{Why Z(BH) = |Z(top)|**2}",
    eprint = "hep-th/0608021",
    archivePrefix = "arXiv",
    month = "8",
    year = "2006"
}

@article{Beasley:2005iu,
    author = "Beasley, Chris and Witten, Edward",
    title = "{New instanton effects in string theory}",
    eprint = "hep-th/0512039",
    archivePrefix = "arXiv",
    reportNumber = "PUPT-2171",
    doi = "10.1088/1126-6708/2006/02/060",
    journal = "JHEP",
    volume = "02",
    pages = "060",
    year = "2006"
}

@article{BenettiGenolini:2026qdm,
    author = "Benetti Genolini, Pietro and Gaar, Florian and Gauntlett, Jerome P. and Sparks, James",
    title = "{Equivariant localization for higher derivative supergravity}",
    eprint = "2604.08656",
    archivePrefix = "arXiv",
    primaryClass = "hep-th",
    month = "4",
    year = "2026"
}

@article{BenettiGenolini:2024lbj,
    author = {Benetti Genolini, Pietro and Gauntlett, Jerome P. and Jiao, Yusheng and L{\"u}scher, Alice and Sparks, James},
    title = "{Equivariant localization for D = 4 gauged supergravity}",
    eprint = "2412.07828",
    archivePrefix = "arXiv",
    primaryClass = "hep-th",
    doi = "10.1007/JHEP08(2025)211",
    journal = "JHEP",
    volume = "08",
    pages = "211",
    year = "2025"
}

@article{BenettiGenolini:2019jdz,
    author = "Benetti Genolini, Pietro and Perez Ipi{\~n}a, Juan Manuel and Sparks, James",
    title = "{Localization of the action in AdS/CFT}",
    eprint = "1906.11249",
    archivePrefix = "arXiv",
    primaryClass = "hep-th",
    doi = "10.1007/JHEP10(2019)252",
    journal = "JHEP",
    volume = "10",
    pages = "252",
    year = "2019"
}

@article{Boruch:2025biv,
    author = "Boruch, Jan and Iliesiu, Luca V. and Murthy, Sameer and Turiaci, Gustavo J.",
    title = "{Multicentered black hole saddles for supersymmetric indices}",
    eprint = "2507.07166",
    archivePrefix = "arXiv",
    primaryClass = "hep-th",
    month = "7",
    year = "2025"
}

@article{Fradkin:1985ys,
    author = "Fradkin, E. S. and Tseytlin, Arkady A.",
    title = "{Quantum String Theory Effective Action}",
    doi = "10.1016/0550-3213(85)90559-0",
    journal = "Nucl. Phys. B",
    volume = "261",
    pages = "1--27",
    year = "1985",
    note = "[Erratum: Nucl.Phys.B 269, 745--745 (1986)]"
}

@article{Fradkin:1985fq,
    author = "Fradkin, E. S. and Tseytlin, Arkady A.",
    title = "{Effective Action Approach to Superstring Theory}",
    reportNumber = "LEBEDEV-85-150",
    doi = "10.1016/0370-2693(85)91468-6",
    journal = "Phys. Lett. B",
    volume = "160",
    pages = "69--76",
    year = "1985"
}

@inproceedings{vanMuiden:2026nsp,
    author = "van Muiden, Jesse",
    title = "{Quantum M2-branes and Holography}",
    eprint = "2603.14544",
    archivePrefix = "arXiv",
    primaryClass = "hep-th",
    month = "3",
    year = "2026"
}

@article{Gautason:2025per,
    author = "Gautason, Fridrik Freyr and van Muiden, Jesse",
    title = "{Localization of the M2-Brane}",
    eprint = "2503.16597",
    archivePrefix = "arXiv",
    primaryClass = "hep-th",
    doi = "10.1103/67bh-xd42",
    journal = "Phys. Rev. Lett.",
    volume = "135",
    number = "10",
    pages = "101601",
    year = "2025"
}

@article{Genolini:2021urf,
    author = "Genolini, Pietro Benetti and Richmond, Paul",
    title = "{Supersymmetry of higher-derivative supergravity in AdS4 holography}",
    eprint = "2107.04590",
    archivePrefix = "arXiv",
    primaryClass = "hep-th",
    doi = "10.1103/PhysRevD.104.L061902",
    journal = "Phys. Rev. D",
    volume = "104",
    number = "6",
    pages = "L061902",
    year = "2021"
}

@article{Hartle:1972ya,
    author = "Hartle, J. B. and Hawking, S. W.",
    title = "{Solutions of the Einstein-Maxwell equations with many black holes}",
    doi = "10.1007/BF01645696",
    journal = "Commun. Math. Phys.",
    volume = "26",
    pages = "87--101",
    year = "1972"
}

@article{Gibbons:1976ue,
    author = "Gibbons, G. W. and Hawking, S. W.",
    title = "{Action Integrals and Partition Functions in Quantum Gravity}",
    reportNumber = "PRINT-76-0995 (CAMBRIDGE)",
    doi = "10.1103/PhysRevD.15.2752",
    journal = "Phys. Rev. D",
    volume = "15",
    pages = "2752--2756",
    year = "1977"
}

@article{Almheiri:2019qdq,
    author = "Almheiri, Ahmed and Hartman, Thomas and Maldacena, Juan and Shaghoulian, Edgar and Tajdini, Amirhossein",
    title = "{Replica Wormholes and the Entropy of Hawking Radiation}",
    eprint = "1911.12333",
    archivePrefix = "arXiv",
    primaryClass = "hep-th",
    doi = "10.1007/JHEP05(2020)013",
    journal = "JHEP",
    volume = "05",
    pages = "013",
    year = "2020"
}

@article{Penington:2019kki,
    author = "Penington, Geoff and Shenker, Stephen H. and Stanford, Douglas and Yang, Zhenbin",
    title = "{Replica wormholes and the black hole interior}",
    eprint = "1911.11977",
    archivePrefix = "arXiv",
    primaryClass = "hep-th",
    doi = "10.1007/JHEP03(2022)205",
    journal = "JHEP",
    volume = "03",
    pages = "205",
    year = "2022"
}

@article{Sen:2005wa,
      author         = "Sen, Ashoke",
      title          = "{Black hole entropy function and the attractor mechanism
                        in higher derivative gravity}",
      journal        = "JHEP",
      volume         = "09",
      year           = "2005",
      pages          = "038",
      doi            = "10.1088/1126-6708/2005/09/038",
      eprint         = "hep-th/0506177",
      archivePrefix  = "arXiv",
      primaryClass   = "hep-th",
      SLACcitation   = "%%CITATION = HEP-TH/0506177;%%"
}

@article{Braden:1990hw,
      author         = "Braden, Harry W. and Brown, J. David and Whiting, Bernard
                        F. and York, Jr., James W.",
      title          = "{Charged black hole in a grand canonical ensemble}",
      journal        = "Phys. Rev.",
      volume         = "D42",
      year           = "1990",
      pages          = "3376-3385",
      doi            = "10.1103/PhysRevD.42.3376",
      reportNumber   = "PRINT-90-0461 (NORTH-CAROLINA)",
      SLACcitation   = "%%CITATION = PHRVA,D42,3376;%%"
}

@inproceedings{Sparks:2026taw,
    author = "Sparks, James",
    title = "{Localization in supergravity}",
    eprint = "2602.21782",
    archivePrefix = "arXiv",
    primaryClass = "hep-th",
    month = "2",
    year = "2026"
}

@article{Iliesiu:2020qvm,
    author = "Iliesiu, Luca V. and Turiaci, Gustavo J.",
    title = "{The statistical mechanics of near-extremal black holes}",
    eprint = "2003.02860",
    archivePrefix = "arXiv",
    primaryClass = "hep-th",
    doi = "10.1007/JHEP05(2021)145",
    journal = "JHEP",
    volume = "05",
    pages = "145",
    year = "2021"
}

@article{Saad:2019lba,
      author         = "Saad, Phil and Shenker, Stephen H. and Stanford, Douglas",
      title          = "{JT gravity as a matrix integral}",
      year           = "2019",
      eprint         = "1903.11115",
      archivePrefix  = "arXiv",
      primaryClass   = "hep-th",
      SLACcitation   = "%%CITATION = ARXIV:1903.11115;%%"
}

@article{H:2023qko,
    author = "H., Anupam A. and Athira, P. V. and Chowdhury, Chandramouli and Sen, Ashoke",
    title = "{Logarithmic correction to BPS black hole entropy from supersymmetric index at finite temperature}",
    eprint = "2306.07322",
    archivePrefix = "arXiv",
    primaryClass = "hep-th",
    doi = "10.1007/JHEP03(2024)095",
    journal = "JHEP",
    volume = "03",
    pages = "095",
    year = "2024"
}

@article{Iliesiu:2021are,
    author = "Iliesiu, Luca V. and Kologlu, Murat and Turiaci, Gustavo J.",
    title = "{Supersymmetric indices factorize}",
    eprint = "2107.09062",
    archivePrefix = "arXiv",
    primaryClass = "hep-th",
    doi = "10.1007/JHEP05(2023)032",
    journal = "JHEP",
    volume = "05",
    pages = "032",
    year = "2023"
}

@article{Anupam:2023yns,
    author = "Anupam, A. H. and Chowdhury, Chandramouli and Sen, Ashoke",
    title = "{Revisiting logarithmic correction to five dimensional BPS black hole entropy}",
    eprint = "2308.00038",
    archivePrefix = "arXiv",
    primaryClass = "hep-th",
    doi = "10.1007/JHEP05(2024)070",
    journal = "JHEP",
    volume = "05",
    pages = "070",
    year = "2024"
}

@article{Denef:2000nb,
    author = "Denef, Frederik",
    title = "{Supergravity flows and D-brane stability}",
    eprint = "hep-th/0005049",
    archivePrefix = "arXiv",
    doi = "10.1088/1126-6708/2000/08/050",
    journal = "JHEP",
    volume = "08",
    pages = "050",
    year = "2000"
}

@article{Bates:2003vx,
    author = "Bates, Brandon and Denef, Frederik",
    title = "{Exact solutions for supersymmetric stationary black hole composites}",
    eprint = "hep-th/0304094",
    archivePrefix = "arXiv",
    reportNumber = "RUNHETC-2003-10",
    doi = "10.1007/JHEP11(2011)127",
    journal = "JHEP",
    volume = "11",
    pages = "127",
    year = "2011"
}

@article{Denef:2007vg,
    author = "Denef, Frederik and Moore, Gregory W.",
    title = "{Split states, entropy enigmas, holes and halos}",
    eprint = "hep-th/0702146",
    archivePrefix = "arXiv",
    doi = "10.1007/JHEP11(2011)129",
    journal = "JHEP",
    volume = "11",
    pages = "129",
    year = "2011"
}

@article{Ooguri:2004zv,
    author = "Ooguri, Hirosi and Strominger, Andrew and Vafa, Cumrun",
    title = "{Black hole attractors and the topological string}",
    eprint = "hep-th/0405146",
    archivePrefix = "arXiv",
    reportNumber = "HUTP-04-A020, CALT-68-2501",
    doi = "10.1103/PhysRevD.70.106007",
    journal = "Phys. Rev. D",
    volume = "70",
    pages = "106007",
    year = "2004"
}

@article{Sen:2009vz,
    author = "Sen, Ashoke",
    title = "{Arithmetic of Quantum Entropy Function}",
    eprint = "0903.1477",
    archivePrefix = "arXiv",
    primaryClass = "hep-th",
    doi = "10.1088/1126-6708/2009/08/068",
    journal = "JHEP",
    volume = "08",
    pages = "068",
    year = "2009"
}

@article{Sen:2008vm,
      author         = "Sen, Ashoke",
      title          = "{Quantum Entropy Function from AdS(2)/CFT(1)
                        Correspondence}",
      booktitle      = "{International conference on string theory (Strings 2008)
                        Geneva, Switzerland, August 18-23, 2008}",
      journal        = "Int. J. Mod. Phys.",
      volume         = "A24",
      year           = "2009",
      pages          = "4225-4244",
      doi            = "10.1142/S0217751X09045893",
      eprint         = "0809.3304",
      archivePrefix  = "arXiv",
      primaryClass   = "hep-th",
      SLACcitation   = "%%CITATION = ARXIV:0809.3304;%%"
}

@article{Aganagic:2004js,
      author         = "Aganagic, Mina and Ooguri, Hirosi and Saulina, Natalia
                        and Vafa, Cumrun",
      title          = "{Black holes, q-deformed 2d Yang-Mills, and
                        non-perturbative topological strings}",
      journal        = "Nucl. Phys.",
      volume         = "B715",
      year           = "2005",
      pages          = "304-348",
      doi            = "10.1016/j.nuclphysb.2005.02.035",
      eprint         = "hep-th/0411280",
      archivePrefix  = "arXiv",
      primaryClass   = "hep-th",
      reportNumber   = "CALT-68-2529, HUTP-04-A0049, UCB-PTH-04-33",
      SLACcitation   = "%%CITATION = HEP-TH/0411280;%%"
}

@article{Strominger:1996sh,
    author = "Strominger, Andrew and Vafa, Cumrun",
    title = "{Microscopic origin of the Bekenstein-Hawking entropy}",
    eprint = "hep-th/9601029",
    archivePrefix = "arXiv",
    reportNumber = "HUTP-96-A002, RU-96-01",
    doi = "10.1016/0370-2693(96)00345-0",
    journal = "Phys. Lett. B",
    volume = "379",
    pages = "99--104",
    year = "1996"
}

@book{Freedman:2012zz,
    author = "Freedman, Daniel Z. and Van Proeyen, Antoine",
    title = "{Supergravity}",
    isbn = "978-1-139-36806-3, 978-0-521-19401-3",
    publisher = "Cambridge Univ. Press",
    address = "Cambridge, UK",
    month = "5",
    year = "2012"
}

@article{Ferrara:1995ih,
    author = "Ferrara, Sergio and Kallosh, Renata and Strominger, Andrew",
    title = "{N=2 extremal black holes}",
    eprint = "hep-th/9508072",
    archivePrefix = "arXiv",
    reportNumber = "CERN-TH-95-211, SU-ITP-95-16",
    doi = "10.1103/PhysRevD.52.R5412",
    journal = "Phys. Rev. D",
    volume = "52",
    pages = "5412--5416",
    year = "1995"
}

@article{Maldacena:1997de,
    author = "Maldacena, Juan Martin and Strominger, Andrew and Witten, Edward",
    title = "{Black hole entropy in M theory}",
    eprint = "hep-th/9711053",
    archivePrefix = "arXiv",
    doi = "10.1088/1126-6708/1997/12/002",
    journal = "JHEP",
    volume = "12",
    pages = "002",
    year = "1997"
}

@article{Sen:2023dps,
    author = "Sen, Ashoke",
    title = "{Revisiting localization for BPS black hole entropy}",
    eprint = "2302.13490",
    archivePrefix = "arXiv",
    primaryClass = "hep-th",
    month = "2",
    year = "2023"
}

@article{Iliesiu:2022kny,
    author = "Iliesiu, Luca V. and Murthy, Sameer and Turiaci, Gustavo J.",
    title = "{Black hole microstate counting from the gravitational path integral}",
    eprint = "2209.13602",
    archivePrefix = "arXiv",
    primaryClass = "hep-th",
    doi = "10.1007/JHEP08(2025)152",
    journal = "JHEP",
    volume = "08",
    pages = "152",
    year = "2025"
}

@article{Dabholkar:2011ec,
    author = "Dabholkar, Atish and Gomes, Joao and Murthy, Sameer",
    title = "{Localization {\&} Exact Holography}",
    eprint = "1111.1161",
    archivePrefix = "arXiv",
    primaryClass = "hep-th",
    doi = "10.1007/JHEP04(2013)062",
    journal = "JHEP",
    volume = "04",
    pages = "062",
    year = "2013"
}

@article{Dabholkar:2010uh,
    author = "Dabholkar, Atish and Gomes, Joao and Murthy, Sameer",
    title = "{Quantum black holes, localization and the topological string}",
    eprint = "1012.0265",
    archivePrefix = "arXiv",
    primaryClass = "hep-th",
    doi = "10.1007/JHEP06(2011)019",
    journal = "JHEP",
    volume = "06",
    pages = "019",
    year = "2011"
}

@article{Maldacena:1999bp,
    author = "Maldacena, Juan Martin and Moore, Gregory W. and Strominger, Andrew",
    title = "{Counting BPS black holes in toroidal Type II string theory}",
    eprint = "hep-th/9903163",
    archivePrefix = "arXiv",
    reportNumber = "IASSNS-HEP-99-16, HUTP-99-A006, YCTP-P5-99",
    month = "3",
    year = "1999"
}

@article{Cotler:2016fpe,
    author = "Cotler, Jordan S. and Gur-Ari, Guy and Hanada, Masanori and Polchinski, Joseph and Saad, Phil and Shenker, Stephen H. and Stanford, Douglas and Streicher, Alexandre and Tezuka, Masaki",
    title = "{Black Holes and Random Matrices}",
    eprint = "1611.04650",
    archivePrefix = "arXiv",
    primaryClass = "hep-th",
    reportNumber = "SU-ITP-16-19, SU-ITP-16/19, YITP-16-124",
    doi = "10.1007/JHEP05(2017)118",
    journal = "JHEP",
    volume = "05",
    pages = "118",
    year = "2017",
    note = "[Erratum: JHEP 09, 002 (2018)]"
}

@article{Balasubramanian:2022gmo,
    author = "Balasubramanian, Vijay and Lawrence, Albion and Magan, Javier M. and Sasieta, Martin",
    title = "{Microscopic Origin of the Entropy of Black Holes in General Relativity}",
    eprint = "2212.02447",
    archivePrefix = "arXiv",
    primaryClass = "hep-th",
    doi = "10.1103/PhysRevX.14.011024",
    journal = "Phys. Rev. X",
    volume = "14",
    number = "1",
    pages = "011024",
    year = "2024"
}

@article{Kiritsis:1997gu,
    author = "Kiritsis, Elias",
    editor = "Falomir, H. and Gamboa Saravi, R. E. and Schaposnik, F. A.",
    title = "{Introduction to nonperturbative string theory}",
    eprint = "hep-th/9708130",
    archivePrefix = "arXiv",
    reportNumber = "CERN-TH-97-217",
    doi = "10.1063/1.54695",
    journal = "AIP Conf. Proc.",
    volume = "419",
    number = "1",
    pages = "265--308",
    year = "1998"
}

@article{Bachas:1997mc,
    author = "Bachas, C. and Fabre, C. and Kiritsis, E. and Obers, N. A. and Vanhove, Pierre",
    title = "{Heterotic / type I duality and D-brane instantons}",
    eprint = "hep-th/9707126",
    archivePrefix = "arXiv",
    reportNumber = "CERN-TH-97-065, CERN-TH-97-65, CPTH-S500-0497",
    doi = "10.1016/S0550-3213(97)00639-1",
    journal = "Nucl. Phys. B",
    volume = "509",
    pages = "33--52",
    year = "1998"
}

@article{Strominger:1996kf,
    author = "Strominger, Andrew",
    title = "{Macroscopic entropy of N=2 extremal black holes}",
    eprint = "hep-th/9602111",
    archivePrefix = "arXiv",
    doi = "10.1016/0370-2693(96)00711-3",
    journal = "Phys. Lett. B",
    volume = "383",
    pages = "39--43",
    year = "1996"
}

@article{Ferrara:1996dd,
    author = "Ferrara, Sergio and Kallosh, Renata",
    title = "{Supersymmetry and attractors}",
    eprint = "hep-th/9602136",
    archivePrefix = "arXiv",
    reportNumber = "UCLA-96-TEP-8, CERN-TH-96-53, SU-ITP-96-8",
    doi = "10.1103/PhysRevD.54.1514",
    journal = "Phys. Rev. D",
    volume = "54",
    pages = "1514--1524",
    year = "1996"
}

@article{Gopakumar:1998ii,
    author = "Gopakumar, Rajesh and Vafa, Cumrun",
    title = "{M theory and topological strings. 1.}",
    eprint = "hep-th/9809187",
    archivePrefix = "arXiv",
    reportNumber = "HUTP-98-A069",
    month = "9",
    year = "1998"
}

@article{Gopakumar:1998jq,
    author = "Gopakumar, Rajesh and Vafa, Cumrun",
    title = "{M theory and topological strings. 2.}",
    eprint = "hep-th/9812127",
    archivePrefix = "arXiv",
    reportNumber = "HUTP-98-A070",
    month = "12",
    year = "1998"
}

@article{LopesCardoso:1998tkj,
    author = "Lopes Cardoso, Gabriel and de Wit, Bernard and Mohaupt, Thomas",
    title = "{Corrections to macroscopic supersymmetric black hole entropy}",
    eprint = "hep-th/9812082",
    archivePrefix = "arXiv",
    reportNumber = "SPIN-1998-11, THU-98-44",
    doi = "10.1016/S0370-2693(99)00227-0",
    journal = "Phys. Lett. B",
    volume = "451",
    pages = "309--316",
    year = "1999"
}

@article{LopesCardoso:1999cv,
    author = "Lopes Cardoso, Gabriel and de Wit, Bernard and Mohaupt, Thomas",
    editor = "Lust, D. and Otto, H. J.",
    title = "{Deviations from the area law for supersymmetric black holes}",
    eprint = "hep-th/9904005",
    archivePrefix = "arXiv",
    reportNumber = "THU-99-09",
    journal = "Fortsch. Phys.",
    volume = "48",
    pages = "49--64",
    year = "2000"
}

@article{LopesCardoso:1999fsj,
    author = "Lopes Cardoso, Gabriel and de Wit, Bernard and Mohaupt, Thomas",
    title = "{Macroscopic entropy formulae and nonholomorphic corrections for supersymmetric black holes}",
    eprint = "hep-th/9906094",
    archivePrefix = "arXiv",
    reportNumber = "AEI-111, THU-99-11",
    doi = "10.1016/S0550-3213(99)00560-X",
    journal = "Nucl. Phys. B",
    volume = "567",
    pages = "87--110",
    year = "2000"
}

@article{deWit:1980lyi,
    author = "de Wit, B. and van Holten, J. W. and Van Proeyen, Antoine",
    title = "{Structure of N=2 Supergravity}",
    reportNumber = "CERN-TH-2985",
    doi = "10.1016/0550-3213(83)90548-5",
    journal = "Nucl. Phys. B",
    volume = "184",
    pages = "77",
    year = "1981",
    note = "[Erratum: Nucl.Phys.B 222, 516 (1983)]"
}

@article{Bergshoeff:1980is,
    author = "Bergshoeff, E. and de Roo, M. and de Wit, B.",
    title = "{Extended Conformal Supergravity}",
    reportNumber = "NIKHEF-H/80-07",
    doi = "10.1016/0550-3213(81)90465-X",
    journal = "Nucl. Phys. B",
    volume = "182",
    pages = "173--204",
    year = "1981"
}

@article{deWit:1984rvr,
    author = "de Wit, B. and Lauwers, P. G. and Van Proeyen, Antoine",
    title = "{Lagrangians of N=2 Supergravity - Matter Systems}",
    reportNumber = "BONN-HE-84-37",
    doi = "10.1016/0550-3213(85)90154-3",
    journal = "Nucl. Phys. B",
    volume = "255",
    pages = "569--608",
    year = "1985"
}

@article{deWit:1999fp,
    author = "de Wit, Bernard and Kleijn, Bas and Vandoren, Stefan",
    title = "{Superconformal hypermultiplets}",
    eprint = "hep-th/9909228",
    archivePrefix = "arXiv",
    reportNumber = "THU-99-23, AEI-1999-18, SWAT-99-239, YITP-99-51",
    doi = "10.1016/S0550-3213(99)00726-9",
    journal = "Nucl. Phys. B",
    volume = "568",
    pages = "475--502",
    year = "2000"
}

@article{deWit:1998zg,
    author = "de Wit, Bernard and Kleijn, Bas and Vandoren, Stefan",
    editor = "Wess, J.",
    title = "{Rigid N=2 superconformal hypermultiplets}",
    eprint = "hep-th/9808160",
    archivePrefix = "arXiv",
    reportNumber = "THU-98-30",
    doi = "10.1007/BFb0104585",
    journal = "Lect. Notes Phys.",
    volume = "524",
    pages = "37",
    year = "1999"
}

@article{York:1972sj,
    author = "York, Jr., James W.",
    title = "{Role of conformal three geometry in the dynamics of gravitation}",
    doi = "10.1103/PhysRevLett.28.1082",
    journal = "Phys. Rev. Lett.",
    volume = "28",
    pages = "1082--1085",
    year = "1972"
}

@article{Israel:1972vx,
    author = "Israel, W. and Wilson, G. A.",
    title = "{A class of stationary electromagnetic vacuum fields}",
    doi = "10.1063/1.1666066",
    journal = "J. Math. Phys.",
    volume = "13",
    pages = "865--871",
    year = "1972"
}

@article{Perjes:1971gv,
    author = "Perjes, Zoltan",
    title = "{Solutions of the coupled Einstein Maxwell equations representing the fields of spinning sources}",
    doi = "10.1103/PhysRevLett.27.1668",
    journal = "Phys. Rev. Lett.",
    volume = "27",
    pages = "1668",
    year = "1971"
}

@article{Behrndt:1997ny,
    author = "Behrndt, Klaus and Lust, Dieter and Sabra, Wafic A.",
    title = "{Stationary solutions of N=2 supergravity}",
    eprint = "hep-th/9705169",
    archivePrefix = "arXiv",
    reportNumber = "HUB-EP-97-30",
    doi = "10.1016/S0550-3213(97)00633-0",
    journal = "Nucl. Phys. B",
    volume = "510",
    pages = "264--288",
    year = "1998"
}

@article{Gibbons:1979xm,
    author = "Gibbons, G. W. and Hawking, S. W.",
    title = "{Classification of Gravitational Instanton Symmetries}",
    doi = "10.1007/BF01197189",
    journal = "Commun. Math. Phys.",
    volume = "66",
    pages = "291--310",
    year = "1979"
}

@article{Wald:1993nt,
    author = "Wald, Robert M.",
    title = "{Black hole entropy is the Noether charge}",
    eprint = "gr-qc/9307038",
    archivePrefix = "arXiv",
    reportNumber = "EFI-93-42",
    doi = "10.1103/PhysRevD.48.R3427",
    journal = "Phys. Rev. D",
    volume = "48",
    number = "8",
    pages = "R3427--R3431",
    year = "1993"
}

@article{Iyer:1994ys,
    author = "Iyer, Vivek and Wald, Robert M.",
    title = "{Some properties of Noether charge and a proposal for dynamical black hole entropy}",
    eprint = "gr-qc/9403028",
    archivePrefix = "arXiv",
    doi = "10.1103/PhysRevD.50.846",
    journal = "Phys. Rev. D",
    volume = "50",
    pages = "846--864",
    year = "1994"
}

@article{Dabholkar:2010rm,
    author = "Dabholkar, Atish and Gomes, Joao and Murthy, Sameer and Sen, Ashoke",
    title = "{Supersymmetric Index from Black Hole Entropy}",
    eprint = "1009.3226",
    archivePrefix = "arXiv",
    primaryClass = "hep-th",
    reportNumber = "TIFR-TH-10-27",
    doi = "10.1007/JHEP04(2011)034",
    journal = "JHEP",
    volume = "04",
    pages = "034",
    year = "2011"
}

@article{Marino:1998pg,
    author = "Marino, Marcos and Moore, Gregory W.",
    title = "{Counting higher genus curves in a Calabi-Yau manifold}",
    eprint = "hep-th/9808131",
    archivePrefix = "arXiv",
    reportNumber = "YCPT-P23-98, YCPT-23",
    doi = "10.1016/S0550-3213(98)00847-5",
    journal = "Nucl. Phys. B",
    volume = "543",
    pages = "592--614",
    year = "1999"
}

@article{Faber:1998gsw,
    author = "Faber, C. and Pandharipande, R.",
    title = "{Hodge integrals and Gromov-Witten theory}",
    eprint = "math/9810173",
    archivePrefix = "arXiv",
    doi = "10.1007/s002229900028",
    journal = "Inventiones mathematicae",
    volume = "139",
    number = "1",
    pages = "173--199",
    year = "2000"
}

@article{Dabholkar:2005dt,
    author = "Dabholkar, Atish and Denef, Frederik and Moore, Gregory W. and Pioline, Boris",
    title = "{Precision counting of small black holes}",
    eprint = "hep-th/0507014",
    archivePrefix = "arXiv",
    reportNumber = "LPTHE-05-14, LPTENS-05-21, TIFR-TH-05-27",
    doi = "10.1088/1126-6708/2005/10/096",
    journal = "JHEP",
    volume = "10",
    pages = "096",
    year = "2005"
}

@article{Dabholkar:2005by,
    author = "Dabholkar, Atish and Denef, Frederik and Moore, Gregory W. and Pioline, Boris",
    title = "{Exact and asymptotic degeneracies of small black holes}",
    eprint = "hep-th/0502157",
    archivePrefix = "arXiv",
    reportNumber = "LPTHE-05-04, LPTENS-05-09, TIFR-TH-05-07",
    doi = "10.1088/1126-6708/2005/08/021",
    journal = "JHEP",
    volume = "08",
    pages = "021",
    year = "2005"
}

@article{Pestun:2007rz,
    author = "Pestun, Vasily",
    title = "{Localization of gauge theory on a four-sphere and supersymmetric Wilson loops}",
    eprint = "0712.2824",
    archivePrefix = "arXiv",
    primaryClass = "hep-th",
    reportNumber = "ITEP-TH-41-07, PUTP-2248",
    doi = "10.1007/s00220-012-1485-0",
    journal = "Commun. Math. Phys.",
    volume = "313",
    pages = "71--129",
    year = "2012"
}

@article{Jeon:2018kec,
    author = "Jeon, Imtak and Murthy, Sameer",
    title = "{Twisting and localization in supergravity: equivariant cohomology of BPS black holes}",
    eprint = "1806.04479",
    archivePrefix = "arXiv",
    primaryClass = "hep-th",
    doi = "10.1007/JHEP03(2019)140",
    journal = "JHEP",
    volume = "03",
    pages = "140",
    year = "2019"
}

@article{Dabholkar:2004yr,
    author = "Dabholkar, Atish",
    title = "{Exact counting of black hole microstates}",
    eprint = "hep-th/0409148",
    archivePrefix = "arXiv",
    reportNumber = "SLAC-PUB-10717, SU-ITP-04-36, TIFR-TH-04-23",
    doi = "10.1103/PhysRevLett.94.241301",
    journal = "Phys. Rev. Lett.",
    volume = "94",
    pages = "241301",
    year = "2005"
}

@article{Dabholkar:2004dq,
    author = "Dabholkar, Atish and Kallosh, Renata and Maloney, Alexander",
    title = "{A Stringy cloak for a classical singularity}",
    eprint = "hep-th/0410076",
    archivePrefix = "arXiv",
    reportNumber = "SLAC-PUB-10790, SU-ITP-04-38, TIFR-TH-04-27",
    doi = "10.1088/1126-6708/2004/12/059",
    journal = "JHEP",
    volume = "12",
    pages = "059",
    year = "2004"
}

@article{Sen:2004dp,
    author = "Sen, Ashoke",
    title = "{How does a fundamental string stretch its horizon?}",
    eprint = "hep-th/0411255",
    archivePrefix = "arXiv",
    doi = "10.1088/1126-6708/2005/05/059",
    journal = "JHEP",
    volume = "05",
    pages = "059",
    year = "2005"
}

\end{document}